\documentclass[a4paper, 12pt]{article}
\usepackage[T1]{fontenc}
\usepackage[latin1]{inputenc}
\usepackage{appendix}
\usepackage{amsmath}
\usepackage{amsthm}
\usepackage{authblk}
\usepackage{bm}

\newtheorem{lemma}{Lemma}
\newtheorem{proposition}{Proposition}
\newtheorem{corollary}{Corollary}
\newtheorem*{definition}{Definition}
\newtheorem{theorem}{Theorem}
\newtheorem{observation}{Observation}

\usepackage{natbib}
\usepackage{amsfonts}
\usepackage{amsmath}
\usepackage{amssymb}
\usepackage[top=3cm,bottom=3cm,left=3cm,right=3cm]{geometry}
\usepackage{setspace}
\usepackage{xcolor}
\usepackage{graphicx}
\usepackage{authblk}
\usepackage{hyperref}
\usepackage{comment}
\usepackage{subfig}
\usepackage{tikz}
\usepackage[normalem]{ulem}
\usetikzlibrary{decorations.pathreplacing,angles,quotes,calligraphy}
\makeatletter
\renewcommand\@biblabel[1]{}

\makeatother

\begin{document}

\title{Optimally Biased Expertise\footnote{We are grateful to Jean-Michel Benkert, Gregorio Curello, Mogens Fosgerau, Alessandro Ispano, Ole Jann, Navin Kartik, Eugen Kov\'{a}\v{c}, Matthias Lang, Stephan Lauermann, Filip Mat\v{e}jka, Jeffrey Mensch, Meg Meyer, Fabio Michelucci, Volker Nocke, Nicolas Schutz, Ludvig Sinander, Peter Norman S{\o}rensen, Allen Vong, Jan Z{\'a}pal, seminar participants at Bern, Copenhagen, EPFL, CERGE-EI, NRU HSE, and THEMA Cergy, as well as the participants of 12th Conference on Economic Design, Oligo Workshop 2022, SAET 2022 and YEM 2022 conferences, CBS Microeconomics workshop, and Nordic Theory Group seminar for helpful comments and valuable feedback.
This project has received funding from the European Research Council under the European Union's Horizon 2020 research and innovation programme (grant agreement No. 740369) and from the Grant Agency of Charles University (GAUK project No. 323221). Funding by the German Research Foundation (DFG)
through CRC TR 224 (project B03) and by the Czech Academy of Sciences through the Lumina
Quaeruntur fellowship (LQ300852101) is gratefully acknowledged.  } }
\author[1]{Pavel Ilinov}
\author[2]{Andrei Matveenko}
\author[3]{Maxim Senkov}
\author[4]{Egor Starkov\footnote{Ilinov: pavel.ilinov@epfl.ch; Matveenko: matveenko@uni-mannheim.de; Senkov: msenkov@ub.edu; Starkov: egor.starkov@econ.ku.dk.}}
\affil[1]{Ecole Polytechnique F\'{e}d\'{e}rale de Lausanne, School of Architecture, Civil and Environmental Engineering}
\affil[2]{University of Mannheim, Department of Economics}
\affil[3]{University of Barcelona, Faculty of Economics and Business}
\affil[4]{University of Copenhagen, Department of Economics}

\maketitle

\vspace{-20pt}
\begin{abstract}
	We show that in delegation problems, a principal benefits from belief misalignment vis-{\`a}-vis an agent when the latter can flexibly acquire costly information. The agent optimally succumbs to confirmatory learning, leading him to favor the ex ante optimal action. We show that the principal prefers to mitigate this by hiring an agent who is ex ante more uncertain about which action is optimal. This is optimal even when the principal is herself biased towards some action: the benefit always outweighs the cost of a small misalignment. Optimally misaligned agent considers weakly more actions than an aligned agent. All results continue to hold when delegation is replaced by communication.
	\\
    
    \noindent {\textbf{JEL-codes}}: D82, D83, D91, M51.\\
	
\end{abstract}
\newpage

\section{Introduction}

Why can it be beneficial for a partisan principal to hire an agent with a misaligned vision? At first sight, such a choice looks counterintuitive -- e.g., \citet{holmstrom1980} suggests that misalignment between a principal and an agent leads to a conflict of interest, since from the principal's point of view, the agent then makes suboptimal decisions.\footnote{See \citealp{holmstrom1980,crawford_strategic_1982,prendergast_theory_1993,alonso2008,egorov_dictators_2011,che_pandering_2013} for various other arguments against misalignment. This point is also supported by the empirical evidence: e.g., \citet{bellodi_personnel_2025} provide evidence that U.S. politicians prefer to appoint aligned bureaucrats, conditional on expertise. \citet{malmendier2023managerial} show that overconfident CEOs tend to hire overconfident CFOs. \citet*{hoffman2018discretion} find that inefficiencies in HR managers' hiring decisions can be a result of their biased preferences. \citet{kennedy2016you} presents evidence that the principals take the conflict of interest into account when selecting an expert.} 
However, we show in this paper that misaligned delegation can be optimal when: (1) the principal is ex ante biased towards one of the options at hand, and (2) the agent needs to invest in acquiring or processing costly information about the optimal course of action.

We demonstrate that in such a setting, an aligned agent succumbs to \emph{confirmatory learning}: given the uncertainty over what the best action is, the agent prefers to err on the side of the ex ante optimal action. The principal is harmed by the agent's confirmatory learning and seeks a more uniform distribution of mistakes in the agent's decisions. We find that this can be achieved by hiring a slightly misaligned agent, who is ex ante more uncertain about the optimal course of action and, therefore, is less prone to confirmatory learning.

To show this, we consider a delegation model in which a principal (she) faces a problem of choosing the best alternative: she needs to pick the unique correct action out of $N$ available alternatives. She delegates this decision to an agent (he), who shares the same objective, but has a misaligned prior belief about which option is likely optimal.
The agent does not have any preexisting private knowledge about the case he is asked to consider, but can use his expertise to acquire the information and make the best decision. To model learning, we adopt a model of rational inattention, in which the agent can acquire any signal about the state of the world subject to a cost. The agent's cost of learning is not internalized by the principal, and the principal's own cost of learning is prohibitively high. 

We show that when the principal is ex ante \emph{biased} towards some action (in the sense of having a non-uniform prior belief over which of the actions is optimal), it is optimal for her to hire a \emph{misaligned} agent.
In particular, she benefits the most from delegating to an agent who is ex ante biased towards the same action as she is, but to a smaller extent, making such an agent more uncertain than she is about what the best course of action is (Proposition \ref{prop:binarydelegationri}, Theorem \ref{prop:opt_choice_beliefs}).
The optimal degree of misalignment is non-monotone in the strength of principal's own bias, since an extremely biased principal does not value the agent's information, and the unbiased principal strictly prefers an unbiased agent (Corollary \ref{corr:nonmonmisal}).
These results hold regardless of who has the final decision rights: the optimal delegation strategy is the same whether the principal delegates the decision rights to the agent or merely expects a recommendation on the optimal course of action (Proposition \ref{prop:comm}).

The effect is due to agent's learning being confirmatory. Suppose, for example, that the agent must decide whether to invest in a risky project. Due to the information being costly, the agent does not learn the state perfectly, leaving scope for false positives and false negatives, i.e., investing in a bad project or rejecting a good project, respectively. If the agent believes ex ante that the project has a positive expected payoff, he will choose to err on the side of investing in it. This gives rise to (rational) confirmatory learning: the rate of false positives will be higher than the rate of false negatives.\footnote{\citet{mackowiak2018lack} show in a similar framework that rationally inattentive agents optimally underprepare for events they consider rare.}
The principal with the same prior belief would then find it optimal to reduce the degree of confirmatory learning. She would prefer to hire an agent who is ex ante more uncertain about whether the project has a positive or a negative payoff (but still leans on the positive side). This misalignment is optimal in spite of the principal also favoring false positives over false negatives.

We extend the intuition above to multiple states and actions. We show that the principal in such a setting prefers to hire an agent who shares ex ante her ranking over which states are likely (and which actions are likely to be optimal), but who is more uncertain than she is, in the sense of the agent's prior belief being majorized by that of the principal.

Our conclusion has implications in various settings. One relates to bilateral relationships -- e.g., when an authority in a public organization wants to find the best expert to delegate a decision to. Our findings provide conditions for when delegation can be beneficial, as well as an upper bound for the expected gains from such a delegation. Moreover, we provide a useful \emph{directional} behavioral tool: the authority should look for an expert who shares similar views but is more uncertain or moderate.
Our second interpretation covers large organizations. We take heterogeneous priors as different views of the people in organizations such as research teams, firms, and political parties. Our results speak in favor of supporting a diversity of views in such organizations. We characterize a useful diversity strategy for the leader: she benefits from having workers with slightly more \emph{moderate} views. Although the optimal agent is unique in our model, our setup captures only a singular interaction. For other decision problems, the leader may have different opinions and, therefore, benefits from having workers with different views in the organization.\footnote{\cite{banerjee2001simple} and \cite{li2004delegating} present a counterargument to the benefits of diversity, arguing that if a decision needs to be made by a collective, diverse collectives may have a harder time agreeing on a decision and may produce worse outcomes. For a recent review on diversity in organizations, see \citet*{shore2009diversity}.}

Our paper mainly connects to the literature on delegation. Most papers on delegation follow \cite{holmstrom1980} in assuming that the agent has preexisting private information relevant to the decision. We adopt instead the ``delegated expertise'' setting of \citet{demski_delegated_1987}, where the agent has no information advantage over the principal ex ante, but rather has to collect information, and the expertise grants him a \emph{learning} advantage over the principal.\footnote{\citet*{graham2015capital} show that delegation tends to be used when the decision-making demands more evidence that the delegatee can provide. Alternatively, the choice to delegate a decision is often associated with a volatile environment that a delegator faces \citep{foss2005performance,ekinci2021determinants}, where any knowledge quickly becomes obsolete.} 
\citet{demski_delegated_1987} explore a contracting problem in a setting in which the agent chooses between a finite number of signal structures. \citet{Lindbeck2017} extend this analysis to a rationally inattentive agent (who can acquire any information subject to entropy costs). 
Instead of contracts, our paper explores instead a different tool available to the principal, which is selecting an agent with misaligned prior beliefs. We show in the Online Appendix (Section \ref{sec:othertools}) that this tool can be as powerful as contracting in some settings.

The closest to our paper is work by \citet{dur2005producing} who show in a delegated expertise setting that hiring an agent who is less biased (in the absolute sense, but more misaligned relative to the principal) is beneficial for the principal, since such an agent exerts more effort in learning the state. Our problem is different in two dimensions: first, we have no ex post conflict between the principal and the agent, with the agent having the decision rights and the whole initial conflict being due to the agent's information costs. Second, we allow for flexible information acquisition by the agent and show that in this case, an agent who is misaligned with the principal acquires not more (in the Blackwell sense), but rather \emph{different} information.
We show that the difference in the rates of false positives and false negatives chosen by different agents also leads the principal to select a misaligned agent. In fact, we show that learning flexibility is necessary for our channel to apply (see Online Appendix Section \ref{sec:symm}).
Further, our setup allows us to obtain novel comparative statics results and show that the principal-optimal amount of misalignment is non-monotone in her initial bias.\footnote{We also show that misalignment performs weakly better than other delegation tools like restricting the agent's action set and contracting, and provide novel results establishing the equivalence of misalignment in beliefs and in preferences (see Online Appendix Section \ref{sec:othertools}).}

A separate literature has investigated the delegated expertise problem in settings where the agent needs to communicate their findings to the principal, as opposed to having the final decision rights.
\cite{che2009opinions} show that the need to communicate may incentivize a misaligned agent to acquire more information than an aligned one, in order to more effectively persuade the principal about which action needs to be taken, as well as to avoid the punishment for concealing evidence. As we show in Section \ref{sec:comm}, both channels they identify (persuasion and prejudice avoidance) are absent from our model, so our argument regarding the desirability of misalignment is completely separate from theirs.
On the other hand, \cite{deimen2019delegated} study a version of the delegated expertise problem, in which the agent's bias is also unknown ex ante, and show that communication outperforms delegation in such a problem. Our results imply that this conclusion depends on the model specifics, as in the setting we consider, communication and delegation yield identical results.

\citet{szalay2005economics} shows that restricting the agent's action set could be a useful tool in the context of a delegated expertise problem, since banning an ex ante optimal ``safe'' action can nudge the agent to acquire more information about which of the risky actions is the best. 
\cite{ball2021benefitting} explore a model similar to that of \cite{szalay2005economics} but allow for the preferences of the principal and the agent to be misaligned ex post. They show that the principal may benefit from some misalignment between her preferences and those of the agent. In their setting, misalignment makes it less costly for the principal to ban the agent's ex ante preferred action (in order to incentivize the agent to acquire information about which of the remaining options is best). Our paper suggests a different channel: using a flexible information acquisition framework, we show that by hiring an agent with a different prior belief, the principal can trade off mistakes in different states of the world, which is typically beneficial. 
Further, we show that restricting the action set is not a useful tool in our setting if the principal can select an optimally misaligned agent (Proposition \ref{prop:restriction_no_good}).

Finally, there exists a literature that argues in favor of misaligned delegation in strategic settings, as a way to commit to a certain strategy. Examples include \citet{rogoff1985optimal,vickers1985delegation,segendorff_delegation_1998,kockesen_strategic_2004,stepanov_biased_2020}, and \citet{ispano2022good}. We differ from that literature in focusing on delegation of \emph{non-strategic} decisions, showing how misalignment may be beneficial even in the absence of a strategic counterparty.
Additionally, a separate literature argues that people may have intrinsic preferences for potentially biased information, see \citet*{masatlioglu2023intrinsic} for a recent example. We abstract from such considerations and focus on a setting in which information has purely instrumental value.

The remainder of the paper is organized as follows: Section \ref{sec:model} formulates the main model for a general family of information costs. It is then analyzed in Section \ref{sec:binary} for the case of binary states and actions. Section \ref{sec:gencase} analyzes the problem with (finitely) many states and actions while restricting attention to a specific information cost function. Section \ref{sec:disc} discusses various assumptions made in our analysis and alternative modeling approaches, while Section \ref{sec:conc} concludes. Proofs of all results are relegated to the Appendix. The Online Appendix contains supplementary results and robustness checks.

\section{Model} \label{sec:model}

\subsection{The Setup} \label{sec:modelmodel}

Consider the following game between a principal (she) and a population of agents (all referred to as ``he'').
Let $\Omega$ denote the finite set of states with a typical element $\omega$. The principal has a prior belief $\mu_p \in \varDelta(\Omega)$, where $\varDelta(\Omega)$ denotes the set of all probability distributions on $\Omega$. Every agent in the population has some prior belief $\mu \in \varDelta(\Omega)$.\footnote{To clarify, we work with a model of non-common prior beliefs about $\omega$, where agents ``agree to disagree''. This and other assumptions are discussed in Section \ref{sec:disc}.} 
In what follows, we refer to an agent according to his prior belief. Let $\mathcal{M}$ denote the set of prior beliefs of all agents in the population. We assume throughout that the population of agents is rich enough to represent the whole spectrum of viewpoints: $\mathcal{M} = \varDelta(\Omega)$. 

The principal needs to select an agent $\mu \in \mathcal{M}$, to whom a given decision problem is delegated. Let $\mathcal{A}$ denote the finite set of actions available in the decision problem, with a typical element $a$. The selected agent would need to choose an action, $a \in \mathcal{A}$.
Prior to making this decision, the selected agent can acquire additional information about the realized state. We assume that the agent can choose any signal structure $\phi: \Omega \to \varDelta(\mathcal{S})$, which prescribes a distribution over signal realizations $s \in \mathcal{S}$ for every state $\omega \in \Omega$, where the signal space $\mathcal{S}$ is fixed but arbitrarily rich. We further use $\sigma : \mathcal{S} \to \varDelta(\mathcal{A})$ to denote the agent's choice rule that maps signal realizations to a distribution over actions.

The payoff that both the principal and the selected agent receive when action $a$ is chosen in state $\omega$ is given by $u(a,\omega) = \mathbb{I}\{a=\omega\}$, where $\mathbb{I}\{\cdot\}$ is the indicator function.\footnote{This state-matching utility function is adopted for simplicity. We show in Online Appendix \ref{sec:genprefs} that our results continue to hold with a quadratic-loss payoff function, common in delegation models.} 
The agent additionally incurs cost $c(\phi,\mu)$ from choosing signal structure $\phi$, which also depends on his prior belief $\mu$.\footnote{Similar to, e.g., \cite{alonso2016bayesian}, we assume that the agent and the principal share the understanding of the signal structure. Combined with them having different (subjective) prior beliefs over states, this implies they would also have different (subjective) posterior beliefs if both could observe the signal realization.} 
We consider a family of uniformly posterior-separable cost functions \citep{caplin2022rationally, miao2024dynamic} given by\footnote{The class of UPS cost functions we consider is also known as Bregman information cost \citep{banerjee2005clustering,fosgerau2023equilibrium}. This class includes the Shannon entropy cost, which is commonly used in rational inattention models \citep{MM}, as a special case when $\hat{c}$ is negative Shannon entropy: $\hat{c}(\eta)= \sum_{\omega \in \Omega} \eta(\omega) \ln \eta(\omega)$, with the convention $0 \cdot \ln 0 = 0$. We use the Shannon entropy cost for most of the figures in this paper.}
\begin{align*}
	c(\phi, \mu) \equiv \lambda \Big[\mathbb{E}[\hat{c}(\eta(s))|\phi] - \hat{c}(\mu)\Big],
\end{align*}
where $\eta: \mathcal{S} \to \varDelta(\Omega)$ denote the agent's posterior belief system, obtained from $\mu$ and $\phi$ using the Bayes' rule, the expectation is taken over signal realizations $s$, and function $\hat{c}(\mu): \varDelta(\Omega) \to \mathbb{R}$ is strictly convex, twice continuously differentiable, symmetric, and $\lim\limits_{\mu(\omega)\to 0+} \hat{c}(\mu) = \infty$ for all $\omega \in \Omega$.\footnote{Symmetry requires that $\hat{c}(\mu)$ is unaffected by permutations of $\mu$: for any $\mu',\mu''$, if there exists a bijection $\psi:\Omega \to \Omega$ such that $\mu''=\psi(\mu')$, then $\hat{c}(\mu') = \hat{c}(\mu'')$. The final ``Inada condition'' requires that ruling out any single state (or learning state with certainty) is prohibitively costly.}

To summarize, the game proceeds as follows. In the first stage, the principal selects an agent from the population based on the agent's prior belief $\mu$. In the second stage, the selected agent chooses signal structure $\phi$ and pays cost $c(\phi,\mu$). In the third stage, the agent observes signal realization $s$ according to the chosen signal structure $\phi$ and selects action $a$ given $s$. Payoffs $u(a,\omega)$ are then realized for the principal and the agent.
We will be looking for a Subgame Perfect Nash Equilibrium of the game, hereinafter referred to simply as ``equilibrium'', defined as follows.

\begin{definition}[Equilibrium]
	An equilibrium of the game is given by $(\mu^*, \{\phi^*_\mu, \sigma^*_\mu \}_{\mu \in \mathcal{M}})$: the principal's choice $\mu^* \in \mathcal{M}$ of the agent who the task is delegated to and a collection of the agents' signal structures $\phi^*_\mu : \Omega \to \varDelta(\mathcal{S})$ and choice rules $\sigma^*_\mu : \mathcal{S} \to \varDelta(\mathcal{A})$ for all $\mu\in \mathcal{M}$, such that:
	\begin{enumerate}
		\item $\phi^*_\mu$ and $\sigma^*_\mu$ constitute a solution to the agent's problem for every $\mu \in \mathcal{M}$:
		\begin{align} \label{eq:agentprb}
			(\phi^*_\mu, \sigma^*_\mu) \in 
			&\ \arg \max_{\phi, \sigma} \left\{ \sum_{\omega \in \Omega} \mu(\omega) \sum_{(s,a) \in \mathcal{S} \times \mathcal{A}} \phi(s|\omega) u \left( \sigma(a|s),\omega \right) - c(\phi,\mu) \right\};
		\end{align}
		\item $\mu^*$ solves the principal's problem given $(\phi^*_\mu, \sigma^*_\mu)$:
		\begin{align} \label{eq:principal_problem}
			\mu^* \in 
			&\arg \max_{\mu} \left\{\sum_{\omega \in \Omega}\mu_p(\omega)\sum_{(s,a) \in \mathcal{S} \times \mathcal{A}}\phi^*_\mu(s|\omega) u \left( \sigma^*_\mu(a|s),\omega \right)\right\}.
		\end{align}
	\end{enumerate}
\end{definition}

\subsection{Preliminary Analysis} \label{sec:preliminary_analysis}

It is known in the literature on rational inattention \citep[c.f.][]{MM,caplin2022rationally,mackowiak2023rational,miao2024dynamic} that the agent's two-step problem -- first choosing the signal structure and then selecting the choice rule -- can be reduced to a single problem of choosing the conditional action probabilities.
The idea is akin to a revelation principle: it is without loss of generality to restrict attention to such signal structures that every signal realization $s$ prescribes a specific action $a$.\footnote{The recommendation principle from Bayesian Persuasion/information design literatures is also closely related, see, e.g., \citet{bergemann2019information}.} 
This is because it is never optimal for the agent to learn in a way that leads to the same action $a$ being taken after different signal realizations $s,s'$, since then it is cheaper to pool $s,s'$ into a single realization, and neither is it optimal to mix over distinct actions $a,a'$ after any signal realization $s$, since then acquiring a marginally more informative signal structure (splitting $s$ into two realizations $s,s'$) helps break the tie between $a$ and $a'$.

Formally, the agent's problem can be rewritten as that of choosing a decision rule $\pi : \Omega \to \varDelta(\mathcal{A})$, i.e., a single state-contingent action distribution, as opposed to the combination of a signal structure $\phi : \Omega \to \varDelta(\mathcal{S})$ and a choice rule $\sigma : \mathcal{S} \to \varDelta(\mathcal{A})$:
\begin{equation}\label{eq:ri_problem_full}
	\max_{\pi} \Bigg\{\sum_{\omega \in \Omega} \mu(\omega) \left( \sum_{a \in \mathcal{A}} \pi(a|\omega) u(a,\omega) \right) - c(\pi,\mu) \Bigg\},
\end{equation}
where $c(\pi,\mu)$ denotes, with abuse of notation, the information cost induced by the decision rule $\pi$, defined as cost $c(\phi,\mu)$ of the cheapest strategy $(\phi,\sigma)$ that generates decision rule $\pi$. 

The principal's problem can then be rewritten as choosing $\mu \in \mathcal{M}$ that solves
\begin{equation}\label{eq:principal_full_prb}
	\begin{aligned}
		&\max_{\mu} \left\{ \sum_{\omega \in \Omega} \mu_p(\omega) \left( \sum_{a \in \mathcal{A}} \pi^*_\mu(a|\omega) u(a,\omega) \right) \right\},
		\\
		&\text{s.t. } \pi^*_\mu \text{ solves \eqref{eq:ri_problem_full}} \text{ given } \mu.
	\end{aligned}
\end{equation}
Our main interest in what follows lies in the properties of the solution $\mu^*$ to the principal's problem above and of the optimal strategy $\pi_{\mu^*}$ of the chosen agent $\mu^*$.

\section{Binary Case} \label{sec:binary}

This section presents our main result in a binary setting with $|\Omega|=|\mathcal{A}|=2$.\footnote{Such a binary model is common in the delegation literature, see, e.g., \cite{li2004delegating}.} 
We show that when choosing a delegate, the principal has to balance off the error probabilities in the two states, since agents with different prior beliefs direct their learning towards different states to different extents. This makes the principal favor agents who are somewhat more uncertain than her regarding the state, but who do not necessarily have a uniform prior belief (Proposition \ref{prop:PScost}).

For this section, assume that the state space is $\Omega = \{l,r\}$, and with abuse of notation, let us represent beliefs $\mu$ by the probability they assign to state $r$, so $\mu \in [0,1]$. Suppose without loss that the principal's prior belief is $\mu_p \geq \frac{1}{2}$. Assume further that the action set is $\mathcal{A} = \{L,R\}$, and the common payoff is given by $u(L,l) = u(R,r) = 1$ and $u(L,r) = u(R,l) = 0$. We proceed by backward induction, looking at the agent's problem first, and then using the agent's optimal behavior to solve the principal's problem of choosing the best agent.

\subsection{The Agent's Problem} \label{sec:agents_prb_binary}

With binary states and actions, the agent's problem \eqref{eq:ri_problem_full} is equivalent to
\begin{align}\label{eq:ri_problem}
	&\max_{\pi} \Big\{ \mu \pi(R|r) + (1-\mu) \pi(L|l) - c(\pi,\mu) \Big\}. 
\end{align}
The solution of this problem is presented in the following lemma.
\begin{lemma}\label{lem:agents_soln_ups}
	The solution to the agent's problem \eqref{eq:ri_problem} is such that there exists a threshold $\bar{\mu} \in \left[\frac{1}{2},1\right]$, and
	\begin{enumerate}
		\item if $\mu \in \left( 1-\bar{\mu}, \bar{\mu} \right)$, then $\phi^*_{\mu}$ is binary, with the two signal realizations $s \in \{\underline{s}, \bar{s}\}$ inducing respective posteriors $\eta \in \left\{ 1-\bar{\mu}, \bar{\mu} \right\}$, and $\sigma^*_{\mu}(L|\underline{s}) = 1 = \sigma^*_{\mu}(R|\bar{s})$;
		\item if $\mu \leq 1-\bar{\mu}$, then the agent acquires no information and selects $a=L$;
		\item if $\mu \geq \bar{\mu}$, then the agent acquires no information and selects $a=R$.
	\end{enumerate}
\end{lemma}
The lemma states that the agent's optimal strategy either is trivial (no information), or produces a binary signal. In the latter case, the optimal signal structure $\phi^*$ generates the same two posterior beliefs, $\bar{\mu}$ and $1-\bar{\mu}$, regardless of the agent's prior belief $\mu$. Intuitively, any learning agent requires the same standard of proof (posterior belief) to take either action, but different agents acquire different information in order to arrive at these posteriors.\footnote{\citet{caplin2022rationally} call this the Locally Invariant Posteriors (LIP) property and show that it is a characteristic property of learning problems with UPS information cost functions. While it is convenient to explain the intuition behind our results using the LIP property, we argue below that our results are not specific to UPS information costs, and we show in Appendix \ref{sec:cost} that they continue to hold with other (non-UPS) information cost functions.}

\begin{figure} 
	\centering 
	\subfloat[][Action precisions.]{
		\centering \includegraphics[width=0.45\textwidth]{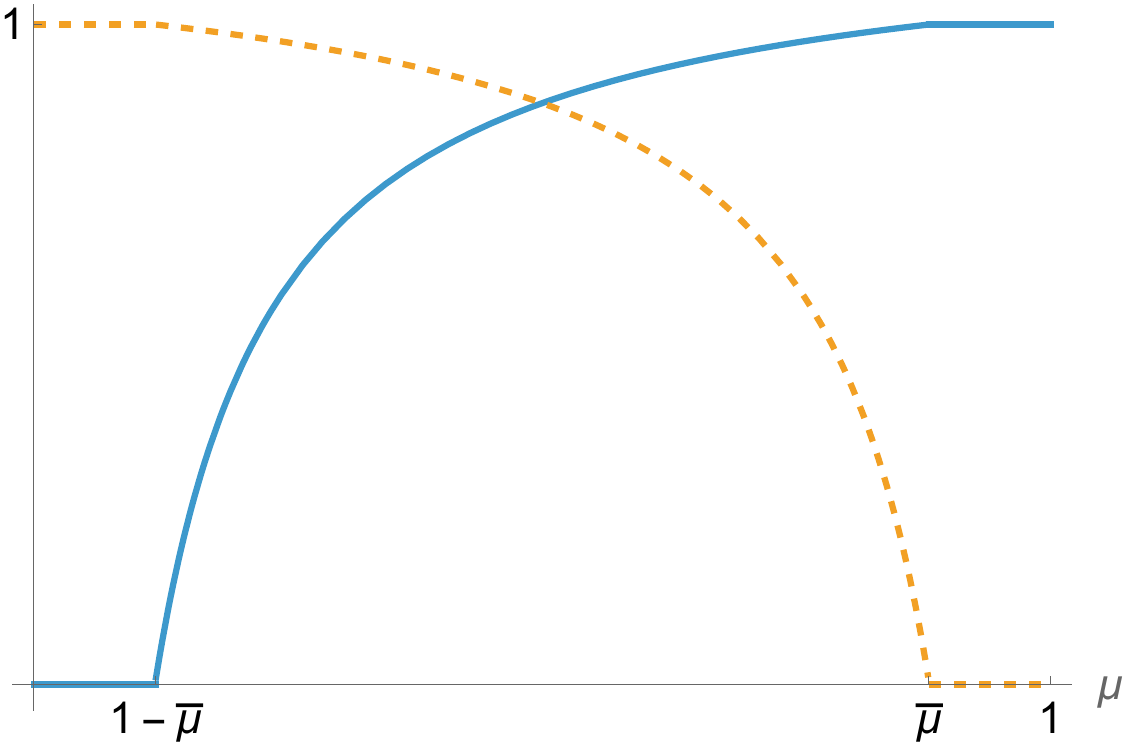}
		\label{fig:precisions}
	}
	\hfill 
	\subfloat[][Unconditional action probabilities.]{
		\centering \includegraphics[width=0.45\textwidth]{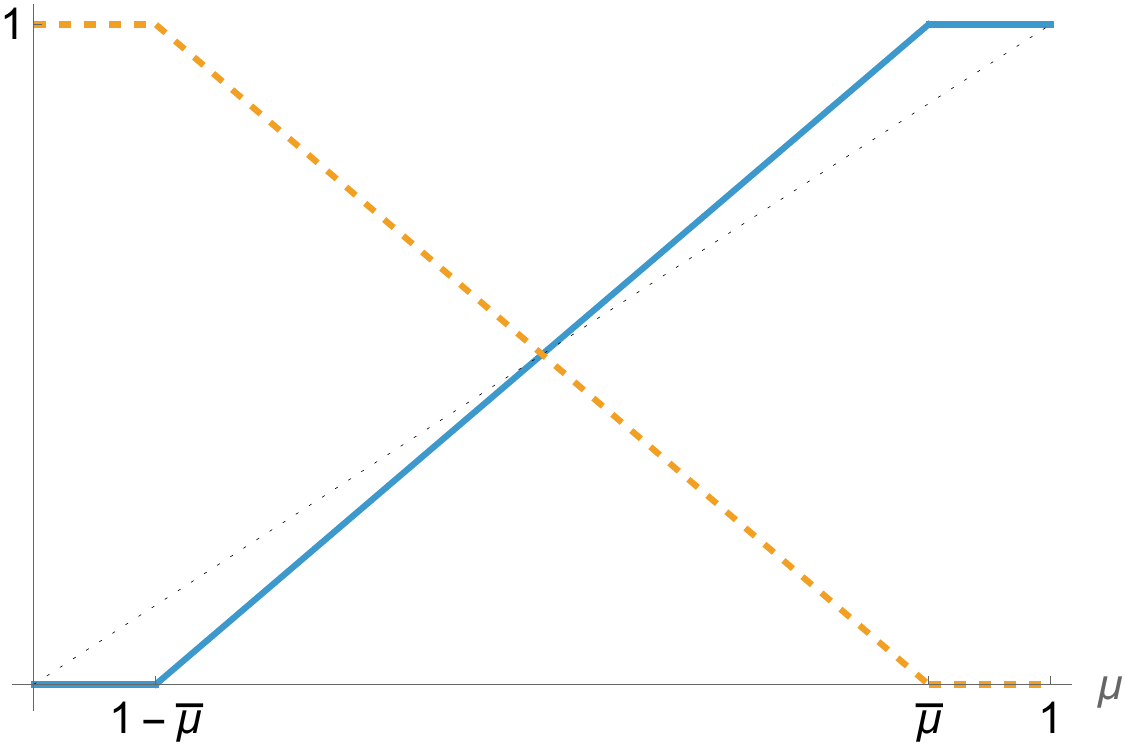}
		\label{fig:beta}
	}
	\caption{Solution of the agent's problem \eqref{eq:ri_problem_full} for different prior beliefs $\mu$. Panel (a): $\pi_{\mu}^{*}(R|r)$ in solid, $\pi_{\mu}^{*}(L|l)$ in dashed. Panel (b): $\beta_{\mu}(R)$ in solid, $\beta_{\mu}(L)$ in dashed, and the $45^\circ$ line in dotted. \\
	} 
	\label{fig:A_soln} 
\end{figure}

In order to understand the trade-off that will be involved in the principal's decision later, let us now explore how the agent's prior belief $\mu$ affects the decision rules $\pi : \Omega \to \varDelta(\mathcal{A})$ induced by the optimal strategy $(\phi^*_{\mu}, \sigma^*_{\mu})$. 
When solving problem \eqref{eq:ri_problem}, the agent faces a trade-off between the \emph{action precisions} $\pi(R|r)$ and  $\pi(L|l)$ and the cost of information. The agent can increase both precisions only by incurring a higher information cost. Alternatively, the agent can keep the information cost the same and trade off one precision against the other. The higher is the probability $\mu$ that the agent's prior belief assigns to state $\omega=r$, the more important is precision $\pi(R|r)$ for his payoff, compared to $\pi(L|l)$. Hence, at the optimum, the agent prefers to minimize the rate of mistakes in the more probable event. 
Figure \ref{fig:precisions} demonstrates how the agent's optimal choice of action precisions depends on his prior belief. It shows that $\pi^*_\mu(R|r) > \pi^*_\mu(L|l)$ when $\mu > 0.5$, and vice versa.

From the Bayes' rule $\mathbb{P} (a | \omega) = \frac{ \mathbb{P} (\omega | a) \mathbb{P} (a) }{ \mathbb{P} (\omega) }$ and Lemma \ref{lem:agents_soln_ups}, we have the following identities for all learning agents $\mu \in (1-\bar{\mu}, \bar{\mu})$:
\begin{equation}\label{eq:precisions_Bayes}
\begin{aligned}
	\pi^*_\mu(R|r) &= \bar{\mu} \cdot \frac{ \beta_{\mu}(R) }{\mu}
	&= \frac{\bar{\mu}}{\bar{\mu} - (1-\bar{\mu})} \cdot \left( 1 - \frac{1-\bar{\mu}}{\mu} \right),
	\\
	\pi^*_\mu(L|l) &= \bar{\mu} \cdot \frac{ 1-\beta_{\mu}(R) }{1-\mu}
	&= \frac{\bar{\mu}}{\bar{\mu} - (1-\bar{\mu})} \cdot \left( 1 - \frac{1-\bar{\mu}}{1-\mu} \right),
\end{aligned}
\end{equation}
where $\beta_{\mu}(a) \equiv \mathbb{E}_\mu \pi(a|\omega)$ is the unconditional probability of choosing action $a$ (given the agent's prior belief $\mu$). 
The Bayes' rule, given by the first (left) pair of equalities in \eqref{eq:precisions_Bayes}, implies that action precisions $\pi$ can be described via an interaction of three components:
\begin{enumerate}
	\item ``standards of proof'' $\mathbb{P}(\omega|a)$, which, according to Lemma \ref{lem:agents_soln_ups}, are constant at $\bar{\mu}$ regardless of the agent's prior $\mu$;
	\item probabilities $\beta_{\mu}(a)$ with which the agent takes either action, which can be calculated using the law of total probability as the probabilities of the respective posteriors $\eta \in \left\{ \bar{\mu}, 1-\bar{\mu} \right\}$: $\beta_{\mu}(R) = \frac{\mu - (1-\bar{\mu})}{\bar{\mu} - (1-\bar{\mu})}$, $\beta_{\mu}(L) = 1-\beta_{\mu}(R)$;
	\item state distribution characterized by $\mu$.
\end{enumerate}

We see from the second (right) pair of equalities in \eqref{eq:precisions_Bayes} that due to constant standards of proof, the precisions simplify to $\pi^*_\mu(R|r) = C_1 - \frac{C_2}{\mu}$ and $\pi^*_\mu(L|l) = C_1 - \frac{C_2}{1-\mu}$, respectively, for some $C_1, C_2$ that depend on $\bar{\mu}$. Therefore, the higher is the agent's belief $\mu$,  the faster precision $\pi^*_\mu(L|l)$ changes  with $\mu$, and  the slower precision $\pi^*_\mu(R|r)$ changes with $\mu$. 
This can be seen in Figure \ref{fig:precisions}.
An alternative graphical intuition is captured in Figure \ref{fig:geom_intuition}: when $\mu > \frac{1}{2}$, precision $\pi^*_\mu(R|r) \propto \frac{\beta_{\mu}(R)}{\mu}$ changes slowly with $\mu$, while $\pi^*_\mu(L|l) \propto \frac{1-\beta_{\mu}(R)}{1-\mu}$ changes rapidly.

\begin{figure}
	\begin{center}
	\begin{tikzpicture}[scale=11]
		\draw (0,0) -- (1,0);
		\draw (0,-0.01) node[below]{$0$} -- (0,0.01);
		\draw (1,-0.01) node[below]{$1$} -- (1,0.01);
		
		\draw (0.2,-0.01) node[below]{$1-\bar{\mu}$} -- (0.2,0.01);
		\draw (0.8,-0.01) node[below]{$\bar{\mu}$} -- (0.8,0.01);
		\draw (0.6,-0.01) node[below]{$\mu$} -- (0.6,0.01);
		
		\draw[decoration={brace,mirror,raise=0.8cm},decorate,thick,color=orange!50!yellow!80!black] (0,0) -- node[below=0.8cm] {$\mu$} (0.59,0);
		\draw[decoration={brace,mirror,raise=0.8cm},decorate,thick,color=orange!50!yellow!80!black] (0.61,0) -- node[below=0.8cm] {$1-\mu$} (1,0);
		\draw[decoration={brace,raise=0.3cm},decorate,thick,color=cyan!85!blue!80!black] (0.2,0) -- node[above=0.3cm] {$\propto \beta_{\mu}(R)$} (0.59,0);
		\draw[decoration={brace,raise=0.3cm},decorate,thick,color=cyan!85!blue!80!black] (0.61,0) -- node[above=0.3cm] {$\propto 1-\beta_{\mu}(R)$} (0.8,0);
	\end{tikzpicture}
	\caption{State probabilities and action probabilities.}
	\label{fig:geom_intuition}
	\end{center}
\end{figure}

Further, the unconditional action probabilities $\beta_\mu(a)$, as presented in Figure \ref{fig:beta}, suggest that the agent succumbs to \emph{confirmatory learning}: he optimally learns in such a way as to err on the side of the ex ante optimal action. Whenever $\mu > \frac{1}{2}$, action $a=R$ is ex ante more likely to be optimal. At the optimum, the agent chooses this action with probability $\beta_{\mu}(R) > \mu$, which is higher than the probability with which this action is optimal ex post, $\mu = \mathbb{P}(\omega=r)$. Therefore, the agent engages in confirmatory learning: he primarily seeks information that justifies the originally favored action, rather than opposes it.\footnote{While this is also sometimes called ``confirmation bias'', we avoid labeling this behavior as such, since it is fully rational in our setting, stemming from the agent's information costs, and is not a ``bias'' per se. }
The following section demonstrates that the agent's tendency towards confirmatory learning is what drives the principal's optimal delegation strategy.

\subsection{The Principal's Problem}

The principal's problem \eqref{eq:principal_full_prb} in the binary case is given by
\begin{equation}\label{eq:principal_problem_binary}
	\max_{\mu } \left\{\mu_p \pi^*_\mu(R|r)+(1-\mu_p) \pi^*_\mu(L|l) \right\}.
\end{equation}
To remind, we assume without loss of generality that $\mu_p \geq \frac{1}{2}$.
It is easy to see that the principal benefits from higher action precisions $\pi(R|r)$ and $\pi(L|l)$, same as the agent. By choosing agents with different prior beliefs $\mu$, the principal can trade off action precisions in one state versus in other state: as argued above, an agent with higher $\mu$ chooses a higher $\pi^*_\mu(R|r)$ but lower $\pi^*_\mu(L|l)$, and vice versa. The optimal choice of $\mu$, therefore, depends on how the principal trades off these two precisions (as given by her marginal rate of substitution, MRS) and how they change with the agent's prior belief $\mu$ (given by the marginal rate of transformation, MRT). 

The principal's MRS of precision $\pi(R|r)$ for precision $\pi(L|l)$ is given by $MRS_{RL} = \frac{\mu_p}{1-\mu_p}$. This is a measure of the principal's preferences: $MRS_{RL}$ captures the amount by which we would need to increase $\pi(L|l)$ in order for the principal to forgo one ``unit'' of $\pi(R|r)$. Since $\mu_p \geq \frac{1}{2}$, the principal assigns a higher weight to $\pi(R|r)$ than to $\pi(L|l)$, implying that $MRS_{RL} \geq 1$.

In turn, the MRT is the ratio that captures how the two precisions, as chosen optimally by the agent, change with the agent's belief $\mu$:
\begin{equation*}
	MRT_{RL}(\mu) \equiv \left| \frac{\frac{d}{d\mu} \pi^*_\mu(L|l)}{\frac{d}{d\mu} \pi^*_\mu(R|r)} \right|
	= \left( \frac{\mu}{1-\mu} \right)^2.
\end{equation*}
In words, $MRT_{RL}(\mu)$ captures how many ``units'' of $\pi_{\mu}^{*}(L|l)$ the principal has to sacrifice in order to increase $\pi_{\mu}^{*}(R|r)$ by one ``unit''. Graphically, one can think of it as the slope of the boundary of the principal's ``production possibility frontier'', PPF, in the space of $\pi(L|l), \pi(R|r)$.
We know from the agent's problem that $\pi^*_\mu(R|r) = C_1 - \frac{C_2}{\mu}$ and $\pi^*_\mu(L|l) = C_1 - \frac{C_2}{1-\mu}$ for some $C_1, C_2$ that depend on $\bar{\mu}$. This implies that $MRT_{RL}(\mu)$ increases with $\mu$, as captured by the expression above.

Suppose now that the principal is biased: $\mu_p > \frac{1}{2}$.
For an aligned agent $\mu = \mu_p$, we have $MRT_{RL}(\mu_p) > MRS_{RL}$: the rate at which the principal can transform $\pi(R|r)$ into $\pi(L|l)$ by reducing $\mu$ is higher than the rate that makes her indifferent. 
We conclude that it is beneficial for the principal to decrease $\mu$ from $\mu_p$ -- i.e., to hire a \emph{misaligned} agent with $\mu < \mu_p$.
Conversely, we can see that for the most uncertain agent, $\mu=1/2$, we have $MRT_{RL}\left(\frac{1}{2}\right) = 1 < MRS_{RL}$, so the principal prefers an agent who is \emph{partially aligned}, $\mu > \frac{1}{2}$.

\begin{figure}[ht]
  \centering
  \subfloat[][Indifference curve and ``PPF''.]{
    \includegraphics[width=0.48\linewidth]{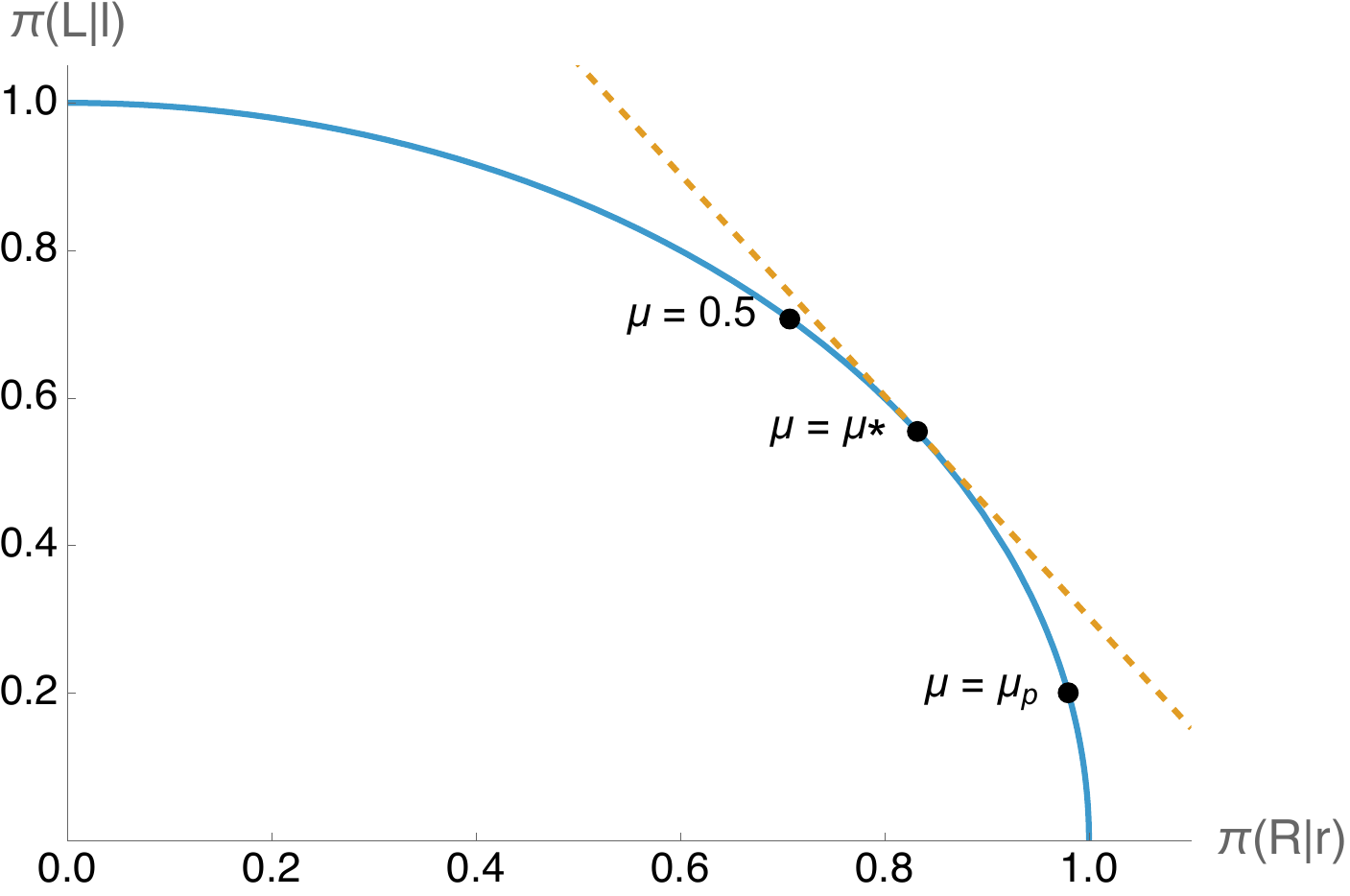}
    \label{fig:P_util_ppf}
  }%
  \hfill
  \subfloat[][The optimal delegation strategy $\mu^*$.]{
    \includegraphics[width=0.48\linewidth]{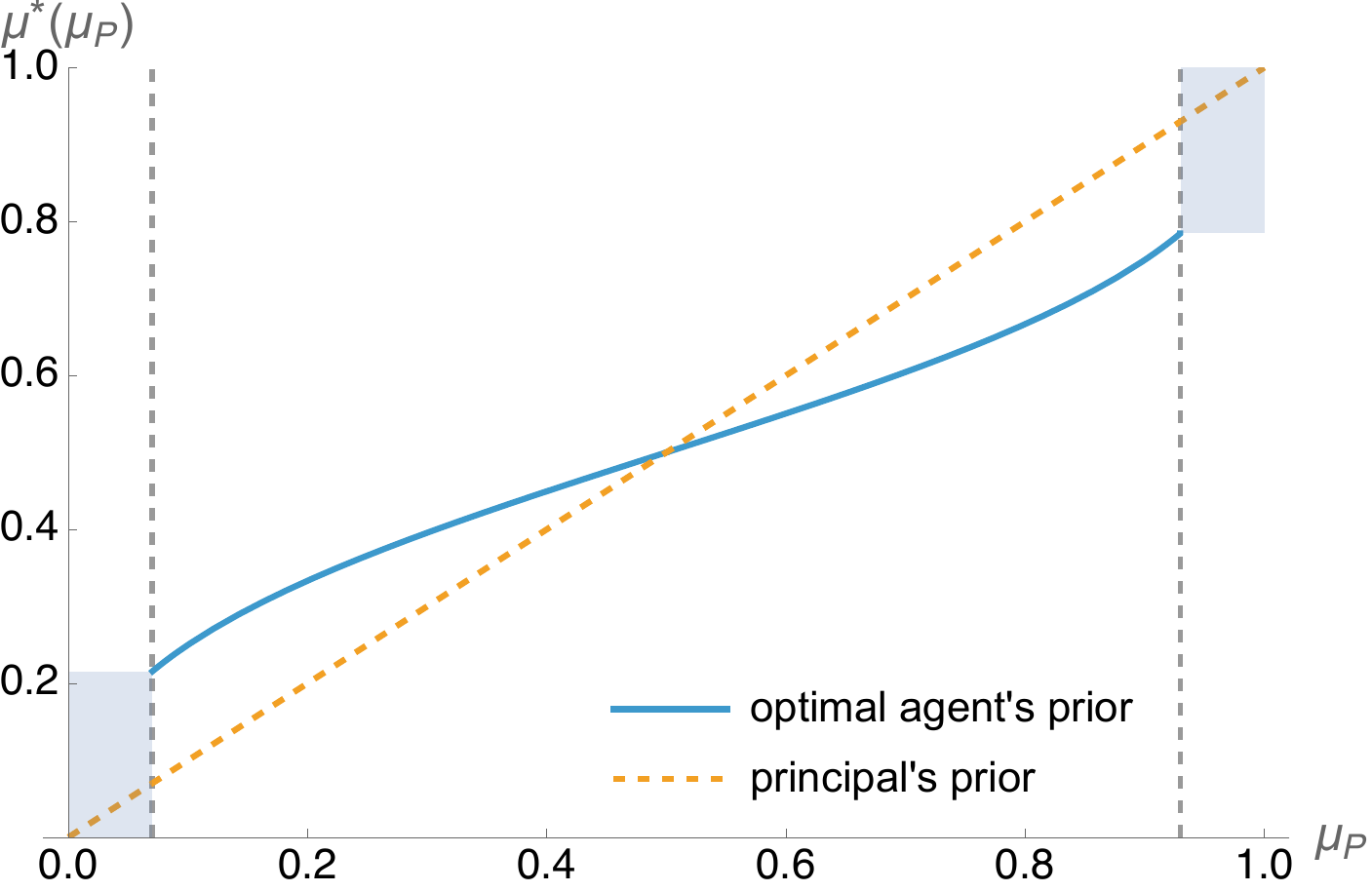}
    \label{fig:opt_deleg_strat}
  }
  \caption{Principal's optimal choice of agent.}
\end{figure}

\begin{figure}[ht]
  \centering
  \subfloat[][Principal's expected utility from delegation and from the ex ante optimal action.]{
    \includegraphics[width=0.48\linewidth]{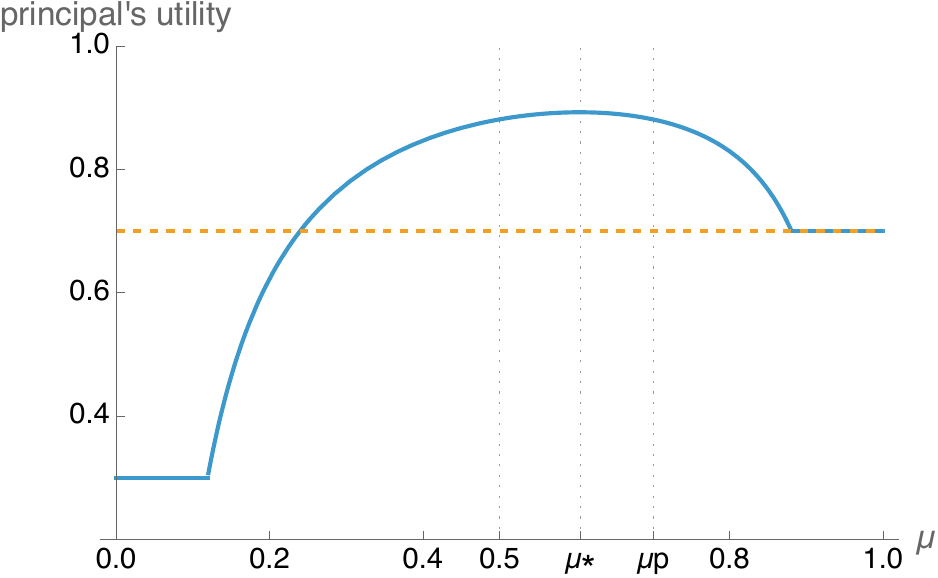}
    \label{fig:P_util}
  }%
  \hfill
  \subfloat[][Principal's gain from optimal delegation compared to aligned delegation.]{
 \includegraphics[width=0.48\linewidth]{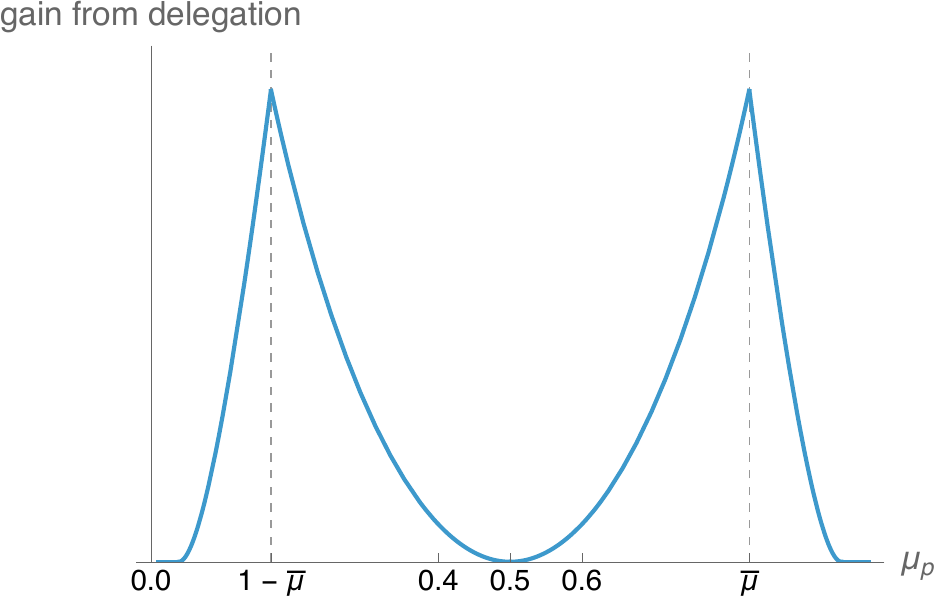}
    \label{fig:P_util_gain}
  }
  \caption{Principal's gain from delegation.}
\end{figure}

In order to find $\mu^*$, the optimal agent that the principal should hire, we can simply find such $\mu$ that $MRT_{RL}(\mu) = MRS_{RL}$, i.e., the principal's indifference curve is tangent to the PPF. Figure \ref{fig:P_util_ppf} presents the principal's problem and her optimal choice visually. It plots the principal's PPF $\left( \pi^*_\mu(R|r), \pi^*_\mu(L|l) \right)_{\mu \in [0,1]}$ (solid blue line), as well as the highest indifference curve the principal can achieve (orange dashed line).
Figure \ref{fig:P_util} plots the principal's expected utility from hiring an agent as a function of the agent's belief $\mu$ (solid blue line) and her expected utility from the ex ante optimal action $R$ (dashed orange line). We can see that the effect described above leads to an increase in the principal's utility as we decrease $\mu$ from $\mu_p$ in the direction of $\mu=\frac{1}{2}$, up to some optimal $\mu^* > \frac{1}{2}$.\footnote{Figure \ref{fig:P_util}, suggests that the principal is better off choosing the ex ante optimal action $R$ rather than delegating to a learning agent with sufficiently low $\mu < \frac{1}{2}$. Thus, when facing a set of agents with a sufficiently strong misalignment, the principal would find it optimal to abstain from delegation altogether and to take the ex ante optimal action.}

The principal's optimal choice is summarized in Proposition \ref{prop:PScost}, which states that the principal's most preferred agent is always more uncertain than her ex ante, but is partially aligned with her: $\mu^* \in \left(\frac{1}{2}, \mu_p \right)$. Figure \ref{fig:opt_deleg_strat} visualizes the optimal delegation strategy as a function of $\mu_p$.

\begin{proposition} \label{prop:binarydelegationri} \label{prop:PScost}
	The principal's optimal delegation strategy is given by
	\begin{align}
		\mu^*(\mu_p) = \frac{\sqrt{\mu_p}}{\sqrt{\mu_p} + \sqrt{1-\mu_p}}.
		\label{eq:opt_deleg_binary}
	\end{align}
	Therefore, if $\mu_p\in \left( \frac{1}{2},1 \right)$, the principal optimally delegates to an agent with belief $\mu \in \left( \frac{1}{2},\mu_p \right)$.
\end{proposition}

In order to understand Proposition \ref{prop:binarydelegationri}, we can view the principal's problem through the lens of \emph{confirmatory learning}. Section \ref{sec:agents_prb_binary} suggests that the agent optimally learns in such a way as to err on the side of safety. 
We can argue that this kind of behavior is not fully desirable for the principal, who, in turn, aims to attenuate the degree of this confirmatory learning and bring the action distribution $\beta$ more in line with the state distribution $\mu_p$.
Let $\beta^P_\mu \in \varDelta(\mathcal{A})$ denote the probability distribution over actions that the principal induces by hiring an agent with prior belief $\mu$: for any $a \in \mathcal{A}$,
\begin{align} \label{eq:beta_P}
	\beta^P_\mu(a) \equiv \mu_p \pi^*_\mu(a|r)+(1-\mu_p) \pi^*_\mu(a|l).
\end{align}
Note that $\beta^P_\mu = \beta_\mu$ when $\mu = \mu_p$, but the two distributions differ otherwise, since they weigh the two precisions differently. By varying the agent's belief $\mu \in \varDelta(\Omega)$, the principal can implement any $\beta^P \in \varDelta(\mathcal{A})$.\footnote{This is immediate in the binary setting, since since both distributions ($\beta$ and $\mu$) only have one degree of freedom. However, Lemma \ref{lemma:optimal_choice} in Appendix \ref{sec:Ncase_prbs} shows that the same is true with $N$ states and actions.} 
Hence, the principal's problem \eqref{eq:principal_problem_binary} is equivalent to a problem of choosing $\beta^P$. 

Proposition \ref{prop:binarydelegationri} then implies that the principal's optimal delegation strategy $\mu^*$ is such that $\mu_p \leq \beta^P_{\mu^*}(R) \leq \beta^P_{\mu_p}(R)$ whenever $\mu_p \geq \frac{1}{2}$, which can be readily verified by plugging \eqref{eq:precisions_Bayes} and \eqref{eq:opt_deleg_binary} into \eqref{eq:beta_P}. In words, an agent who is aligned with the principal engages in too much confirmatory learning for the principal's taste. Instead, the signal structure of the principal's optimal agent is less confirmatory than that of an aligned agent. 
However, the principal does not want to eliminate confirmatory learning completely: whenever $\mu_p > \frac{1}{2}$, it is true that $\beta^P_{\mu^*}(R) > \mu_p$. The reason for this is that the principal also prefers the agent to err on the side of action $a=R$, since $\mu_p > \frac{1}{2}$ and this action is ex ante more appealing to her, same as to the agent. Therefore, the principal always prefers an agent who shares her ex ante preference: $\mu^*(\mu_p) > \frac{1}{2}$ when $\mu_p > \frac{1}{2}$. 

It should be noted that a \emph{more uncertain} agent does not acquire \emph{more information} in the sense of the Blackwell order, which would require that both action precisions $\pi(R|r), \pi(L|l)$ are higher in one signal structure than another. It is evident from Figure \ref{fig:precisions} that the signal structures chosen by agents with different priors $\mu$ are Blackwell-incomparable. Instead, the optimal agent acquires \emph{different} information relative to an aligned one.
However, it is true that a more uncertain agent acquires ``more information'' in terms of $c(\phi^*_\mu,\mu_p)$, the cost of the principal-optimal signal structure $\phi^*_\mu$ to an aligned agent.

Further, the optimally misaligned agent may not acquire any information whatsoever: if $\mu_p$ is very high, it may be that $\mu^*(\mu_p) \geq \bar{\mu}$. In this case, the principal prefers to simply take action $R$ without making the agent acquire any additional information. This can be achieved by selecting an action directly (without delegation) or by delegating to \emph{any} agent $\mu > \bar{\mu}$. This indifference is captured in Figure \ref{fig:opt_deleg_strat} by the blue shaded area. Proposition \ref{prop:binarydelegationri} then states one optimal delegation strategy, but it is not strictly unique in the sense described above. This multiplicity is, however, immaterial, and is ignored in what follows.

Figure \ref{fig:P_util_gain} plots the difference between the principal's expected payoff under delegation to the optimal agent with prior $\mu^*$ and to an aligned agent with prior $\mu_p$. From Proposition \ref{prop:binarydelegationri} and Figures \ref{fig:opt_deleg_strat} and \ref{fig:P_util_gain}  we can immediately see that misalignment is most beneficial to a moderately-biased principal, while if $\mu_p$ is close to either $\frac{1}{2}$ or $1$, then it is best to hire an (almost-)aligned agent. This is summarized by Corollary \ref{corr:nonmonmisal} below.

\begin{corollary} \label{corr:nonmonmisal}
	Both the optimal misalignment $|\mu_p - \mu^*(\mu_p)|$ and the benefit of misalignment to the principal are single-peaked in $\mu_p\in \left( \frac{1}{2},1 \right)$.
\end{corollary}

The intuition behind Corollary \ref{corr:nonmonmisal} is that an unbiased principal ($\mu_p \approx \frac{1}{2}$) already faces balanced learning from an aligned agent, $\pi^*_{\mu}(L|l) = \pi^*_{\mu}(R|r)$, and hence cannot profitably trade one precision for another by hiring a misaligned agent. A very biased principal ($\mu_p \approx 1$) cares only about precision $\pi(R|r)$, and is not willing to trade it for precision $\pi(L|l)$, since she believes state $\omega=l$ is extremely unlikely, making a highly-aligned agent the preferred choice. In contrast, a moderately-biased principal values both precisions but faces confirmatory learning by an aligned agent. This confirmatory learning can and should be curbed by hiring a misaligned agent instead.

One thing to note about Proposition \ref{prop:binarydelegationri} is that the optimal delegation strategy \eqref{eq:opt_deleg_binary} does not depend on the agent's information cost factor, $\lambda$. While it is immediate that the higher is $\lambda$, the less information the agent with any given prior $\mu$ collects, Proposition \ref{prop:binarydelegationri} serves to show that the relative trade-off between the two action precisions remains the same for different $\lambda$ and does not depend on the absolute quantity of information the agent acquires.
Online Appendix Section \ref{sec:cost} does, however, suggest that this specific conclusion relies on the information cost function being UPS.

\section{Many States and Actions} \label{sec:gencase}

In this section, we extend the analysis to a general problem of finding the best alternative, allowing for $N>2$ actions and states while restricting the cost function to Shannon entropy. We show that the principal's optimal delegation strategy is qualitatively the same as in the binary case, i.e., it is optimal to hire an agent who is ``more uncertain'' (in some sense), since such an agent is less prone to confirmatory learning and investigates more actions in search of the best one than a fully aligned agent.

\subsection{Problem Setup} \label{sec:Ncase_setup}

We are now looking at the model with $\mathcal{A} \equiv \{ a_1, ..., a_N \}$ and $\Omega \equiv \{ \omega_1, ..., \omega_N \}$ for some $N>2$, and the payoffs for the principal and the agent are given by $u(a_i,\omega_i) = 1$ and $u(a_i,\omega_j) = 0$ if $i \neq j$. Without loss of generality, we assume that the principal's belief $\mu_p \in \varDelta(\Omega)$ is such that $\mu_p(\omega_{1}) \geq \mu_p(\omega_{2}) \geq \ldots \geq \mu_p(\omega_{N})$ (otherwise states and actions can be relabeled as necessary). 
Further, we assume that the agent's information cost function is given by the Shannon entropy cost function, meaning the cost is proportional to the expected reduction in entropy of the agent's belief from observing the signal realization:
\begin{align}
	c_E(\phi,\mu) \equiv \lambda \Bigg( &- \sum_{\omega \in \Omega} \mu(\omega) \ln \mu(\omega) + \nonumber
	\\
	& + \sum_{\omega \in \Omega} \sum_{s \in \mathcal{S}} \left( \sum_{\omega' \in \Omega} \mu(\omega') \phi(s|\omega') \right) \eta(\omega | s) \ln \eta(\omega | s) \Bigg),
	\label{eq:c_entropy}
\end{align}
where $\lambda \in \mathbb{R}_{++}$ is a cost parameter.\footnote{We follow the standard convention and let $0 \ln 0 = 0$.}
This is a particular example of a UPS cost function that we adopt for the sake of tractability.\footnote{The entropy parametrization has been rationalized in information theory as a cost function arising from the optimal encoding problem \citep[see ][]{cover}, and has been shown to work as a microfoundation of the logit choice rule commonly used in choice estimation \citep{MM}. See \citet{mackowiak2023rational} for a survey of the literature using Shannon entropy information costs.}

The revelation principle from Section \ref{sec:preliminary_analysis} continues to apply, meaning that the agent's problem is equivalent to choosing the decision rule $\pi : \Omega \to \varDelta(\mathcal{A})$ that maximizes \eqref{eq:ri_problem_full}, and the principal selects an agent according to his prior $\mu \in \mathcal{M}$ to maximize \eqref{eq:principal_full_prb}.

\subsection{Properties of the Optimal Delegation Strategy} \label{sec:soln_properties}

The agent's optimal strategy with $N$ states and actions is similar to the one captured by Lemma \ref{lem:agents_soln_ups}. It is stated formally in the Appendix (Section \ref{sec:Ncase_prbs}), and here we only describe it verbally. If his prior belief assigns too low of a probability to some state $\omega_i$, he never takes action $a_i$. For all other actions, which have a high enough ex ante probability to be optimal, there exists some standard of proof $\bar{\mu}$ that the agent requires in order to justify a given action. I.e., when the agent takes action $a_i$, his posterior belief assigns probability of at least $\bar{\mu} \in (0.5, 1)$ to that action being optimal: $\eta(\omega_i|s_i) \geq \bar{\mu}$, where signal realization $s_i$ is the recommendation to take action $a_i$.\footnote{In this general model, the standard of proof $\bar{\mu}$ depends on the agent's prior belief $\mu$ through the size of the consideration set -- i.e., the number of actions that the agent plays with positive probability at the optimum.}
Since $\bar{\mu} < 1$, this again means that the agent optimally succumbs to a form of ``confirmatory learning'', preferring to err on the side of the ex ante optimal action(s). The main driving force behind the principal's choice of agent is then the desire to mitigate this confirmatory learning to some extent. 

However, some of the statements above are ambiguous in the multidimensional world. Firstly, what exactly does confirmatory learning mean in this case? Does the agent lean towards a single action that is the best ex ante, or are errors somehow proportional to all actions' ex ante appeal? Secondly, which aspect of the agent's learning is the principal trying to fix? Is she more concerned that the ex ante optimal action is chosen too often by an aligned agent at the expense of other actions, or that too little attention is devoted to the ex ante worst action? In a world with more than two actions, these ``biases'' are no longer directly connected. Finally, how big is the principal's toolbox when she attempts to fix the agent's learning? Can she choose any action distribution $\beta^P \in \varDelta(\mathcal{A})$ by varying the agent's belief $\mu \in \varDelta(\Omega)$ appropriately, or is she restricted in her choice? Alternatively, how does her preferred distribution $\beta^P$ translate into the choice of agent's belief $\mu$?

The key to answering all the questions posed above lies in additive separability of both the principal's and the agent's problems, which allows us to consider one pair of states and actions at a time and apply the lessons from the binary model. Specifically, for a given principal's belief $\mu_p \in \varDelta(\Omega)$, fix some $i,j \in \{1,...,N\}$ with $i>j$ and probabilities $\{\mu(\omega_l)\}$ for $l \neq i,j$. Then among the two actions $a_i$ and $a_j$, the agent would exhibit confirmatory learning towards $a_i$ whenever $\mu_i > \mu_j$. This is in the sense of $a_i$ being chosen by the agent with higher (relative) probability than what is warranted by the prior belief, i.e.,
\begin{align*}
	\frac{\beta_{\mu}(a_i)}{\beta_{\mu}(a_j)} &> \frac{\mu_i}{\mu_j}.
\end{align*}
This is then exactly the ``bias'' that the principal seeks to correct: our findings from Section \ref{sec:binary} suggest that the principal prefers -- and is able to -- choose such agent's belief $\mu_i,\mu_j$ given $\{\mu(\omega_l)\}_{l \neq i,j}$ that confirmatory learning within this pair of actions is mitigated, though not eliminated. We can then apply the same logic to any pair of states, since both the agent's and the principal's problems are additively separable with respect to different states. The resulting characterization is captured in Theorem \ref{prop:opt_choice_beliefs} below.

\begin{theorem} \label{prop:opt_choice_beliefs}
	The principal's equilibrium delegation strategy $\mu^*$ is such that for all $i,j \in \{1,...,N\}$:
	\begin{align} \label{eq:opt_belief_N}
		\frac{\mu^*(\omega_i)}{\mu^*(\omega_j)} = \frac{\sqrt{\mu_p(\omega_i)}}{\sqrt{\mu_p(\omega_j)}}.
	\end{align}
\end{theorem}

The characterization in Theorem \ref{prop:opt_choice_beliefs} suggests that the principal seeks an agent who is more uncertain than her \emph{across any pair of states}, to the exact extent captured by Proposition \ref{prop:binarydelegationri}. This implies that the optimal agent ranks the states' probabilities in the same way as the principal, but in order to conclude that the optimal agent is more uncertain than the principal, we need to define what ``\emph{more uncertain}'' means with $N$ states. It is immediate that a uniform prior belief is the most uncertain about which state is the true state, while a degenerate prior belief (one that assigns probability one to some state) is the most certain, but it is not clear how to compare the beliefs in between. 
Here, we present two different definitions, according to which the optimal agent described in Theorem \ref{prop:opt_choice_beliefs} is ``more uncertain'' than the principal. 
One option is to use Shannon entropy, which is a common measure of the amount of information contained in a probability distribution \citep{cover}. The characterization in Theorem \ref{prop:opt_choice_beliefs} implies that indeed, the optimal agent's prior belief $\mu^*$ has higher Shannon entropy than that of the principal, $\mu_p$, and is hence more uncertain.

Another way to arrive at this conclusion is to see that the optimal agent's prior belief $\mu^*$ in our setting is majorized by that of the principal, $\mu_p$. The majorization order is defined as follows \citep[p.45]{hardy1934inequalities}.
\begin{definition}
	A discrete probability distribution $\mu_{p} \in \varDelta(\Omega)$ majorizes another distribution $\mu^{*} \in \varDelta(\Omega)$ if both probability vectors are sorted in a descending order and for all $k\in\{1,...,N\}$:
	\begin{equation*}
		\sum_{i=1}^{k}\mu_{p}(\omega_{i}) \geq \sum_{i=1}^{k}\mu^{*}(\omega_{i}).
	\end{equation*}
\end{definition}
Intuitively, $\mu_p$ majorizing $\mu^*$ means that $\mu^*$ lies in the convex hull of permutations of $\mu_p$. We interpret this as saying that $\mu^*$ is a ``more uniform'' distribution than $\mu_p$ and, hence, more uncertain. To our knowledge, this is a novel way to rank beliefs according to the amount of uncertainty associated with them.\footnote{It is known that majorization order can be linked to uncertainty: e.g., a random variable $X$ second-order stochastically dominates $Y$ if and only if its c.d.f. $F_X$ majorizes c.d.f. $F_Y$ (see, e.g., \citealp[p.1560]{kleiner2021extreme}). However, our interpretation of the majorization order in application to probability distributions is different.}

We conclude that while $\mu^*$ is not strictly between $\mu_p$ and a uniform belief over $\Omega$, $\mu^*$ has a higher Shannon entropy than $\mu_p$ and is majorized by it, and hence the optimal agent $\mu^*$ is indeed ``more uncertain'' than the principal $\mu_p$. 
Majorization further implies that the optimal agent assigns a higher probability to the state the principal considers the least likely and, conversely, a lower probability to the principal's most likely state  -- so the optimal agent is also more uncertain regarding the most extreme states. 
All of the properties mentioned above are formalized in Corollary \ref{cor:opt_beliefs_properties} below.
\begin{corollary} \label{cor:opt_beliefs_properties}
	Characterization \eqref{eq:opt_belief_N} in Theorem \ref{prop:opt_choice_beliefs} implies that $\mu^{*}$ is majorized by $\mu_{p}$ and has higher Shannon entropy than $\mu_{p}$. Further, $\mu^*(\omega_1) \geq ... \geq \mu^*(\omega_N)$, $\mu^*(\omega_1) \leq \mu_p(\omega_1)$, and $\mu^*(\omega_N) \geq \mu_p(\omega_N)$, with equalities if and only if $\mu_p(\omega_1) = ... = \mu_p(\omega_j)$.
\end{corollary}

\subsection{Properties of the Delegation Set} \label{sec:restriction}

The conflict of interest in delegation problems often leads the principal to restrict the set of actions available to the agent \citep{holmstrom1980}. We show in this section that the same is not true in the delegated expertise problem that we consider. The converse is true: optimal delegation expands the set of actions taken by the agent in equilibrium, as compared to delegating to a fully aligned agent. We then show that the principal in our problem does not benefit from restricting the agent's action set.

The former claim -- the one that optimal delegation expands the equilibrium action set -- is captured by Proposition \ref{prop:consideration} below. In order to state it formally, we have to introduce additional notation.
Let $\bar{\beta}$ denote the choice probabilities, as perceived by the agent, that would be generated under aligned delegation ($\mu = \mu_p$), and $\beta^*$ denote the choice probabilities that arise under optimal delegation ($\mu = \mu^*(\mu_p)$). Let then $\bar{C}$ and $C^*$ denote the two respective consideration sets:
\begin{align*}
	\bar{C} & \equiv \left\{ a \in \mathcal{A} \mid \bar{\beta}(a) > 0 \right\},
	\\
	C^* & \equiv \left\{ a \in \mathcal{A} \mid \beta^*(a) > 0 \right\}.
\end{align*}
In words, $\bar{C}$ and $C^*$ are the sets of actions taken with positive probabilities by an aligned agent and by an optimal agent, respectively. Now we can state the result.

\begin{proposition}\label{prop:consideration}
	Optimal delegation weakly expands the consideration set relative to aligned delegation: $\bar{C} \subseteq C^*$.
\end{proposition}

In words, delegating to an optimally misaligned agent leads to a wider range of actions taken in equilibrium. This is a direct consequence of delegating to a more uncertain agent. Every action has some positive probability of being optimal ex post, but due to information costs, an agent may only consider some of them worth investigating ex ante. The optimally chosen agent is more uncertain than the principal (or an aligned agent) what the optimal action is ex ante, so he deems (weakly) more actions worth investigating. 

Whether the effect is strict -- whether the optimal agent ever chooses an action that an aligned agent never would -- depends on the degree of optimal misalignment, discussed in Corollary \ref{corr:nonmonmisal} above. If the principal is ex ante uncertain ($\mu_p$ is uniform), she prefers to appoint an agent who is almost perfectly aligned to her, and so the misalignment does not result in actions being substantially different. If the principal is ex ante highly certain that a certain action is optimal ($\mu_p(\omega_1) \approx 1$), then the same is true: she prefers an aligned agent, exactly because such an agent would not investigate anything and would simply take her preferred action. However, if a principal is ex ante moderately biased towards some action but retains some uncertainty whether this action is optimal, an aligned agent would choose this action too frequently for her taste. In this case, the principal would prefer a misaligned agent, who would possibly consider a wider set of actions.

This result leads us to the question of whether the principal can ever benefit in our setting from restricting the delegation set -- i.e., limiting the set of actions that the agent may take \citep[see, e.g.,][]{holmstrom1980}. In the context of ``delegated expertise'' problems, \cite{szalay2005economics} and \cite{ball2021benefitting} show that it may be optimal to rule out an ex ante optimal \emph{safe} action in order to force the agent to exert effort and learn which of the ex post optimal (but ex ante risky) actions is best. \cite{lipnowski2019} show a similar result in a Bayesian Persuasion setting in which the receiver is rationally inattentive to the sender's message.

In our setting, however, due to the state-matching utility assumption, there are no ``safe'' actions that the principal could rule out. Assuming that the principal and the agent hold the same prior belief $\mu_p$, and $\mu_p(\omega_1) > ... > \mu_p(\omega_N)$, action $a_1$ is the ``safest'' in the sense of being the most likely ex ante to be optimal. However, it could not be optimal for the principal to ban $a_1$ -- since, indeed, this is the action that is ex ante the most likely to be ex post optimal! 
In other words, while excluding $a_1$ from the delegation set would induce more learning about other actions, it would also lead to larger ex post losses due to the agent being unable to select action $a_1$ in cases in which it would be optimal to do so. Thus while the principal still wants to nudge the agent to acquire more information about ex ante suboptimal actions in our setting, restricting the delegation set is not an instrument that lends any value to her.

Further, even if the principal wanted to prevent the agent from taking certain actions, she could do this by hiring an agent who assigns zero ex ante probability to these actions being optimal. In other words, in our setting, the principal's ability to select a misaligned agent is a strictly more powerful tool than restricting the delegation set. 
Proposition \ref{prop:restriction_no_good} below summarizes this logic. It claims that in a problem where the principal can choose both the agent's belief $\mu$ and the delegation set $A^*$, it is optimal for her to offer unrestricted delegation to the optimally misaligned agent as described in Theorem \ref{prop:misaligned_prefs_bride}.\footnote{A working paper version of this article also showed that if the principal can only delegate to an aligned agent, unrestricted delegation is still optimal.} 

Consider the agent's problem as given by
\begin{align}\label{eq:agents_problem_restr}
	\max_{\pi} \Bigg\{\sum_{j=1}^{N} \mu(\omega_{j}) \sum_{i=1}^{N} \pi(a_{i}|\omega_{j}) u(a_{i},\omega_{j}) - c_E(\phi,\mu) \Bigg\},
\end{align}
for a given prior belief $\mu$ and a fixed delegation set $A^* \subseteq \mathcal{A}$ (where the maximization is w.r.t. a mapping $\pi: \Omega \to \varDelta(A^*)$). Consider then the principal's \textbf{restriction problem} as the problem of choosing the agent's prior $\mu \in \varDelta(\Omega)$ together with a delegation set $A^* \subseteq \mathcal{A}$:
\begin{equation} \label{eq:principal_problem_restr}
\begin{aligned}
	\max_{\mu,A^*} &\left\{\sum_{j=1}^N \mu_p(\omega_j) \sum_{i=1}^N \pi(a_i|\omega_j) u(a_i,\omega_j) \right\},
	\\
	\text{s.t. } &\pi: \Omega \to \varDelta(A^*) \text{ solves \eqref{eq:agents_problem_restr} given } \mu \text{ and } A^*.
\end{aligned}
\end{equation}
Then we can state the result as follows.

\begin{proposition} \label{prop:restriction_no_good}
	The principal's restriction problem \eqref{eq:principal_problem_restr} is solved by the unrestricted delegation set $A^* = \mathcal{A}$ and the optimally misaligned agent's prior belief \eqref{eq:opt_belief_N}.
\end{proposition}

\subsection{Communication} \label{sec:comm}

In this section, we examine the importance of decision rights in our model. In particular, we juxtapose the \emph{delegation} scheme explored so far, under which the agent has the power to make the final decision, to \emph{communication}, where an agent must instead communicate his findings to the principal, who then  chooses the action. A large literature in organizational economics is devoted to comparing delegation and communication in various settings (see \citealp*{dessein_authority_2002,alonso_centralization_2008,rantakari_governing_2008} for some examples). In particular, \citet{deimen2019delegated} explore this question in the context of a delegated expertise problem similar to ours and show that communication outperforms delegation in the class of problems they consider. In contrast, we show that in our setting, communication performs exactly as well as delegation -- i.e., the principal will always find it optimal to follow the agent's recommendation.\footnote{The different conclusion is mainly driven by the assumption that states and actions are discrete in our model and continuous in \citet{deimen2019delegated}. This implies that the scope for the agent to deviate from the principal's preferred action under delegation is smaller in our model.}

Although the principal and the agent have the same preferences, it is generally unclear whether it is optimal for the principal to follow the agent's recommendation due to the misalignment in their beliefs. Namely, since the principal and the agent start from different prior beliefs, the same is true for posteriors: if the principal could observe the information that the agent obtained, her posterior belief would be different from that of the agent. This implies that ex interim, the principal could prefer a different action compared to the agent, and could benefit from overruling the agent's decision if she had the power to do so. However, this would mean that the agent's incentives to acquire information are different from the baseline model, and the principal could have some influence over the agent's choice of the signal structure via her final choice rule.\footnote{\citet*{argenziano_strategic_2016} provide one example of how the principal can manipulate the agent's information acquisition incentives under cheap talk communication.}
We show below that none of these effects are relevant in our setting, and there exists a communication equilibrium that replicates the delegation equilibrium.

In the ``communication game'', the final stage of our ``delegation game'' (``agent selects an action $a \in \mathcal{A}$'') is replaced by two. First, after observing signal realization $s \in \mathcal{S}$ generated by his signal structure $\phi$, the agent selects a recommendation (message) $\tilde{a} \in \mathcal{A}$ to the principal. After that, the principal observes the recommendation $\tilde{a}$, uses it to update her belief $\mu_p(\omega|\tilde{a})$ about the state of the world, and then selects an action $a \in \mathcal{A}$ that determines both parties' payoffs.
The equilibrium of the communication game is then defined as follows.\footnote{Given that message labels are arbitrary, we focus w.l.o.g. on ``direct'' equilibria, in which the agent's message corresponds to an action recommendation. Further, for simplicity we assume that the principal only observes the recommendation made by the agent, and not the signal realization he observed or the signal structure he requested. Finally, to economize on notation, we restrict attention to deterministic communication strategies and choice rules.}
\begin{definition}[Communication Equilibrium]
	An equilibrium of the cheap talk game is characterized by $(\mu^*, \{\phi^*_\mu, \tilde{\sigma}^*_\mu \}_{\mu \in \mathcal{M}}, \sigma^*, \mu_p)$, which consists of the following:
	\begin{enumerate}
		\item the principal's posterior beliefs $\mu_p : \mathcal{A} \to \varDelta(\Omega)$ that are consistent with $(\phi^*_\mu, \sigma^*_\mu)$ (i.e., satisfy Bayes' rule on the equilibrium path);
		\item the principal's choice rule $\sigma^* : \mathcal{A} \to \mathcal{A}$, which solves the following for every $\tilde{a} \in \mathcal{A}$, given the posterior $\mu_p$:
		\begin{align*}
			\max_{\sigma(\tilde{a})} \left\{ \sum_{\omega \in \Omega} \mu_p(\omega|\tilde{a}) u(\sigma(\tilde{a}), \omega) \right\};
		\end{align*}
		\item a collection of the agents' signal structures $\phi^*_\mu : \Omega \to \varDelta(\mathcal{S})$ and communication strategies $\tilde{\sigma}^*_\mu : \mathcal{S} \to \mathcal{A}$ that solves the following given $\sigma$ for every $\mu \in \mathcal{M}$:
		\begin{align*}
			\max_{\phi, \tilde{\sigma}} \left\{ \sum_{\omega \in \Omega} \mu(\omega) \sum_{s \in \mathcal{S}} \phi(s|\omega) u( \sigma( \tilde{\sigma}(s)) , \omega) - c_E(\phi,\mu) \right\};
		\end{align*}
		\item the principal's choice $\mu^* \in \mathcal{M}$ of the agent to whom the task is delegated, which solves the following given $(\phi^*_\mu, \tilde{\sigma}^*_\mu)$, $\sigma^*$, and $\mu_p$:
		\begin{align*}
			\max_{\mu} &\left\{\sum_{\omega \in \Omega}\mu_p(\omega)\sum_{s\in \mathcal{S}}\phi(s|\omega)u( \sigma( \tilde{\sigma}(s)) ,\omega)\right\}.
		\end{align*}
	\end{enumerate}
\end{definition}

We can then state the result as follows.

\begin{proposition} \label{prop:comm}
	There exists a communication equilibrium $(\mu^*, \{\phi^*_\mu, \tilde{\sigma}^*_\mu \}_{\mu \in \mathcal{M}}, \sigma^*, \mu_p)$ that is outcome-equivalent to the equilibrium $(\mu^*, \{\phi^*_\mu, \sigma^*_\mu \}_{\mu \in \mathcal{M}})$ of the original game, in the sense that $\mu^*$ and $\phi^*_{\mu^*}$ coincide across the two equilibria, $\tilde{\sigma}^*_{\mu^*} = \sigma^*_{\mu^*}$, and $\sigma^*$ is the identity mapping.
\end{proposition}

The result is, perhaps, unsurprising, since \cite{holmstrom1980} showed that communication is equivalent to restricting the agent's action set, and this latter instrument was shown in Section \ref{sec:restriction} to be irrelevant in our setting, as long as the principal can select an agent with the prior belief she prefers.
The result in Proposition \ref{prop:comm}, however, is subject to a few caveats. First, cheap talk models are plagued by equilibrium multiplicity: for any informative equilibrium, there exist equilibria with less informative communication, up to completely uninformative equilibria.\footnote{If an agent makes uninformed recommendations, it is optimal for the principal to ignore it. If the principal ignores the recommendation, it is optimal for the agent to not acquire any information. Neither player can unilaterally deviate to informative communication in this situation.} 
This means that, under communication, there is a risk of miscoordinating on an uninformative equilibrium, whereas under delegation the equilibrium is unique. The same force may also work the other way, and there may be equilibria that are preferred by the principal to the delegation equilibrium, that can only be sustained under cheap talk (see \citealp{argenziano_strategic_2016} for an example of how such equilibria may arise). However, the question of whether such equilibria exist is beyond the scope of this paper.

The second caveat lies in the fact that Proposition \ref{prop:comm} relies on state-matching preferences. In our setting, any action is either ``right'' or ``wrong'', without any degrees of correctness. The misalignment of beliefs across the principal and the agent is thus small enough to not warrant the principal overriding the agent's suggested action. In contrast, in the settings of \citet{che2009opinions}, \citet{argenziano_strategic_2016}, and \citet{deimen2019delegated}, both states and actions lie in a continuum, and the principal's loss is proportional to the distance between the realized state and the chosen action. In such a setting, any misalignment between the principal and the agent would lead to the principal being willing to override the agent's recommendation, leading to the delegation equilibrium being no longer directly sustainable under communication. 
This ability to exploit interim misalignment is also what drives the persuasion and prejudice avoidance channels that underlie the result of \cite{che2009opinions}. By shutting these channels down we provide a novel explanation for the desirability of bias in delegation.

On a separate note, it is immediate from Proposition \ref{prop:comm} that the same equilibrium would survive in a setting with \emph{verifiable communication} a l{\`a} \citet{dur2005producing} and \cite{che2009opinions}, where an agent chooses between disclosing a signal realization that he received and disclosing nothing -- as opposed to cheap talk communication assumed above, where the agent can send any message. Since in the cheap talk equilibrium described in Proposition \ref{prop:comm}, the principal always follows the (optimally chosen) agent's recommendation and takes the agent's most preferred action, the agent would never have an incentive to conceal evidence from the principal.

\section{Discussion} \label{sec:disc}

In this section, we discuss some of the assumptions behind our model.

We capture the problem of a principal who needs to make a decision but lacks information to do so. For simplicity, we assume that the principal is unable to acquire the information on her own. Our results extend naturally to the case when the principal can learn, but her information costs are larger than the agent's.
We further assume that the principal does not internalize the agent's information cost. The leading interpretation \citep*[shared by, e.g.,][]{lipnowski2019} of this assumption is that the cost reflects the cognitive process of the agent. Information acquisition costs thus lead to moral hazard, with the agent potentially not willing to acquire the amount of information desired by the principal. This is the main conflict between the two parties in our model.

The principal in our model is concerned with choosing the best agent for the job, where all experts have a common interest with the principal ex post, but differ in their ex ante opinions on the issue (prior beliefs about the state). Our results provide insights into the comparative statics of a principal-agent w.r.t. the degree of agent's ex ante misalignment. However, we believe that exploring the principal's problem of selecting one agent from a population with heterogeneous beliefs is valid as well.\footnote{\citet*{kahneman2021noise} survey a large body of evidence suggesting that similar experts and decision-makers in similar conditions make extremely different judgements and predictions, with a large share of these differences attributable to the interpersonal heterogeneity (and a smaller share being due to intra-personal noise in decision-making). We argue that this heterogeneity can be leveraged by the principal through selecting an agent whose bias fits a given problem the most.}
The agents' beliefs may be observable due to their reputation concerns, i.e., the need to publicly establish a particular stance on a broad policy question for sake of earning, and subsequently capitalizing on, a specific reputation

We allow the agents to have heterogeneous prior beliefs, and thus to ``agree to disagree''. Such settings are not uncommon in economic theory: see \citet{morris1995common,che2009opinions,alonso2016bayesian} for some examples and discussion. It is well known \citep[see ][]{aumann_agreeing_1976,bonanno_agreeing_1997} that agents starting with a common prior can not commonly know that they hold differing beliefs. We allow the agents to have heterogeneous prior beliefs, and thus to ``agree to disagree''.  While it may be possible to replicate our results in a common-prior model with asymmetric information, where an agent's ex ante belief is affected by some private information not observed by the principal, such a model would feature signaling concerns. E.g., an agent could learn something about the principal's prior from the fact that he was chosen for the job, which would introduce signaling concerns to the principal's problem. We prefer to abstract from such concerns and simply assume non-common priors from the start, but exploring such a signaling problem could be an interesting direction for future work.
We also show in the Online Appendix (Section \ref{sec:diff}) how our results translate to the case of common beliefs but misaligned preferences.

In line with the delegation literature, we assume that the principal cannot use monetary or other kinds of transfers to manage the agent's incentives. This is primarily because learning is non-contractible in most settings -- indeed, it is difficult to think of a setting, in which a learning-based contract could be enforceable, meaning either the principal or the agent could demonstrate beyond reasonable doubt exactly how much effort the agent has put into learning the relevant information, and what kind of conclusions he has arrived at. A simpler justification of the no-transfer assumption could be that such transfers are institutionally prohibited in some settings (see \citealp{laffont_politics_1990,armstrong_recent_2007,alonso2008} for some examples and a discussion of such settings). However, we provide results in the Online Appendix (Section \ref{sec:othertools}) suggesting that even when contracting is feasible, it may not improve upon hiring an agent with a misaligned belief.

Our results rely on information costs nudging the agent towards confirmatory learning, as well as the fact that the principal can exploit this learning bias. Specifically, confirmatory learning arises due to the flexibility of the agent's learning technology. We capture this flexibility by using a mode of rational inattention with uniformly posterior-separable information cost function, which allows the agent to acquire arbitrary signal structures (see \citealp{mackowiak2023rational} for a recent survey of the literature on rational inattention and \citealp*{caplin2022rationally} for a discussion of UPS costs).
The agent's optimal choice of a signal structure in this model depends on his prior belief: an agent whose prior is skewed towards some state of the world chooses a signal structure which is relatively more informative regarding that state and thus allows him to make a better decision in that state. 
We show in the Online Appendix Section \ref{sec:cost} that this dimension of flexibility is crucial, and the takeaways are different if one restricts attention to signal structures that are symmetric across states.
Our results, however, are \emph{not} specific to the UPS information costs and continue to hold with other information cost specifications that allow for flexible learning, as shown in the Section \ref{sec:cost} for the channel capacity cost \citep{woodford2012inattentive} and the log-likelihood ratio cost \citep*{pomatto2018cost}.
These robustness exercises also confirm that our results are driven by the fact that agents with different prior beliefs seek out different information, and are not purely driven by the artifact that the cost of information is dependent on the agent's prior belief (and so different agents have different concepts of what a \emph{cheaper} signal structure is), which is inherent to UPS cost functions.

Due to the inherent complexity of rational inattention models, we mostly confine our exploration to a discrete state-matching model, which strays away from the continuous models more commonly used in delegation problems. In a model with a continuum action space, the scope for the effects of misalignment to manifest is much larger, and hence the trade-off between the agent's information acquisition and decision-making would supposedly be different. However, we show in the Online Appendix (Section \ref{sec:genprefs}) that our results extend verbatim to a quadratic-loss framework common in literature on delegation. Due to tractability concerns, we confine the analysis there to a binary-quadratic model, as opposed to the more commonly encountered uniform-quadratic \citep{holmstrom1980} or normal-quadratic \citep{che2009opinions} models, but there are few reasons to believe that this dimension of richness would produce results that are substantially different.

\section{Conclusion}\label{sec:conc}

We show that hiring an agent with beliefs that are misaligned with those of the principal can be beneficial for the principal, though only when the principal herself is ex ante biased. We show this in a model where the agent can flexibly acquire costly information before making a decision. More specifically, a biased principal prefers to delegate to an agent who is ex ante more uncertain about what the best action is but is somewhat biased towards the same action as her. This is mainly due to agent's confirmatory learning: information costs drive the agent to err in favor of the ex ante optimal action. A more uncertain agent is less prone to it and makes mistakes more evenly across states and actions, which benefits the principal -- despite her also favoring the ex ante optimal action.
 
As we show, exploiting belief misalignment can be a valid instrument that the principal can use in delegation. The value of this instrument is highest to a moderately-biased principal, whereas both an unbiased and an extremely biased principals would optimally select an aligned agent. We show that in our setting, misalignment is strictly more helpful to the principal than the restriction of the agent's action set. On the contrary, the principal wants to nudge the agent to consider more actions: the set of actions the agent takes under optimal delegation is larger than under aligned delegation. We further show that communication performs on par with delegation in our setting. Finally, we show in the Online Appendix that misalignment performs on par with -- or better than -- contingent transfers (contracts) in some settings.

An assumption that may feel excessively strong in our analysis is the common knowledge of all agents' and the principal's prior beliefs. On the one hand, agents' viewpoints may be private, and agents may be strategic in presenting them to the principal. On the other hand, an agent could be making inferences about the principal's belief about the problem she is facing from the fact that she chose him for the job. Such signaling concerns could yield an economically meaningful effect, yet they are omitted from our model and could present a promising direction for future work.

\appendix

\section{Appendix: Proofs}

\subsection{Proof of Lemma \ref{lem:agents_soln_ups}}

We solve the agent's problem using the so-called posterior approach, where instead of a signal structure $\phi$, the agent maximizes over a distribution of posterior beliefs.\footnote{This approach is popular in the Bayesian Persuasion literature \citep{kamenica2011bayesian}. For some examples of this approach being used in problems with rationally inattentive agents, see, e.g., \cite{jain2021search} and \cite{matyskova2023bayesian}, with \cite{caplin2022rationally} presenting a general treatment.\label{foot:posterior_approach}}
In particular, consider an agent that chooses a signal structure $\phi$ with signal realizations in $S\in \mathcal{S}$. Each signal realization $s\in S$ is associated with a corresponding posterior belief $\eta \in [0,1]$ (as with the priors, we represent all posterior beliefs $\eta \in \varDelta(\Omega)$ in terms of the probability they assign to state $\omega=r$). Hence, a signal structure $\phi$ induces a distribution over posterior beliefs. It is commonly known from the literature on rational inattention (see, e.g., \citealp{caplindean2013}) that instead of considering the set of all signal structures $\phi$, one can consider the set of distributions of posterior beliefs that average out to the prior. 

In order to avoid introducing new notation for such distributions, we instead consider ``direct'' signal structures, in which signal realizations are labeled according to the posterior belief they induce. Formally, for this proof, let $\mathcal{S} \equiv \varDelta(\Omega)$ and restrict attention to the \emph{set of feasible direct signal structures} $\phi$:
\begin{equation*}
	\Phi_{\mu} \equiv \left\{ \phi \in \varDelta(\varDelta (\Omega)):\ \forall s \in \varDelta(\Omega): \eta(s)=s;\ \mathbb{E}[s|\phi]= \mu \right\}.
\end{equation*}
In other words, any signal structure in $\Phi_{\mu}$ prescribes some distribution over signal realizations $s \in \varDelta(\Omega)$ such that any $s$ induces a posterior belief $\eta(s)=s$, and signal realizations average out to the prior belief $\mu$. An agent with prior belief $\mu$ is limited to choosing a signal structure $\phi \in \Phi_{\mu}$. In what follows, we suppress signal realizations $s$ and refer to them according to the posterior beliefs $\eta(s)$ that they induce.

With state-matching preferences, the agent's optimal choice rule as a function of his posterior belief $\eta$ is given by (up to a tie-breaking rule)
\begin{equation*}
	\sigma^*(\eta) = \begin{cases}
		R & \text{ if } \eta \geq 0.5,
		\\
		L & \text{ if } \eta < 0.5.
	\end{cases}
\end{equation*}
The agent's expected decision payoff is hence $\mathbb{E}[u(\sigma^*(\eta),\omega) | \eta] = \max\{\eta,1-\eta\}$. Since the information cost (of a given direct signal structure $\phi$) is given by
\begin{align*}
	c(\phi, \mu) \equiv \lambda \Big[\mathbb{E}[\hat{c}(\eta)|\phi] - \hat{c}(\mu)\Big],
\end{align*}
the agent's problem then amounts to
\begin{equation}\label{eq:pscf_agent_problem}
	\begin{gathered}
		\max_{\phi \in \Phi_{\mu}} \mathbb{E}\Big[ \max \{\eta, 1-\eta\} - \lambda \hat{c}(\eta) + \lambda \hat{c}(\mu) \mid \phi \Big].    
	\end{gathered}
\end{equation}
Since the last term $\lambda \hat{c}(\mu)$ does not affect the maximization, it can be safely ignored.
Then we can define the agent's \emph{net utility} as $v(\eta) \equiv \max \{\eta, 1-\eta\} - \lambda\hat{c}(\eta)$. Problem \eqref{eq:pscf_agent_problem} is then equivalent to $\max_{\phi \in \Phi_{\mu}} \mathbb{E} [v(\eta) | \phi]$, which can be solved using a concavification approach \citep{kamenica2011bayesian,caplin2022rationally}. 
In particular, let $\hat{v}(\eta)$ denote the \emph{concavified net utility function} (also known as the concave closure of $v(\eta)$; \citealp{rockafellar1970convex}), which is defined as the minimal concave function that majorizes all net utilities. 
We characterize $\hat{v}(\eta)$ by establishing some properties of $v(\eta)$ and its derivatives $v',v''$. 

Note first that since $\hat{c}(\eta)$ is assumed to be symmetric around $0.5$, $v(\eta)$ is also symmetric around $0.5$: for any $\eta_1, \eta_2$ s.t. $\eta_1 + \eta_2 = 1$, it is true that $v(\eta_1) = v(\eta_2)$, $v'(\eta_1) = -v'(\eta_2)$, $v''(\eta_1) = v''(\eta_2)$. We thus limit our analysis to the interval $\eta \in [0,0.5)$, and the analysis for $\eta \in (0.5,1]$ is analogous.
Clearly, if $\eta\in [0,0.5)$ then $v(\eta)=1-\eta-\lambda \hat{c}(\eta)$. Function $\hat{c}(\eta)$ is assumed to be convex, thus $v''(\eta) = -\lambda \hat{c}''(\eta)<0$ for $\eta\in[0,0.5)$. Therefore, the derivative $v'(\eta)$ is decreasing for $\eta \in [0,0.5)$. 
Convexity and symmetry of $\hat{c}(\eta)$ also imply that it attains minimum at $\eta=0.5$, so $\hat{c}'(0.5)=0$ and $\lim\limits_{\eta\to 0.5-} v'(\eta) < 0$. On the other hand, from the Inada properties of $\hat{c}(\eta)$, we also have that $\lim\limits_{\eta\to 0+} v'(\eta) = +\infty$. Since derivative $v'(\eta)$ is continuous and decreasing for all $\eta\in [0,0.5)$, there exists a unique root of the equation $v'(\eta)=0$. Denote this root as $\eta_{L}$ and note that it is then a maximizer of $v(\eta)$ on the interval $[0,0.5)$.
Symmetry of $v(\eta)$ implies that there exists a symmetric maximizer $\eta_{R}\in(0.5,1]$ of $v(\eta)$ on $(0.5,1]$. Moreover, $\eta_{L}+\eta_{R}=1$ and $v(\eta_{L}) = v(\eta_{R})$.

It further follows from the monotonicity of $v'(\eta)$ on each interval that $v(\eta)$ is concave on $\eta \in [0,\eta_{L}] \cup [\eta_{R},1]$, hence $\hat{v}(\eta) = v(\eta)$ for such $\eta$.
Conversely, if $\eta \in (\eta_{L},\eta_{R})$, then $\hat{v}(\eta)$ is the straight line connecting points $(\eta_{L},f(\eta_{L}))$ and $(\eta_{R},f(\eta_{R}))$. 
Thus, if the agent's prior is $\mu\in [0,\eta_{L}] \cup [\eta_{R},1]$, then tangent lines to $v(\eta)$ and $\hat{v}(\eta)$ coincide, and the agent does not acquire any information; if $\mu \in (\eta_{L},\eta_{R})$, then the agent optimally chooses signal structure $\phi$ with two posterior beliefs in the support, $\eta_{L}$ and $\eta_{R}$.
Denoting $\bar{\mu} \equiv \eta_R$ (so $\eta_{L} = 1-\bar{\mu}$), we get the statement of the Lemma.

We proceed to explicitly calculate the optimal signal structure $\phi^*$ in the latter case (i.e., when the agent acquires information). 
First, by the law of total probability:
\begin{align} \label{eq:bq_phistar_uncond}
	\phi^*(\eta_R) \eta_R + \phi^*(\eta_L) \eta_L &= \mu
	& \Rightarrow &&
	\phi^*(\eta_R) &= \frac{\mu-\eta_L}{\eta_R-\eta_L}.
\end{align} 
Further, from the Bayes' rule, we have that $\eta_R = \frac{\mu \cdot \phi^*(\eta_R|1)}{\mu \cdot \phi^*(\eta_R|1) + (1-\mu) \cdot \phi^*(\eta_R|0)}$. From the law of total probability, $\mu \cdot \phi^*(\eta_R|1) + (1-\mu) \cdot \phi^*(\eta_R|0) = \phi^*(\eta_R)$. Combining the two, we get
\begin{equation} \label{eq:bq_phistar_cond}
	\begin{aligned}
		\phi^*(\eta_R|1) &= \frac{\eta_R}{\mu} \cdot \frac{\mu - \eta_L}{\eta_R - \eta_L} = 1-\phi^*(\eta_L|1),
		\\
		\phi^*(\eta_R|0) &= \frac{1-\eta_R}{1-\mu} \cdot \frac{\mu - \eta_L}{\eta_R - \eta_L} = 1-\phi^*(\eta_L|0),
	\end{aligned}
\end{equation}
which completes the characterization of $\phi^*$. 
However, it will also prove useful to note that while not reflected in the notation, $\phi^*$ does depend on the prior, $\mu$, and this dependence is given by
\begin{equation} \label{eq:bq_dphistar}
	\begin{aligned}
		\frac{\partial \phi^*(\eta_R|1)}{\partial \mu} &= \frac{\eta_R \cdot \eta_L}{\mu^2 \cdot (\eta_R-\eta_L)} = -\frac{\partial \phi^*(\eta_L|1)}{\partial \mu},
		\\
		\frac{\partial \phi^*(\eta_R|0)}{\partial \mu} &= \frac{(1-\eta_R) \cdot (1-\eta_L)}{(1-\mu)^2 \cdot (\eta_R-\eta_L)} = -\frac{\partial \phi^*(\eta_L|0)}{\partial \mu}.
	\end{aligned}
\end{equation}

\subsection{Proof of Proposition \ref{prop:PScost}}

Using the notation from the proof of Lemma \ref{lem:agents_soln_ups} ($\bar{\mu} = \eta_R$, $\underline{\mu} = \eta_L$), the principal's expected utility from hiring a learning agent with prior belief $\mu \in (\eta_L, \eta_R)$ is given by
\begin{align} \label{eq:pscf_principal_obj}
	\mathbb{E}[u(\sigma^*(\eta),\omega) | \mu_p, \phi^*] = 
	\mu_p \phi^*(\eta_R|r) + (1-\mu_p) \phi^*(\eta_L|l).
\end{align}
If it is optimal to hire a learning agent, then the optimal agent's prior $\mu^*$ must maximize \eqref{eq:pscf_principal_obj}, hence $\mu^*$ must solve the FOC given by
\begin{align*}
	\mu_p \frac{\eta_R \cdot \eta_L}{(\mu^*)^2 \cdot (\eta_R-\eta_L)} &= (1-\mu_p) \frac{(1-\eta_R) \cdot (1-\eta_L)}{(1-\mu^*)^2 \cdot (\eta_R-\eta_L)}
	\\ \iff
	\frac{\mu^*}{1-\mu^*} &= \sqrt{ \frac{\mu_p}{1-\mu_p} },
\end{align*}
where the first line uses \eqref{eq:bq_dphistar}, and then the second line follows from $\eta_L = 1-\eta_R$. It is trivial to verify that the SOC also hold at this point.
Representation \eqref{eq:opt_deleg_binary} therefore applies conditional on the principal hiring a learning agent.
To fully solve the principal's problem it is then left to characterize her choice between a learning and a non-learning agent.

Suppose w.l.o.g. $\mu_p \geq 0.5$ (the logic for $\mu_p < 0.5$ is analogous).
We first characterize the principal's optimal strategy for $\mu_p \in \left[ 0.5, (\mu^*)^{-1}(\eta_R) \right)$ -- i.e., such $\mu_p$ that an agent with $\mu^*(\mu_p)$ is learning. Here $(\mu^*)^{-1}(\mu)$ is the inverse of function $\mu^*(\mu_p)$ given by \eqref{eq:opt_deleg_binary}, so $(\mu^*)^{-1}(\eta_R) = \frac{\eta_R^2}{\eta_R^2 + (1-\eta_R)^2}$.
Conditional on hiring a non-learning agent, the principal is indifferent between all agents $\mu \in [\eta_R,1]$, since any such non-learning agent chooses $a=R$ and yields the principal an expected payoff equal to $\mu_p$.
If $\mu_p \in \left[ 0.5, (\mu^*)^{-1}(\eta_R) \right)$, hiring an optimal learning agent $\mu^*(\mu_p)$ yields
\begin{equation} \label{eq:pscf_princ_U_learn}
	\mu_p \phi^*(\eta_R|r) + (1-\mu_p) \phi^*(\eta_L|l) = \frac{\eta_R}{\eta_R-\eta_L} \left( \eta_R - 2 \eta_L \sqrt{\mu_p(1-\mu_p)} \right).
\end{equation}
Routine algebraic manipulations help establish that \eqref{eq:pscf_princ_U_learn} is convex in $\mu_p$, and it is tangent to $\mu_p$ at $\mu_p = (\mu^*)^{-1}(\eta_R)$. Therefore, hiring agent $\mu^*(\mu_p)$ is better than hiring a non-learning agent for all $\mu_p \in \left[ 0.5, (\mu^*)^{-1}(\eta_R) \right)$.

In turn, for $\mu_p \geq (\mu^*)^{-1}(\eta_R)$, maximizing \eqref{eq:pscf_principal_obj} yields a corner solution $\mu=\eta_R$, hence hiring a learning agent is not optimal, and the principal prefers instead to hire any non-learning agent $\mu \in [\eta_R,1]$ -- hence representation \eqref{eq:opt_deleg_binary} prescribes one optimal strategy for such $\mu_p$. This concludes the proof.

\subsection{Proof of Corollary \ref{corr:nonmonmisal}}

We start by showing single-peakedness of the optimal misalignment. Proof of Proposition \ref{prop:binarydelegationri} shows that $\mu^*(\mu_p)<\mu_p$ for all $\mu_p \in (0.5, 1)$, hence we can ignore the absolute value operator. Using expression \eqref{eq:opt_deleg_binary} we then obtain
\begin{align*}
	\frac{d}{d\mu_p} \left( \mu_p - \mu^*(\mu_p) \right) 
	&= \frac{ 4\mu_p(1-\mu_p) + 2\sqrt{\mu_p(1-\mu_p)}-1 }{ 2\sqrt{\mu_p(1-\mu_p)} \cdot \left( \sqrt{\mu_p} + \sqrt{1-\mu_p} \right)^2 },
\end{align*}
where the denominator is weakly positive for all $\mu_p \in (0.5, 1)$, and the numerator is positive if and only if $\sqrt{\mu_p(1-\mu_p)} \notin \left( \frac{-1-\sqrt{5}}{4}, \frac{-1+\sqrt{5}}{4} \right)$, which is equivalent to $\mu_p \leq \frac{1}{2} + \frac{1}{2} \cdot \sqrt{ \frac{\sqrt{5}-1}{2} } \approx 0.893$. Then $|\mu_p - \mu^*(\mu_p)|$ is increasing for these values of $\mu_p$ and decreasing otherwise, meaning it satisfies single-peakedness.

We now show the single-peakedness of the principal's benefit of misalignment. Let $U^*(\mu_p)$ denote the principal's expected payoff \eqref{eq:principal_problem_binary} from hiring the optimal agent $\mu=\mu^*(\mu_p)$ as a function of $\mu_p$. Similarly, let $U^A(\mu_p)$ denote the principal's expected utility from hiring the aligned agent $\mu=\mu_p$. Finally, let $\varDelta U(\mu_p) \equiv U^*(\mu_p) - U^A(\mu_p)$ denote the benefit of delegation. We want to show that $\varDelta U(\mu_p)$ is single-peaked on $\mu_p \in (0.5,1)$. To do so, we consider three separate intervals for $\mu_p$:
\begin{enumerate}
	\item $\mu_p \in (0.5, \bar{\mu})$. Then both the optimal and the aligned agent acquire information. Hence, $U^*(\mu_p)$ is given by \eqref{eq:pscf_princ_U_learn}, which is convex in $\mu_p$ and increasing on the defined interval. In turn, $U^A(\mu_p)$ is given by
	\begin{align*}
		\mu_p \pi^*_{\mu_p}(R|r) + (1-\mu_p) \pi^*_{\mu_p}(L|l)
		&= \bar{\mu},
	\end{align*}
	where $\pi^*_{\mu_p}$ are given by \eqref{eq:precisions_Bayes}. We can see that $U^A(\mu_p)$ is independent of $\mu_p$, meaning that $\varDelta U(\mu_p)$ is increasing on this interval.
	
	\item $\mu_p \in \left[\bar{\mu}, \left(\mu^*\right)^{-1}(\bar{\mu}) \right)$. Then the aligned agent takes action $a=R$ without learning, hence $U^A(\mu_p) = \mu_p$. The optimal agent acquires information, so $U^*(\mu_p)$ is still given by \eqref{eq:pscf_princ_U_learn}, which, from the proof of Proposition \ref{prop:binarydelegationri}, is convex in $\mu_p$ and is tangent to $\mu_p$ at $\mu_p = (\mu^*)^{-1}(\bar{\mu})$. It follows that $\varDelta U(\mu_p)$ is decreasing on this interval.
	
	\item $\mu_p \in \left[ \left(\mu^*\right)^{-1}(\bar{\mu}), 1 \right)$. Then both the optimal and the aligned agents take action $a=R$ without learning, so $U^*(\mu_p) = U^A(\mu_p) = \mu_p$ and $\varDelta U(\mu_p)=0$ is constant on this interval.
\end{enumerate}
This concludes the proof.

\subsection{Preliminaries for the Model with $N$ States and Actions} \label{sec:Ncase_prbs}

Let $\beta_{\mu} (a_{i})$ denote the respective unconditional probability of choosing alternative $a_i$, calculated using the agent's own prior belief $\mu$:
\begin{equation} \label{eq:choice_prob_uncond}
	\beta_{\mu}(a) \equiv \sum_{\omega \in \Omega} \mu(\omega) \pi(a|\omega).
\end{equation}

In what follows, we refer to problem \eqref{eq:principal_full_prb} as the principal's \textbf{full problem}, in contrast to the relaxed problem introduced further.

\paragraph*{The Agent's Problem}
Proceeding by backward induction, we start by looking at the problem of an agent with some prior belief $\mu$. Invoking Theorem 1 from \cite{MM}, we obtain that the agent's optimal decision rule satisfies:
\begin{align} \label{eq:choice_prob_cond_N}
	\pi(a_i|\omega_j) = \frac{ \beta_{\mu}(a_i) e^{\frac{u(a_i,\omega_j)}{\lambda}} }{ \sum_{k=1}^N \beta_{\mu}(a_k) e^{\frac{u(a_k,\omega_j)}{\lambda}} } = \frac{ (1+\delta) \beta_{\mu}(a_i) }{ 1 + \delta \beta_{\mu}(a_i) },
\end{align}
where $\delta \equiv e^{\frac{1}{\lambda}} - 1$, and $\beta_{\mu}(a_i)$ is defined by \eqref{eq:choice_prob_uncond} and itself depends on $\{\pi(a_i|\omega_j)\}_{j=1}^N$. 
While \eqref{eq:choice_prob_cond_N} does not provide a closed-form solution for the decision rule $\pi(a_i|\omega_j)$, it implies that the conditional choice probabilities $\pi$ are uniquely determined given the unconditional choice probabilities $\beta$, and this mapping depends solely on the agent's payoffs and not on his prior belief. 
In what follows, we use the implication that a collection of the unconditional choice probabilities $\beta$ pins down the whole decision rule $\pi$ and use $\beta$ to summarize the agent's chosen decision rule. 

The above is not to say that closed-form expressions cannot be obtained. \cite{caplin2019rational} show (see their Theorem 1) that in the setting of \cite{MM}, an agent with a prior belief $\mu$ optimally chooses a decision rule that generates unconditional choice probabilities
\begin{align} \label{eq:p3_2}
	\beta_{\mu} (a_{i}) = \max \left\{ 0, \frac{1}{\delta} \left( \frac{(K(\beta_{\mu})+\delta) \mu(\omega_{i}) }{ \sum\limits_{j\in C(\beta_{\mu})} \mu(\omega_{j}) } -1 \right) \right\},
\end{align}
where $C(\beta) \equiv \left\{i \in \{1,...,N\}: \beta(a_i)>0 \right\}$ denotes the \textbf{consideration set} implied by $\beta$, i.e., the set of actions that are chosen with strictly positive probabilities, and $K(\beta) \equiv |C(\beta)|$ denotes the power of (number of actions in) this set.

\paragraph*{The Principal's Problem}
As mentioned previously, \eqref{eq:choice_prob_cond_N} implies that a collection of unconditional choice probabilities $\beta$ pins down the whole decision rule $\pi$. Let us then consider a \textbf{relaxed problem} for the principal, in which instead of choosing the agent's prior $\mu$, she is free to select the unconditional choice probabilities $\beta \in \varDelta(\mathcal{A})$ directly:
\begin{align}
	&\max_{\beta} \left\{ \sum_{j=1}^{N} \mu_p(\omega_{j}) \left( \sum_{i=1}^{N} \frac{ (1+\delta) \beta(a_i) }{ 1 + \delta \beta(a_i) } u(a_{i},\omega_{j}) \right) \right\}
	\label{eq:principal_relaxed_prb}
	\\
	\iff 
	& \max_{\beta} \left\{ \sum_{j=1}^{N} \mu_p(\omega_{j}) \frac{ (1+\delta) \beta(a_j) }{ 1 + \delta \beta(a_j) } \right\}.
	\label{eq:principal_relaxed_prb2}
\end{align}
In the above, we used \eqref{eq:choice_prob_cond_N} to represent the conditional probabilities $\pi(a_i|\omega_j)$ in \eqref{eq:principal_full_prb} in terms of the unconditional probabilities $\beta(a_i)$. In what follows, we show that the solution to this relaxed problem is implementable in the full problem -- i.e., that there exists an agent's belief $\mu$ that generates the principal-optimal choice probabilities $\beta$. 

Note that $\beta(a_i)$ in the above represents the probability with which \emph{an agent} expects to select action $a_i$. The principal's expected probability of $a_i$ being selected, $\sum_{j=1}^N \mu_p(\omega_j) \pi(a_i|\omega_j)$, would be different if $\mu \neq \mu_p$. 
Despite the potential confusion this enables, analyzing the principal's problem through the prism of choosing $\beta$ is the most convenient approach due to the RI-logit structure of the solution to the agent's problem.

We can now state the solution to the principal's problem as follows.

\begin{lemma} \label{lemma:optimal_choice}
	The solution to the principal's relaxed problem \eqref{eq:principal_relaxed_prb2} is given by
	\begin{align*}
		\beta^*(a_{i}) = \max \left\{ 0, \frac{1}{\delta} \left( \frac{ (K(\beta^*)+\delta)\sqrt{\mu_p(\omega_{i})} }{ \sum\limits_{j\in C(\beta^*)} \sqrt{\mu_p(\omega_{j})}} -1 \right) \right\},
	\end{align*}
	where $\delta \equiv e^{\frac{1}{\lambda}} - 1$.
\end{lemma}

Lemma \ref{lemma:optimal_choice} describes the solution in terms of the action choice probabilities, which do not necessarily give the reader a good idea of its features and the intuition behind this solution. We explore these in more detail in Section \ref{sec:soln_properties}. Before that, however, we need to ensure that this solution is attainable in the principal's full problem, which is done in the following section.

The question then is: can the principal generate choice probabilities $\beta^*$ by appropriately choosing the agent's prior belief $\mu$? In the binary case, the answer was trivially ``yes'': due to continuity of the agent's strategy, by varying the agent's belief $\mu(r)$ between $0$ and $1$, the principal could induce any unconditional probability $\beta(R)$. In the multidimensional case, this is not immediate. However, the following result shows that the result still holds with $N$ actions and states under state-matching preferences. 

\begin{lemma}\label{lem:equivalence}
	In the principal's full problem \eqref{eq:principal_full_prb}, any vector $\beta \in \varDelta(\mathcal{A})$ of unconditional choice probabilities is \textbf{implementable}: there exists a prior belief $\mu \in \varDelta(\Omega)$ such that $\beta(a_i) = \sum_{j=1}^N \mu(\omega_j) \pi^*_\mu (a_i|\omega_j)$, where $\pi^*_\mu$ solves the agent's problem \eqref{eq:ri_problem_full} given $\mu$.
\end{lemma}

The lemma states that if $\mathcal{M} = \varDelta(\Omega)$, then the principal can generate any vector of unconditional action probabilities. Note that this does not imply that she is able to select any decision rule $\pi(a_i|\omega_j)$ -- if this were the case, under the state-matching preferences she would simply choose to have $\pi(a_i|\omega_i)=1$ for all $i$.
However, Lemma \ref{lem:equivalence} does imply that the choice probabilities described in Lemma \ref{lemma:optimal_choice} -- those that solve the principal's relaxed problem, -- are implementable and thus also solve her full problem.
The result does, however, rely on the state-matching preferences: we show in Section \ref{sec:diff} that it does not hold for arbitrary payoff functions.

\subsection{Proof of Lemma \ref{lemma:optimal_choice}}

The goal is to find the optimal choice probabilities $\beta^* \in \varDelta(\mathcal{A})$ which maximize the principal's expected utility \eqref{eq:principal_relaxed_prb2}. First, let us rewrite expression \eqref{eq:principal_relaxed_prb2} using $\delta \equiv e^{\frac{1}{\lambda}} - 1$:
\begin{align*}
	\sum_{j=1}^{N} \mu_p(\omega_j) \frac{ \beta(a_j) e^{\frac{1}{\lambda}} }{ 1 + \delta \beta(a_j) }
	&= \sum_{j \in C(\beta)} e^{\frac{1}{\lambda}} \frac{ \frac{\mu_p(\omega_j)}{\delta} (1 + \delta \beta(a_j)) - \frac{\mu_p(\omega_j)}{\delta} }{ 1 + \delta \beta(a_j) }
	\\
	&= \sum_{j \in C(\beta)} e^{\frac{1}{\lambda}} \left( \frac{\mu_p(\omega_j)}{\delta} - \frac{\mu_p(\omega_j)}{\delta \left( 1 + \delta \beta(a_j) \right)} \right).
\end{align*}
The first term in the brackets above is independent of $\beta$, so the principal's maximization problem is equivalent to 
\begin{equation}\label{principal_problem}
	\min_{\beta} \sum_{j \in C(\beta)} \frac{ \mu_p(\omega_{j}) }{ 1 + \delta \beta(a_{j})}.
\end{equation}
Let $\xi$ denote the Lagrange multiplier corresponding to the constraint $\sum\limits_{j=1}^{N} \beta(a_{j})=1$. Then the first-order condition for $\beta(a_i)$ with $i \in C(\beta)$ is
\begin{align} \label{eq:l1_1}
	(1 + \delta \beta(a_{i}))^{2} = - \frac{\mu_p(\omega_{i})}{\xi}.
\end{align}
Summing up these equalities over all $j \in C(\beta)$, we get that 
\begin{align} \label{eq:l1_2}
	\sum\limits_{j\in C(\beta)} (1 + \delta \beta(a_{j}))^{2} = -\frac{ \sum_{j\in C(\beta)}\mu_p(\omega_{j}) }{\xi}
\end{align}
Combining \eqref{eq:l1_1} and \eqref{eq:l1_2}:
\begin{align} \label{eq:l1_3}
	1 + \delta \beta(a_i) = \frac{ \sqrt{\mu_p(\omega_{i})} }{ \sqrt{\sum_{j\in C(\beta)}\mu_p(\omega_{j})} } \sqrt{\sum\limits_{j\in C(\beta)} (1 + \delta \beta(a_{j}))^{2}}
\end{align} 
Once again summing up these equalities over all $j \in C(\beta)$, we get that 
\begin{align*}
	K(\beta) + \delta = \frac{ \sum_{j\in C(\beta)}\sqrt{\mu_p(\omega_{j})} }{ \sqrt{\sum_{j\in C(\beta)}\mu_p(\omega_{j})} } \sqrt{\sum\limits_{j\in C(\beta)} (1 + \delta \beta(a_{j}))^{2}}.
\end{align*}
Expressing $\sqrt{\sum_{j\in C(\beta)} (1 + \delta \beta(a_{j}))^{2}}$ from this expression and plugging it into \eqref{eq:l1_3} allows us to express $\beta(a_i)$ (for $i \in C(\beta)$) in closed form as
\begin{align} \label{eq:l1_4}
	\beta(a_i) = \frac{1}{\delta} \left( \frac{(K(\beta) + \delta) \sqrt{\mu_p(\omega_{i})}}{\sum_{j\in C(\beta)}\sqrt{\mu_p(\omega_{j})}} -1 \right).
\end{align}
The necessary condition for option $i$ to be in a consideration set ($i \in C(\beta)$) is $\beta(a_i) \geq 0$ or, equivalently,
\begin{align*}
	\sqrt{\mu_p(\omega_{i})} > \frac{1}{K(\beta)+\delta}\sum\limits_{j\in C(\beta)}\sqrt{\mu_p(\omega_{j})}.
\end{align*}

Now let $\xi_{k}$ denote the Lagrange multiplier for the constraint $\beta(a_{k})\geq 0$. Then the first-order condition for an alternative $k \notin C(\beta)$ that is not chosen is
\begin{align*}
	\mu_p(\omega_{k}) &= -\xi-\xi_{k} & \Rightarrow && \mu_p(\omega_{k}) &\leq -\xi.
\end{align*}
Plugging in $\xi$ from \eqref{eq:l1_1} into the inequality above yields 
\begin{align*}
	\mu_p(\omega_{k}) &\leq \frac{\sum_{j\in C(\beta)}\mu_p(\omega_{j})}{\sum_{j\in C(\beta)}(1+\delta \beta(a_{j}))^{2}} 
	& \Leftrightarrow && 
	\sqrt{\mu_p(\omega_{k})} &\leq \frac{1}{K(\beta)+\delta} \sum\limits_{j\in C(\beta)} \sqrt{\mu_p(\omega_{j})}
\end{align*}
for all $k \notin C(\beta)$.

Since the minimization problem has a convex objective function and linear constraints, the Kuhn-Tucker conditions are necessary and sufficient. Thus the necessary and sufficient conditions that the solution $\beta^*$ must satisfy are given by:
\begin{align*}
	\begin{cases}
		\sqrt{\mu_p(\omega_{i})} > \frac{1}{K(\beta^*)+\delta} \sum\limits_{j\in C(\beta^*)}\sqrt{\mu_p(\omega_{j})} & \text{ for all } i \in C(\beta^*),
		\\
		\sqrt{\mu_p(\omega_{k})} \leq \frac{1}{K(\beta^*)+\delta} \sum\limits_{j\in C(\beta^*)}\sqrt{\mu_p(\omega_{j})} & \text{ for all } k \notin C(\beta^*).
	\end{cases}
\end{align*}

Recall that we assumed, without loss of generality, that $\mu_p(\omega_{1}) \geq \mu_p(\omega_{2}) \geq \ldots \geq \mu_p(\omega_{N})$. Suppose that the solution $\beta^*$ is such that $K(\beta^*) = K'$. Clearly then, in the optimum, the consideration set $C(\beta^*)$ will consist of the first $K'$ alternatives.

Denote $\Delta_{L} \equiv (L+\delta) \sqrt{\mu_p(\omega_{L})} - \sum\limits_{j=1}^{L} \sqrt{\mu_p(\omega_{j})}$. Notice that for all $L>1$:
\begin{align*}
	\Delta_{L} \equiv & (L+\delta) \sqrt{\mu_p(\omega_{L})} - \sum\limits_{j=1}^{L} \sqrt{\mu_p(\omega_{j})}
	\\
	=& (L-1+\delta) \sqrt{\mu_p(\omega_{L-1})} - \sum\limits_{j=1}^{L-1} \sqrt{\mu_p(\omega_{j})} - \sqrt{\mu_p(\omega_{L})} 
	\\
	&- (L-1+\delta) \sqrt{\mu_p(\omega_{L-1})} + (L+\delta) \sqrt{\mu_p(\omega_{L})}
	\\
	=& \Delta_{L-1} - (L-1+\delta) \left( \sqrt{\mu_p(\omega_{L-1})} - \sqrt{\mu_p(\omega_{L})} \right).
\end{align*}
Therefore, $\Delta_{L}$ decreases in $L$. Since $\Delta_{1}>0$, there either exists unique $K'$ such that $\Delta_{K'}>0$ and $\Delta_{K'+1}\leq 0$, or $\Delta_{L}>0$ for all $L$. In the former case, $K(\beta^*) = K'$, and in the latter case, $K(\beta^*) = N$. 

In the end, the solution to the principal's problem is given by $\beta^*(a_i)$ as in \eqref{eq:l1_4} if $i \in C(\beta^*)$, $\beta^*(a_i)=0$ if $i \notin C(\beta^*)$, and $C(\beta^*) = {1,...,K(\beta^*)}$, where $K(\beta^*)$ is as described above.

\subsection{Proof of Lemma \ref{lem:equivalence}}

Corollary 2 from \cite{MM} shows that a vector of the unconditional choice probabilities $\beta_{\mu} \in \varDelta(\mathcal{A})$ solves \eqref{eq:choice_prob_cond_N} only if it solves the system of equations given by
\begin{align} \label{eq:agent_cond_N}
	\sum_{j=1}^N \mu(\omega_j) \frac{ e^{\frac{u(a_i,\omega_j)}{\lambda}} }{ \sum_{k=1}^N \beta_{\mu}(a_k) e^{\frac{u(a_k,\omega_j)}{\lambda}} } = 1,
\end{align}
for every $i \in \{1,...,N\}$ such that $\beta_{\mu}(a_i) > 0$.

The question then is: given a vector $\beta \in \Delta(\mathcal{A})$ of unconditional choice probabilities, can we find $\mu \in \varDelta(\Omega)$ that solves the following system of equations:
\begin{equation} \label{eq:t1_2}
\begin{cases}
	\mu(\omega_{1})+\mu(\omega_{2}) +\ldots+ \mu(\omega_{N})=1, & 
	\\
	\sum\limits_{j=1}^{N}\mu(\omega_{j})\frac{e^\frac{u(a_{i},\omega_{j})}{\lambda}}{\sum\limits_{k=1}^{N} \beta(a_k) e^\frac{u(a_{k},\omega_{j})}{\lambda}}=1 &\forall i\in C(\beta).
\end{cases}
\end{equation}
The system above is a linear system of $K(\beta)+1$ equations with $N$ unknowns. To prove the solution exists, we use the Farkas' lemma \citep[][Corollary 5.85]{aliprantis2006infinite}. It states that given some matrix $A \in \mathbb{R}^{m\times n}$ and a vector $b \in \mathbb{R}^{m}$, the linear system $Ax=b$ has a non-negative root $x\in\mathbb{R}^{n}_+$ if and only if there exists no vector $y \in \mathbb{R}^{m}$ such that $A'y\geq 0$ with $b'y<0$. The two latter inequalities applied to our case form the following system:
\begin{equation} \label{eq:t1_3}
	\begin{cases}
		y_{0} \left( \sum\limits_{k=1}^{N} \beta(a_k) e^\frac{u(a_{k},\omega_{j})}{\lambda} \right) + \left( \sum\limits_{i \in C(\beta)} y_{i} e^\frac{u(a_{i},\omega_{j})}{\lambda} \right) \geq 0 & \forall j\in \{1,...,N\}, \\ 
		y_0 + \sum\limits_{i \in C(\beta)} y_{i} < 0. &
	\end{cases}
\end{equation}
We need to show there exists no $y \in \mathbb{R}^{K(\beta)+1}$ that solves the system above.
Let us define $z_{i} \equiv y_{i} + y_{0} \beta(a_i) $ for $i \in C(\beta)$. Then, recalling that $e^\frac{u(a_{i},\omega_{i})}{\lambda} = e^\frac{1}{\lambda}$ and $e^\frac{u(a_{i},\omega_{j})}{\lambda} = 1$ for $i \neq j$, system \eqref{eq:t1_3} transforms to
\begin{equation} \label{eq:t1_1}
	\begin{cases}
		z_j e^\frac{1}{\lambda} + \sum\limits_{i \in C(\beta) \backslash \{j\}} z_{i} \geq 0 & \forall j\in C(\beta), \\ 
		\sum\limits_{i \in C(\beta)} z_{i} \geq 0 & \forall j\in \{1,...,N\} \backslash C(\beta), \\
		\sum\limits_{i \in C(\beta)} z_{i} < 0. &
	\end{cases}
\end{equation}
System \eqref{eq:t1_1} above does not have a solution. Indeed, if $C(\beta) \subsetneq \{1, ..., N\}$, then the middle set of inequalities directly contradicts the latter inequality. If $C(\beta) = \{1, ..., N\}$, then subtracting the latter inequality from the former, for a given $j \in C(\beta)$, yields $z_j \delta \geq 0 \iff z_j \geq 0$. Since this must hold for all $j \in C(\beta)$, we obtain a contradiction with the latter inequality, $\sum\limits_{i \in C(\beta)} z_{i} < 0$.

By the Farkas' lemma, we then conclude that for any vector $\beta \in \Delta(\mathcal{A})$ there exists a belief $\mu \in \varDelta(\Omega)$ that solves system \eqref{eq:t1_2}. This concludes the proof.

\subsection{Proof of Theorem \ref{prop:opt_choice_beliefs}}

Consider an agent with a prior belief
\begin{align} \label{eq:p3_1}
	\mu(\omega_{i}) = \frac{ \sqrt{\mu_p(\omega_{i})} }{ \sum\limits_{j=1}^{N} \sqrt{\mu_p(\omega_{j})} }.
\end{align} 
It is trivial to verify that prior belief $\mu$ defined this way satisfies the candidate strategy in the statement of the theorem, and hence represents the candidate strategy.
Consider an agent hired in accordance with the candidate rule. Substituting \eqref{eq:p3_1} into \eqref{eq:p3_2} yields 
\begin{align}
	\beta_{\mu} (a_{i}) = \max \left\{ 0, \frac{1}{\delta} \left( \frac{(K(\beta^*) + \delta) \sqrt{\mu_p(\omega_{i})} }{ \sum\limits_{j\in C(\beta^*)} \sqrt{\mu_p(\omega_{j})} } - 1 \right) \right\},
\end{align}
which are exactly the probabilities stated in Lemma \ref{lemma:optimal_choice}.
Therefore, an agent hired according to the candidate strategy makes decisions in such a way that generates the principal-optimal unconditional choice probabilities.
Therefore, delegation according to the candidate strategy is indeed optimal for the principal.

\subsection{Proof of Corollary \ref{cor:opt_beliefs_properties}}

\textbf{The third property} (order preservation, $\mu^{*}(\omega_{1})\geq \mu^{*}(\omega_{2}) \geq \ldots \geq \mu^{*}(\omega_{N})$) follows clearly from \eqref{eq:p3_1} and is a necessary condition for the two beliefs to be comparable w.r.t. the majorization order. 

To show \textbf{the first property} (majorization), we want to show that for
all $k\in\{1,...,N\}$, 
\[
	\sum_{i=1}^{k}\mu^{*}(\omega_{i})\leq\sum_{i=1}^{k}\mu_{p}(\omega_{i}).
\]
Using (\ref{eq:p3_1}), this inequality can be rewritten as
\begin{align}
	\nonumber
	\frac{\sum\limits _{i=1}^{k}\sqrt{\mu_{p}(\omega_{i})}}{\sum\limits _{j=1}^{N}\sqrt{\mu_{p}(\omega_{j})}} &\leq \sum_{l=1}^{k}\mu_{p}(\omega_{l}),
	\\ \nonumber \iff 
	1 & \leq\frac{\sum\limits_{l=1}^{k}\mu_{p}(\omega_{l})\sum\limits _{j=1}^{N}\sqrt{\mu_{p}(\omega_{j})}}{\sum\limits _{i=1}^{k}\sqrt{\mu_{p}(\omega_{i})}}
	\\ \iff 
	1 & \leq \sum\limits_{l=1}^{k}\mu_{p}(\omega_{i}) + \sum\limits _{j=k+1}^{N}\sqrt{\mu_{p}(\omega_{j})}\cdot\rho,
	\label{eq:maj_proof1}
\end{align}
where $\rho\equiv\frac{\sum\limits_{l=1}^{k}\mu_{p}(\omega_{l})}{\sum\limits _{i=1}^{k}\sqrt{\mu_{p}(\omega_{i})}}$.
Since $\sum\limits_{l=1}^{N}\mu_{p}(\omega_{i}) = 1$, to verify that \eqref{eq:maj_proof1} holds, it is enough to show that $\rho \geq \sqrt{\mu_{p}(\omega_{k+1})}$ (since $\sqrt{\mu_{p}(\omega_{k+1})} \geq ... \geq \sqrt{\mu_{p}(\omega_{N})}$), with the former inequality being equivalent to
\begin{align*}
	\frac{\sum_{l=1}^{k}\mu_{p}(\omega_{l})}{\sum\limits _{i=1}^{k}\sqrt{\mu_{p}(\omega_{i})}}  \geq & \sqrt{\mu_{p}(\omega_{k+1})}
	\\ \nonumber
	\iff\mu_{p}(\omega_{1})+...+\mu_{p}(\omega_{k})\geq & \sqrt{\mu_{p}(\omega_{1})\mu_{p}(\omega_{k+1})}+...+\sqrt{\mu_{p}(\omega_{k})\mu_{p}(\omega_{k+1})},
\end{align*}
which is true because $\mu_{p}(\omega_{k+1}) \leq \mu_{p}(\omega_{k}) \leq ... \leq \mu_{p}(\omega_{1}).$
We conclude that $\mu_{p}$ majorizes $\mu^{*}$.

To show \textbf{the second property}, observe that the Shannon entropy for some probability vector $\mu$ is given by 
\[
H\left(p\right)\equiv-\sum_{i=1}^{N}\mu\left(\omega_{i}\right)\cdot\log\mu\left(\omega_{i}\right).
\]
\citet*[p.101]{marshall1979inequalities} show that $H\left(p\right)$ is a Schur-concave function (i.e., it is decreasing w.r.t. the majorization order). It therefore holds that if $\mu_{p}$ majorizes $\mu^{*}$, then $H(\mu_{p})\leq H(\mu^{*})$.
\textbf{The fourth property} (more uncertainty at the extremes) also follows immediately from majorization.
This concludes the proof.

\subsection{Proof of Proposition \ref{prop:consideration}}

\citet{caplin2019rational} show (see their Theorem 1) that in our problem, probabilities $\beta_{\mu}$ given agent's prior $\mu$ are given by 
\begin{align} \label{eq:choiceprobs_caplin} 
	\beta_{\mu}(a_i) = \max \left\{ 0, \frac{1}{\delta} \left( \frac{ (K(\beta_{\mu})+\delta) {\mu(\omega_{i})} }{ \sum\limits_{j\in {C(\beta_{\mu})}} {\mu(\omega_{j})}} -1 \right) \right\},
\end{align}
where $\delta \equiv e^{\frac{1}{\lambda}} - 1$, $C(\beta_{\mu})$ is the consideration set corresponding to $\beta_{\mu}$, and $K(\beta_{\mu}) \equiv |C(\beta_{\mu})|$.
Since $\mu_p(\omega_1) > ... > \mu_p(\omega_N)$, the consideration set in the aligned problem is then simply $\bar{C} = C(\beta_{\mu_p}) = \{1,...,\bar{K}\}$, where its size $\bar{K}$ is the unique solution of
\begin{align} 
	\label{eq:K_aligned}
	&\mu_p(\omega_{\bar{K}}) > \frac{1}{\bar{K}+\delta} \sum_{j=1}^{\bar{K}} \mu_p(\omega_j) \geq \mu_p(\omega_{ \bar{K}+1})
	\\
	\iff& 
	\sum\limits_{j=1}^{\bar{K}} \frac{\mu_p(\omega_{j})}{\mu_p(\omega_{\bar{K}})} < \bar{K} + \delta \leq \sum\limits_{j=1}^{\bar{K}} \frac{\mu_p(\omega_{j})}{\mu_p(\omega_{\bar{K}+1})}
\end{align}

Applying the same logic and Lemma \ref{lemma:optimal_choice} to the case of optimal delegation, we can see that size $K^* \equiv |C^*|$ of consideration set $C^* = C(\beta_{\mu^*})$ must satisfy
\begin{align} 
	\label{eq:K_optimal}
	&\sqrt{\mu_p(\omega_{K^*})} > \frac{1}{K^*+\delta} \sum_{j=1}^{K^*} \sqrt{\mu_p(\omega_j)} \geq \sqrt{\mu_p(\omega_{ K^*+1})}
	\\
	\iff &
	\label{eq:p4_3}
	\sum\limits_{j=1}^{K^*} \frac{\sqrt{\mu_p (\omega_{j})}}{\sqrt{\mu_p (\omega_{K^*})}} < K^* + \delta \leq \sum\limits_{j=1}^{K^*} \frac{\sqrt{\mu_p (\omega_{j})}}{\sqrt{\mu_p (\omega_{K^*+1})}}.
\end{align}

Since $\frac{\mu_p(\omega_{i})}{\mu_p(\omega_{\bar{K}})} > 1$ for all $i < \bar{K}$, we have that $\frac{ \mu_p(\omega_{i}) }{ \mu_p(\omega_{\bar{K}}) } > \frac{ \sqrt{\mu_p(\omega_{i})} }{ \sqrt{\mu_p(\omega_{\bar{K}})} } > 1$ for all $i < \bar{K}$. Therefore,
\begin{align} \label{eq:p4_1}
	\sum\limits_{j=1}^{\bar{K}} \frac{ \sqrt{\mu_p (\omega_{j})} }{ \sqrt{\mu_p (\omega_{\bar{K}})}} < \bar{K} + \delta.
\end{align}
Two cases are possible, depending on whether
\begin{align} \label{eq:p4_2}
	\bar{K} + \delta \gtreqless \sum\limits_{j=1}^{\bar{K}} \frac{\sqrt{\mu_p (\omega_{j})}}{\sqrt{\mu_p (\omega_{\bar{K}+1})}}.
\end{align}
If $\bar{K} + \delta \leq RHS$ in \eqref{eq:p4_2}, where RHS refers to the right-hand side, then together with \eqref{eq:p4_1} this implies that $\bar{K}$ solves \eqref{eq:p4_3}. Thus, $\bar{K} = K^*$ and $\bar{C} = C^*$, which satisfies the statement of the proposition. 

If, however, $\bar{K} + \delta > RHS$ in \eqref{eq:p4_2}, then $\bar{K}$ does not solve \eqref{eq:p4_3}. In this case, note that going from $K$ to $K+1$ increases the LHS of \eqref{eq:p4_2} by $1$ and increases the RHS by the amount strictly greater than $1$, since a new term $\frac{ \sqrt{\mu_p(\omega_{K+1})} }{ \sqrt{\mu_p(\omega_{K+2})} } > 1$ is added to the sum, and all existing terms increase because $\mu_p(\omega_{K + 1}) < \mu_p(\omega_K)$. This holds for all $K$, meaning that if $\bar{K} + \delta > RHS$ in \eqref{eq:p4_2}, then $K + \delta > \sum\limits_{j=1}^{K} \frac{\sqrt{\mu_p (\omega_{j})}}{\sqrt{\mu_p (\omega_{K+1})}}$ for all $K < \bar{K}$. Therefore, the unique solution $K^*$ of \eqref{eq:p4_3} must be such that $K^* > \bar{K}$, and so $\bar{C} \subseteq C^*$. 
This concludes the proof.

\subsection{Proof of Proposition \ref{prop:restriction_no_good}}

Consider some restriction set $A^* \subset \mathcal{A}$ and agent's belief $\mu$; refer to the agent's problem \eqref{eq:agents_problem_restr} given these as ``the original problem''. We now show that its solution can also be obtained in a ``the modified problem'' with $\tilde{A}^* \equiv \mathcal{A}$ and $\tilde{\mu} \equiv \mu|A^*$. Specifically, we construct the new belief $\tilde{\mu}$ by conditioning the original belief $\mu$ on the original restriction set $A^*$: $\tilde{\mu}(\omega_i) = \frac{\mu(\omega_i) }{ \mu(A^*) } \mathbb{I}\{\omega_i \in A^*\}$ for all $\omega_i \in \Omega$, where $\mu(A^*) = \sum_{\omega_j \in A^*} \mu(\omega_j)$. 

Let $\Omega^* \equiv \left\{ \omega_i \in \Omega \mid a_i \in A^* \right\}$ denote the set of states that the agent is allowed to match.
Observe from \eqref{eq:c_entropy} and \eqref{eq:agents_problem_restr} that the agent's objective function is additively separable across states: it can be represented as
\begin{align} \label{eq:agents_problem_innerproduct}
	\max_\pi \left\{ \sum_{j=1}^N \mu(\omega_j) \left( U_j(\pi,\phi,\lambda) - \lambda \ln \mu(\omega_j) \right) \right\},
\end{align}
where $U_j(\pi,\phi,\lambda)$ captures the whole expected payoff minus information cost in state $\omega_j$, apart from the $\lambda \ln \mu(\omega_j)$ cost term, which does not affect the maximization and can be ignored. Note that $U_j(\pi,\phi,\lambda)$ does not depend on $\mu(\omega_j)$.

In the original problem, the agent cannot choose any $a \notin A^*$, hence his optimal signal structure is uninformative about all states $\omega \notin \Omega^*$ (since this minimizes the information cost without affecting the payoff). Taking this as granted, all terms related to states $\omega \notin \Omega^*$ can be ignored in \eqref{eq:agents_problem_innerproduct}, since they no longer affect the maximization. The agent's objective function in original problem \eqref{eq:agents_problem_innerproduct} can thus be restated as
\begin{align}
	\max_\pi \left\{ \sum_{\omega_j \in \Omega^*} \mu(\omega_j) U_j(\pi,\phi,\lambda) \right\},
\end{align}
which is trivially equivalent to the agent's objective function in the modified problem. Therefore, if some $\pi^*$ solves the original problem, it also solves the modified problem.\footnote{While not important for the argument, it is worth noting that the modified problem admits many solutions $\pi$ that differ in the choice rules they prescribe for states $\omega_i \notin \Omega^*$. Since the agent treats these states as impossible in the modified problem, he perceives any information that distinguishes these states as free. But equally, he faces no motive to learn about these states. We believe the latter force motivates selecting the $\pi^*$ that solves the original problem.}
We conclude that it is without loss of generality to restrict the principal to offering the unrestricted delegation set $\tilde{A}^* \equiv \mathcal{A}$ to the agent, since any individual action can be ruled out by the appropriate choice of the agent's prior belief $\mu$, meaning that the optimal misalignment $\mu^*$ characterized in Theorem \ref{prop:opt_choice_beliefs} is still optimal in the principal's restriction problem.

\subsection{Proof of Proposition \ref{prop:comm}}

We first show that there exists an equilibrium in the communication game that replicates the delegation equilibrium: the optimal agent acquires the same information, makes a truthful action recommendation, and the principal follows the recommendation.

Suppose that under delegation, the optimally chosen agent follows a decision rule $\beta_{\mu^*}$ that yields a consideration set $C^* \equiv C(\beta_{\mu^*}) = \{1, ... , K^*\}$. By Lemma \ref{lemma:optimal_choice}, we have that
\begin{align}
	\sqrt{\mu(\omega_{K^*})} &\geq \frac{1}{K^*+\delta} \sum_{i=1}^{K^*} \sqrt{\mu(\omega_i)}
	\nonumber
	\\
	\iff
	\delta \sqrt{\mu(\omega_{K^*})} &\geq \sum_{i=1}^{K^*-1} \left( \sqrt{\mu(\omega_i)} - \sqrt{\mu(\omega_{K^*})} \right)
	\label{eq:p9_3}
\end{align}
Suppose the agent reports truthfully. Given the state-matching payoffs, for the principal to follow recommendation $\tilde{a} = \tilde{a}_{K^*}$ whenever it is issued, it must hold that
\begin{align} \label{eq:p9_1}
	\mu_p(\omega_{K^*} | \tilde{a}_{K^*}) = \max_i \mu_p(\omega_i | \tilde{a}_{K^*}),
\end{align}
where $\mu_p(\omega|\tilde{a})$ is the probability that the principal's posterior belief assigns to state $\omega$ after hearing recommendation $\tilde{a}$ from the agent. In equilibrium, the principal's posterior $\mu_p(\omega_{K^*} | \tilde{a}_{K^*})$ must satisfy Bayes' rule:
\begin{align*}
	\mu_p(\omega_{K^*} | \tilde{a}_{K^*}) &= \frac{\pi(a_{K^*} | \omega_{K^*}) \mu_p(\omega_{K^*})}{\sum_{i=1}^N \mu_p(\omega_i) \pi(a_{K^*} | \omega_i)}
	\\
	&= \frac{\beta_{\mu^*}(a_{K^*}) e^\frac{1}{\lambda}}{\beta_{\mu^*}(a_1) + ... + \beta_{\mu^*}(a_{K^*-1}) + \beta_{\mu^*}(a_{K^*}) e^\frac{1}{\lambda}} \cdot \frac{\mu_p(\omega_{K^*})}{\sum_{i=1}^N \mu_p(\omega_i) \pi(a_{K^*} | \omega_i)}
	\\
	&= \frac{\beta_{\mu^*}(a_{K^*}) e^\frac{1}{\lambda}}{1 + \delta \beta_{\mu^*}(a_{K^*})} \cdot \frac{\mu_p(\omega_{K^*})}{\sum_{i=1}^N \mu_p(\omega_i) \pi(a_{K^*} | \omega_i)}
	\\
	&= \frac{ \sum_{i=1}^{K^*} \sqrt{\mu_p(\omega_i)} }{K^*+\delta}  
	\cdot \beta_{\mu^*}(a_{K^*}) e^\frac{1}{\lambda} \cdot \frac{ \sqrt{\mu(\omega_{K^*})} }{\sum_{i=1}^N \mu_p(\omega_i) \pi(a_{K^*} | \omega_i)},
\end{align*}
Where the last line is obtained by plugging the expression for $\beta^*(a_{K^*})$ from Lemma \ref{lemma:optimal_choice} as $\beta_{\mu^*}(a_{K^*})$ in the denominator of the preceding line. Similarly, we can calculate the probability that the principal's posterior assigns to any other state $\omega_j$:
\begin{align*}
	\mu_p(\omega_{j} | \tilde{a}_{K^*}) &=
	\begin{cases}
		\frac{ \sum_{i=1}^{K^*} \sqrt{\mu_p(\omega_i)} }{K^*+\delta}  
		\cdot \beta_{\mu^*}(a_{K^*}) e^\frac{1}{\lambda} \cdot \frac{ \sqrt{\mu(\omega_{j})} }{\sum_{i=1}^N \mu_p(\omega_i) \pi(a_{K^*} | \omega_i)} & \text{ if } j < K^*,
		\\
		0 & \text{ if } j > K^*.
	\end{cases}
\end{align*}
For condition \eqref{eq:p9_1} to hold, it is then enough for
\begin{align} \label{eq:p9_2}
	e^\frac{1}{\lambda} \sqrt{\mu(\omega_{K^*})} &\geq \sqrt{\mu(\omega_1)}
	& \iff && 
	\delta \sqrt{\mu(\omega_{K^*})} &\geq \sqrt{\mu(\omega_{1})} - \sqrt{\mu(\omega_{K^*})},
\end{align}
to be satisfied. Note, however, that it is strictly weaker than \eqref{eq:p9_3}, since
\begin{align*}
	\sqrt{\mu(\omega_{1})} - \sqrt{\mu(\omega_{K^*})} < \sum_{i=1}^{K^*-1} \left( \sqrt{\mu(\omega_i)} - \sqrt{\mu(\omega_{K^*})} \right).
\end{align*}
Therefore, we conclude that \eqref{eq:p9_2} holds, and thus it is optimal for the principal to choose action $a_{K^*}$ when the agent with prior belief $\mu^*$ recommends it. 

Following the same argument, we can show the same for any other recommendation $\tilde{a}_i$ for $i \in C(\beta^*)$: the necessary and sufficient condition for the principal to find it optimal to follow the recommendation would be
\begin{align*}
	e^\frac{1}{\lambda} \sqrt{\mu(\omega_{i})} &\geq \sqrt{\mu(\omega_1)},
\end{align*}
which is implied by \eqref{eq:p9_1}, since $\mu(\omega_i) \geq \mu(\omega_{K^*})$ for $i \in C(\beta^*)$. This concludes the proof.

\bibliographystyle{apalike}
\bibliography{sample}

\newpage

\section{Online Appendix}

\subsection{Misaligned Beliefs Versus Contracting} \label{sec:othertools}

The analysis in the main paper set the foundation for using misalignment in beliefs as an instrument in delegation. We have already shown in Section \ref{sec:restriction} that restricting the delegation set is, in contrast, not a valuable instrument in our setting. In this section, we study the value of contracts between the principal and the agent. We show that contracts cannot improve the outcomes if the principal can find an optimally misaligned agent. This supports the value of misalignment as an alternative delegation tool.

We keep the overall structure of the problem the same as in Section \ref{sec:gencase}, assuming that the information cost is given by Shannon entropy. We then modify the problem to allow the principal to offer additional payments to the agent, and compare the outcomes in these modified problems to those in the baseline problem of choosing an agent with the optimal beliefs.

\subsubsection{Contracting on Actions/Misaligned Preferences} \label{sec:diff}

The most basic delegation tool is contracting: if the principal could offer the agent a contract that specifies contingent payments, this would be the most direct way to provide incentives (see \citealp{laffont_theory_2009} for many examples).
We begin by looking at \emph{action}-contingent contracts $\tau: \mathcal{A} \to \mathbb{R}_+$, which allow the principal to incentivize the agent by offering payments to the agent that depend on the action he selects. This assumes that actions are contractible (i.e., observable and verifiable) and the principal has the institutional power to make such contracts -- either of which may or may not hold in any given setting.
We assume that all agents and the principal have a common prior belief $\mu_p$, all players' preferences are quasilinear in payments, and the principal's marginal utility of money is $\rho$, and the agent's marginal utility of money is $1$.\footnote{In line with the baseline problem, we do not impose any explicit participation constraints on the agent that would impose a lower bound on the transfers. The implicit assumption here is that the agent is being paid some non-negotiable unconditional salary if he is hired, which is sufficient to ensure participation. Payments $\tau$ should then be treated as premia. We do, however, require these payment to be non-negative due to limited liability.}

Note that instead of contracting, we can interpret this setup as a problem of selecting an agent with misaligned preferences by setting $\rho=0$. Schedule $\tau$ then represents not payments, but rather an agent's ``biases'', i.e., inherent preferences towards certain actions on top of the ``unbiased'' utility $u(a,\omega)$. Such a problem of selecting an agent with optimally misaligned preferences is a natural counterpart to our baseline problem of selecting an agent with optimally misaligned beliefs.

The agent's problem (again using the equivalence presented in \ref{sec:preliminary_analysis}) is then given by
\begin{align}\label{eq:agents_problem_biasprefs}
	\max_{\pi} \Bigg\{\sum_{j=1}^{N} \mu(\omega_{j}) \sum_{i=1}^{N} \pi(a_{i}|\omega_{j}) \big( u(a_{i},\omega_{j}) + \tau(a_i) \big) - c(\phi,\mu) \Bigg\},
\end{align}
given $\tau$ and $\mu$, and the principal's \textbf{contracting problem} is
\begin{align} \label{eq:principal_problem_biasprefs}
	\begin{aligned}
		\max_{\tau} &\left\{\sum_{j=1}^N \mu_p(\omega_j) \sum_{i=1}^N \pi(a_i|\omega_j) \big( u(a_i,\omega_j) - \rho \tau(a_i) \big) \right\}
		\\ 
		\text{s.t. } &\pi \text{ solves \eqref{eq:agents_problem_biasprefs} given } \tau \text{ and } \mu.
	\end{aligned}
\end{align}

We start with the case of a common prior/aligned delegation, $\mu = \mu_p$.
Instead of providing a closed-form solution to this problem, we appeal to Lemma \ref{lem:equivalence} to argue that regardless of $\rho$, the principal cannot obtain higher expected utility than in the baseline problem of choosing an agent with a misaligned belief $\mu$. In particular, Lemma \ref{lem:equivalence} implies that any unconditional choice probabilities $\beta \in \varDelta(\mathcal{A})$ generated by an agent, who is incentivized by payments or misaligned preferences, can also be obtained by selecting an agent with appropriately misaligned beliefs. Moreover, by using Proposition 3 of \cite{MATVEENKO2021105356} we can also show the converse -- that any decision rule achievable with misaligned beliefs can be replicated with payments $\tau$, and the two tools are thus equivalent in what they allow to achieve.\footnote{Lemma 1 of \cite{MATVEENKO2021105356} implies that a third equivalent tool is setting the quotas, i.e., imposing specific unconditional choice probabilities for a different action.}
These results are formalized by the following theorem.

\begin{theorem} \label{prop:misaligned_prefs_bride}
	Suppose the information cost is given by Shannon entropy. Then:
	\begin{enumerate}
		\item For any vector $\tau: \mathcal{A} \to \mathbb{R}_+$ of payments/biases and a corresponding $\pi : \Omega \to \varDelta(\mathcal{A})$ that solves \eqref{eq:agents_problem_biasprefs} given $\tau$ and $\mu = \mu_p$, there exists a prior belief $\mu \in \varDelta(\Omega)$ such that $\pi$ also solves \eqref{eq:ri_problem_full} given $\mu$.
		
		\item For any $\mu \in \varDelta(\Omega)$ and the corresponding $\pi_{\mu} : \Omega \to \varDelta(\mathcal{A})$ that solves \eqref{eq:ri_problem_full} given $\mu$, there exist payments $\tau: \mathcal{A} \to \mathbb{R}_+$ such that $\pi_{\mu}$ also solves \eqref{eq:agents_problem_biasprefs} given $\tau$ and $\mu = \mu_p$.
	\end{enumerate}
	As a consequence, the principal's problem of contracting on actions \eqref{eq:principal_problem_biasprefs} with $\rho=0$ and $\mu = \mu_p$ is equivalent to her full (delegation) problem \eqref{eq:principal_full_prb}
\end{theorem}

The theorem above directly implies that neither of the two instruments (contracting on actions or searching for an agent with stronger/weaker preferences for specific actions) can yield strictly better results than hiring an agent with an optimally misaligned belief. On the other hand, neither can misaligned beliefs yield better outcomes than action-contingent contracts. With the limited liability constraint ($\tau(a_i) \geq 0$ for all $i$), it is immediate that contracting on actions is strictly worse when $\rho >0$, since it cannot yield a better decision rule, but requires payments from the principal -- payments which are avoidable if she instead hires an agent who is intrinsically motivated by his beliefs over states or preferences towards specific actions.

Further, our Lemma \ref{lem:equivalence} and the results of \cite{MATVEENKO2021105356} also imply that no combination of misaligned beliefs, misaligned preferences, and payments for actions can perform better than any individual instrument. Moreover, they also imply that suboptimal misalignment along any single dimension can be amended using other instruments. That is, if a given agent holds a non-optimal prior belief (that does not coincide with the principal's either), the optimal behavior might be induced via action-contingent transfers. Conversely, if an agent has biased preference towards certain actions, this misalignment can be compensated for by selecting an agent with an approprite prior belief. Proposition \ref{prop:transfers_n_beliefs} presents an example of such equivalence in the context of a model with $N=2$ and Shannon entropy cost function.

\begin{proposition}\label{prop:transfers_n_beliefs}
	Suppose $N=2$ and the information cost is given by Shannon entropy. Consider the principal's problem of contracting on actions \eqref{eq:principal_problem_biasprefs}, where $\rho=0$ and the agent holds prior belief $\mu \neq \mu_p$. Then:
	\begin{enumerate}
		\item for any $\mu$, there exist payments/biases $\{\tau^*(L), \tau^*(R)\}$ that implement the optimal choice probabilities from Lemma \ref{lemma:optimal_choice};
		\item these payments/biases are such that\footnote{The closed-form expressions are presented in the proof.}
		\begin{align*}
			\tau^*(R) \geq \tau^*(L) \iff \mu \leq \mu^{*}=\frac{\sqrt{\mu_{p}}}{\sqrt{\mu_{p}}+\sqrt{1-\mu_{p}}}.
		\end{align*}
	\end{enumerate}
\end{proposition}

It is easy to see the intuition behind the proposition: if the agent's prior belief $\mu$ assigns lower probability to state $\omega=r$ compared to the principal-optimal prior $\mu^*$ given in Proposition \ref{prop:binarydelegationri}, such an agent is ex ante too biased towards action $a=L$ for the principal's taste, even though he potentially acquires more information than an agent with belief $\mu^*$. Therefore, the principal can nudge the agent towards action $a=R$ by offering higher payment if he selects $R$ (or find an agent whose preference bias towards $R$ offsets his belief bias towards state $l$).\footnote{This is broadly related to the findings of \cite{espitia2023confidence}, who shows that the bias in the agent's preferences can be counteracted by the bias in beliefs (although, the belief biases in his paper are limited to over- and underconfidence).}
This discussion also emphasizes that what matters for our results is not the agent's uncertainty about the state per se, but the agent's uncertainty about \emph{what the optimal action is}. E.g., an agent who assigns very high probability to state $\omega=l$ can be optimal for the principal, as long as the agent's preferences are sufficiently biased in favor of action $a=R$ -- so the agent is actually uncertain about which action to take and chooses to acquire additional information to break the indifference.

The results above also have an implication for the empirical literature estimating discrete choice models. Specifically, Proposition \ref{prop:transfers_n_beliefs} claims that a given action distribution $\beta$ can be obtained using multiple different combinations $(\mu,\tau)$ of beliefs and action preferences. Therefore, the agent's observed action choice probabilities alone do not allow an external observer to jointly identify the decision maker's beliefs and preferences in our setting.

\subsubsection{Contracting on Outcomes}

We now turn to exploring \emph{outcome}-contingent contracts. An outcome in our model can be measured by whether a correct action was chosen ($a=a_j$ when $\omega=\omega_j$) or not. We thus let the principal select payments $\bar{\tau},\underline{\tau}$ that the agent receives, so that $\tau(a_i,\omega_i) = \bar{\tau}$ and $\tau(a_i,\omega_j) = \underline{\tau}$ if $i \neq j$.\footnote{If the principal could contract on both actions and outcomes, she would have the freedom to select any payment schedule $\{\tau(a_i,\omega_j)\}$. \cite{Lindbeck2017} study such a problem with $N$ states and two actions.}
We assume limited liability ($\bar{\tau},\underline{\tau} \geq 0$), quasilinearity of preferences in payments for all agents, and let the agent's marginal utility of money to be $1$, and the principal's marginal utility of money to be $\rho$.

The agent's problem is then choosing $\pi: \Omega \to \varDelta(\mathcal{A})$ that solves\footnote{It is common in the literature to consider an agent who yields no intrinsic utility from actions and is motivated exclusively via payments. For sake of consistency with the remainder of our analysis, we maintain the assumption that the agent enjoys the same intrinsic utility $u(a,\omega)$ as the principal. However, we do allow the agent to trade off this intrinsic utility against monetary transfers at a rate that is different to the principal.} 
\begin{align}\label{eq:agents_problem_contracts}
	\max_{\pi} \Bigg\{\sum_{j=1}^{N} \mu(\omega_{j}) \sum_{i=1}^{N} \pi(a_{i}|\omega_{j}) \big( u(a_{i},\omega_{j}) + \tau(a_i,\omega_j) \big) - c(\phi,\mu) \Bigg\},
\end{align}
given $\tau$, and the principal's \textbf{contracting problem} is
\begin{align} \label{eq:principal_problem_contracts}
	\max_{\bar{\tau},\underline{\tau}} &\left\{\sum_{j=1}^N \mu_p(\omega_j) \sum_{i=1}^N \pi(a_i|\omega_j) \big( u(a_i,\omega_j) - \rho \tau(a_i,\omega_j) \big) \right\},
	\\
	\text{s.t. } & \tau(a_i,\omega_i) = \bar{\tau} \text{ for all } i,
	\nonumber
	\\
	& \tau(a_i,\omega_j) = \underline{\tau} \text{ for all } i, j \neq i,
	\nonumber
	\\
	& \pi \text{ solves \eqref{eq:agents_problem_contracts} given } \bar{\tau},\underline{\tau}.
	\nonumber
\end{align}

It is trivially optimal for the principal to set $\underline{\tau}=0$, since her objective is to provide incentives for the agent to \emph{match} the state. Then the agent's payoff (not including the information cost) becomes $u(a_i,\omega_j) + \tau(a_i,\omega_j) = (1+\bar{\tau}) u(a_i,\omega_j)$, and the principal's payoff is $u(a_i,\omega_j) - \tau(a_i,\omega_j) = (1 - \rho \bar{\tau}) u(a_i,\omega_j)$. In other words, by increasing the incentive payment $\bar{\tau}$, the principal effectively lowers the relative cost of information for the agent, at the cost of decreasing her own payoff. It then appears like an instrument that could be universally useful for the principal -- even when she chooses an agent with the optimal prior belief, she could still benefit from reducing the agent's information cost, which would result in him acquiring more information.
However, when payments are sufficiently costly to the principal, contracting on outcomes cannot improve on delegating to the optimally misaligned agent, or even to the aligned agent. The following proposition states the result for the case when $N=2$.

\begin{proposition} \label{prop:contracts}
	Suppose $N=2$, $\mu_p \geq \frac{1}{2}$, and the information cost is given by Shannon entropy. 
	Consider the principal's contracting problem \eqref{eq:principal_problem_contracts}. 
	Then for any $\rho \geq \min \left\{ 1, \frac{1}{2\lambda} \right\}$ there exist $\bar{\mu}_{L}, \bar{\mu}_{R}$ and $\hat{\mu}_{L},\hat{\mu}_{R}$ such that:
	\begin{enumerate} 
		\item $\hat{\mu}_{L} \leq \bar{\mu}_{L} < \mu^* < \mu_p < \bar{\mu}_{R} \leq \hat{\mu}_{R} $, where $\mu^*$ is given by \eqref{eq:opt_deleg_binary};
		\item the principal's problem \eqref{eq:principal_problem_contracts} is solved by $\bar{\tau}>0$ if $\mu \in (\hat{\mu}_{L}, \bar{\mu}_{L} ) \cup ( \bar{\mu}_{R}, \hat{\mu}_{R} )$;
		\item the principal's problem \eqref{eq:principal_problem_contracts} is solved by $\bar{\tau}=0$ otherwise.
	\end{enumerate} 
\end{proposition}

The proposition states that the principal uses the incentive payments, $\bar{\tau} > 0$, when she has an intermediate degree of misalignment in opinions with the agent. This may happen if the agent is moderately more biased than the principal and acquires too little information for the principal's taste (which is the case when $\mu_{p} < \bar{\mu}_R < \mu < \hat{\mu}_R$). An additional reward for matching the state then incentivizes the agent to acquire more information and improves the principal's payoff, despite her giving a part of it to the agent. If the agent is too biased, however ($\mu > \hat{\mu}_R$), then the incentives become too costly for the principal to provide, and she chooses $\bar{\tau}=0$. The logic is analogous if the agent is sufficiently biased in the opposite direction ($\mu \ll 0.5$). Finally, if the agent is sufficiently aligned with the principal, $\mu \in [\bar{\mu}_L, \bar{\mu}_R]$, then providing bonus payments does not provide enough of an additional incentive to the agent to justify the cost for the principal. This latter case includes both the aligned agent ($\mu = \mu_p$) and the optimally biased agent ($\mu = \mu^*(\mu_p)$). Specifically, it follows that outcome-contingent contracts cannot benefit the principal who can choose an agent with optimally misaligned beliefs.

\subsection{Alternative Preference Specifications} \label{sec:genprefs}

Our analysis is heavily reliant on state-matching preferences, which we assume are shared by both the principal and the agent(s). It is reasonable to ask whether our conclusions hold under other preference specifications. 
We first consider a ``binary-quadratic'' model, which is closely related to the ``uniform-quadratic'' model commonly used in delegation literature starting with \citealp{holmstrom1980} and \citealp{crawford_strategic_1982}. In such a uniform-quadratic model, the state is $\omega \sim U[0,1]$, the action set is $\mathcal{A}=[0,1]$, and the players' preferences are described by $u_i(a,\omega) = -(a - (\omega+b_i))^2$ for $i \in \{P,A\}$ and some ``biases'' $b_A,b_P$. Such a uniform-quadratic model is not tractable with entropy information costs. 
Instead, we consider a related model with binary state and quadratic loss and show that the characterization from Proposition \ref{prop:binarydelegationri} continues to hold in this setting.
We then proceed to explore general preferences with finite states and actions, and whether the results of Section \ref{sec:diff} on the equivalence of misalignment in beliefs and misalignment in preferences carry over to such settings.

\subsubsection{Binary-quadratic model.}
Consider a \emph{binary-quadratic problem} that looks as follows: the state space is $\Omega = \{0,1\}$; all beliefs $\mu$ are represented in terms of probability $\mu \in [0,1]$ assigned to state $\omega=1$. The action set is $\mathcal{A} = [0,1]$; the principal's and the agent's common utility function is $u(a,\omega) = -(a-\omega)^2$; the agent's information cost $c(\phi,\mu)$ is given by Shannon entropy \eqref{eq:c_entropy}. Everything else is analogous to the binary model presented in Section \ref{sec:binary}.

We show in the following proposition that our characterization \eqref{eq:opt_deleg_binary} of the optimal delegation strategy continues to hold in such a model, implying that our result is not specific to the state-matching utility function adopted by the baseline model. 

\begin{proposition} \label{prop:bqp}
	In the binary-quadratic problem, there exists $\hat{\mu} \geq 0.5$ such that the principal's optimal delegation strategy is given by:
	\begin{align*}
		\mu^{**}(\mu_p) = \begin{cases}
			\mu^*(\mu_p) & \text{ as given by \eqref{eq:opt_deleg_binary} if } \mu_p \in \left( 1-\hat{\mu}, \hat{\mu} \right),
			\\
			\mu_p & \text{ otherwise}.
		\end{cases}
	\end{align*}
\end{proposition}

The reason why we obtain the same characterization as in Section \ref{sec:binary} lies in the solution to the agent's problem. As we showed in Section \ref{sec:binary}, the agent learns in such a way as to meet specific \emph{standards of proof} $\bar{\mu}$ in order to take either action. The fact that these $\bar{\mu}$ are independent of the agent's prior $\mu$ pin down the shape of the production possibility frontier that the principal faces and, by extension, the solution to the principal's problem.
As the proof of Proposition \ref{prop:bqp} shows, the agent's solution in the binary-quadratic problem exhibits the same property. Specifically, the agent aims at some threshold $\hat{\mu}$ that his belief in a given state must achieve in order to warrant an action. This threshold is independent of the agent's prior $\mu$, enabling us to apply the logic from Section \ref{sec:binary}.

\subsubsection{Alternative principal's preferences.}
We now turn to the question of how general the results of Section \ref{sec:diff} are.
We first generalize the principal's utility function $u_p(a,\omega)$ while maintaining the agent's intrinsic preference for matching the state: $u_A(a_i,\omega_i)=1$, $u_A(a_i,\omega_j)=0$ if $i\neq j$. Naturally, the specific functional forms of the optimal delegation strategies (such as those presented in Proposition \ref{prop:binarydelegationri}, Theorem \ref{prop:opt_choice_beliefs}, and Lemma \ref{lemma:optimal_choice}) depend on the specific form of the principal's utility function. However, Lemma \ref{lem:equivalence} only depends on the agent's utility function, meaning that Theorem \ref{prop:misaligned_prefs_bride} still holds: with Shannon entropy cost, any outcome that can be achieved by contracting on actions or hiring an agent with misaligned intrinsic preferences can also be achieved by hiring an agent with misaligned beliefs (and vice versa). This means that regardless of the principal's objective function, hiring an agent with state-matching preferences and a suitable belief is as good as hiring an agent with aligned prior belief, state-matching preferences, and either some additional preference over actions, or action-contingent payments on top of that.

\subsubsection{Alternative aligned preferences.}
The argument above does, however, hinge on the agent having state-matching preferences as a baseline. If we allow arbitrary payoffs that are common for the principal and the agent, then the equivalence stated in Theorem \ref{prop:misaligned_prefs_bride} breaks down. In such a general case, finding an agent with optimally misaligned preferences may yield strictly better results for the principal than hiring an agent with an optimally misaligned belief, and hence contracting on actions may, in principle, yield better results too. 
This is due to the fact that the equivalence between the principal's full and relaxed problems, presented in Section \ref{sec:Ncase_prbs}, breaks down with general preferences. This is captured by the following proposition. 

\begin{proposition}\label{prop:impossibility}
	Suppose the information cost is given by Shannon entropy. 
	There exists a utility function $u(a,\omega)$ such that the solution to the principal's relaxed problem \eqref{eq:principal_relaxed_prb2} cannot be attained as a solution to the full problem \eqref{eq:principal_full_prb}. 
\end{proposition} 
\begin{corollary}
	There exists a non-state-matching utility function $u(a,\omega)$ such that the conclusions of Theorem \ref{prop:misaligned_prefs_bride} do not hold.
\end{corollary}

The proposition above states that with general preferences, the principal is no longer able to implement any vector of unconditional choice probabilities $\beta$ via an appropriate choice of the agent's prior $\mu$ -- which is still possible through the choice of action-contingent contracts as in Section \ref{sec:diff} (see Proposition 3 in \citealp{MATVEENKO2021105356}). 
Therefore, hiring an agent with misaligned preferences can be strictly better for the principal than hiring an agent with misaligned beliefs for general utility functions $u(a,\omega)$. Equivalently, performance of an agent with optimally misaligned beliefs may, in general, be improved upon by providing further incentives through action-contingent contracts.

\subsection{Alternative cost functions} \label{sec:cost}

Our analysis assumed that the information cost function is uniformly posterior separable (UPS). This assumption has an implication that the cost of a given signal structure depends on the agent's prior belief \citep*[see][]{mensch2018cardinal,denti2022experimental}, which may not be fitting in all settings.
To demonstrate that our main result does not hinge on this or any other specific properties of UPS cost functions, this section explores a number of alternative specifications of the cost function in the binary setting of Section \ref{sec:binary}. We show that in all cases, the principal's optimal delegation strategy looks similar to what is obtained in Proposition \ref{prop:binarydelegationri}: if $\mu_p$ is not too extreme, it is optimal for the principal to delegate to an ex ante more unbiased agent: $\mu^*(\mu_p) \in [0.5, \mu_p)$.

To remind, in the binary setting of Section \ref{sec:binary}, we assumed $\Omega = \{l,r\}$, $\mathcal{A} = \{L,R\}$, with $u(L,l) = u(R,r) = 1$ and $u(L,r) = u(R,l) = 0$. In all settings below, we will be looking for an equilibrium as defined in Section \ref{sec:model}, with the $c(\phi,\mu)$ in the agent's problem \eqref{eq:ri_problem_full} replaced by one of the respective cost functions defined below.

\subsubsection{Channel capacity cost function.}
The first cost function we consider is the channel capacity cost proposed by \citet{woodford2012inattentive}. We follow the analysis by \citet{nimark2019inattention}, hereinafter referred to as NS. The channel capacity cost of a given signal structure $\phi: \Omega \to \varDelta(\mathcal{S})$ is given by
\begin{align*}
	c_C(\phi) \equiv \max_{\mu \in \varDelta(\Omega)} c_E(\phi,\mu),
\end{align*}
where $c_E(\phi,\mu)$ is the entropy cost \eqref{eq:c_entropy}. Intuitively, the channel capacity measures the maximum amount of information that can be extracted from signal structure $\phi$ by any agent. 
By definition, cost $c_C(\phi)$ of a given signal structure does not depend on the selected agent's prior $\mu$, unlike the Shannon entropy cost function or other UPS cost functions.

NS show that the argument from Section \ref{sec:preliminary_analysis} continues to hold with the channel capacity cost: the agent optimally selects a ``recommender'' signal structure, where each signal realization is associated with a unique action. Thus we can reduce the agent's problem to that of choosing a decision rule $\pi:\Omega \to \Delta(\mathcal{A})$ which solves
\begin{align}\label{eq:ri_problem_cc}
	&\max_{\pi} \Big\{ \mu \pi(R|r) + (1-\mu) \pi(L|l) - c_C(\pi,\mu) \Big\}, 
\end{align}
where $c_C(\pi,\mu)$ denotes the information cost induced by $\pi$ (which, in this formulation, does depend on $\mu$).

NS show that the agent's problem \eqref{eq:ri_problem_cc} is well-defined and the solution always exists. It shares the same broad features as the solution with entropy costs: an agent with $\mu>0.5$ chooses a signal structure such that $\pi(R|r)>\pi(L|l)$ and vice versa. More broadly, $\pi(R|r)$ is continuous and increasing in $\mu$, while the opposite is true for $\pi(L|l)$; a more uncertain agent also acquires more information as measured by costs (but not more in the Blackwell sense).

\begin{figure} 
	\centering 
	\subfloat[][$\lambda=0.5$]{
		\includegraphics[width=0.48\textwidth]{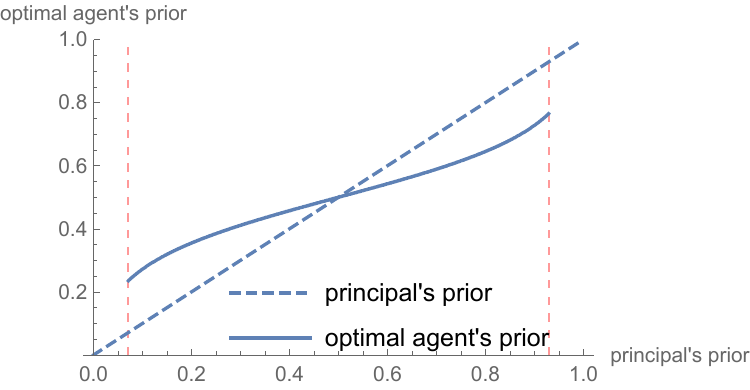}
	}
	\hfill 
	\subfloat[][$\lambda=1$]{
		\includegraphics[width=0.48\textwidth]{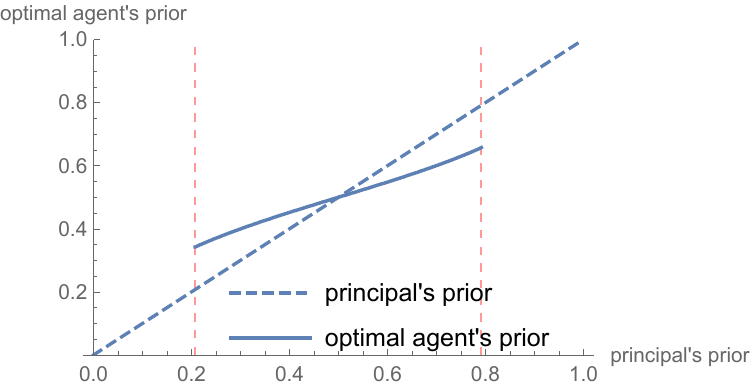}
	}
	\caption{The optimal delegation strategy $\mu^*(\mu_p)$ with channel capacity cost function.}
	\label{fig:opt_deleg_strat_cc} 
\end{figure}

The continuity of $\pi$ w.r.t. $\mu$ implies that the principal's problem \eqref{eq:principal_problem_binary} always has a solution. 
While the principal's problem proved to be analytically intractable, Figure \ref{fig:opt_deleg_strat_cc} plots numerical solutions for two values of $\lambda$ and all $\mu_p$.

\begin{observation}
	Figure \ref{fig:opt_deleg_strat_cc} suggests that the solution of the principal's problem with channel capacity costs $c_C(\phi)$ satisfies the second property in Proposition \ref{prop:binarydelegationri} and the property in Corollary \ref{corr:nonmonmisal}.
\end{observation}

We see that both plots in Figure \ref{fig:opt_deleg_strat_cc} demonstrate a delegation strategy that is qualitatively the same as in Figure \ref{fig:opt_deleg_strat}, which plotted it for entropy costs: if the principal's belief $\mu_p$ is not too extreme, she chooses an agent with belief $\mu$ between $\mu_p$ and $0.5$. For extreme $\mu_p$, same as before, she selects (an agent who acquires no information and chooses) the ex ante optimal action.
However, Figure \ref{fig:opt_deleg_strat_cc} also highlights a difference relative to the model with UPS costs in that the principal's solution now depends on the cost parameter $\lambda$: higher $\lambda$ leads to less delegation under channel capacity costs. This suggests that the independence of the principal's strategy from $\lambda$ is a special feature of UPS cost functions.

\subsubsection{Log-likelihood ratio cost function.}
We now move on to consider the log-likelihood ratio (LLR) cost function proposed by \citet*{pomatto2018cost}, hereinafter referred to as PTS. PTS derive the LLR cost function axiomatically as the cost of \emph{acquiring} information (as opposed to the entropy cost being that of \emph{processing} information, according to their argument) from a set of intuitive cost linearity axioms. 
The LLR cost of a given signal structure $\phi: \Omega \to \varDelta(\mathcal{S})$ is defined as 
\begin{align*}
	c_L(\phi) &\equiv \sum_{\omega_i, \omega_j \in \Omega} \lambda_{ij} \int_{\mathcal{S}} \ln \left( \frac{\phi(s|\omega_i)}{\phi(s|\omega_j)} \right) d\phi(s|\omega_i),
\end{align*}
where $\lambda_{ij}$ are the parameters encoding the ``closeness'' of states $\omega_i$ and $\omega_j$ (how difficult it is to distinguish them). In our binary setting, we assume $\lambda_{LR}=\lambda_{RL}=\lambda$. As in the case of channel capacity costs, PTS' main representation theorem shows that the LLR cost of a given signal structure $\phi$ does not depend on the prior belief $\mu$.

In the binary setting, the agent optimally chooses no more than two signal realizations, because LLR cost is monotone with respect to the Blackwell order. Therefore, we can again invoke the logic of Section \ref{sec:preliminary_analysis} and reduce the agent's problem to that of choosing a decision rule $\pi:\Omega \to \Delta(\mathcal{A})$ subject to cost $c_L(\pi)$.\footnote{PTS provide a representation for $c_L(\pi)$ not presented here.} 

\begin{figure}
	\subfloat[][$\lambda=0.1$]{
		\includegraphics[width=0.48\linewidth]{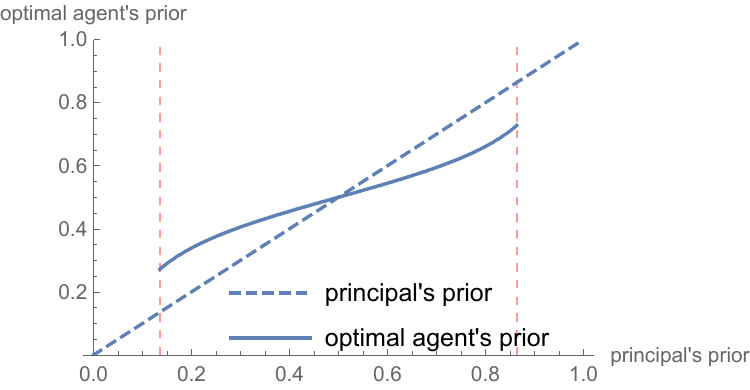}
	}
	\hfill 
	\subfloat[][$\lambda=0.5$]{
		\includegraphics[width=0.48\linewidth]{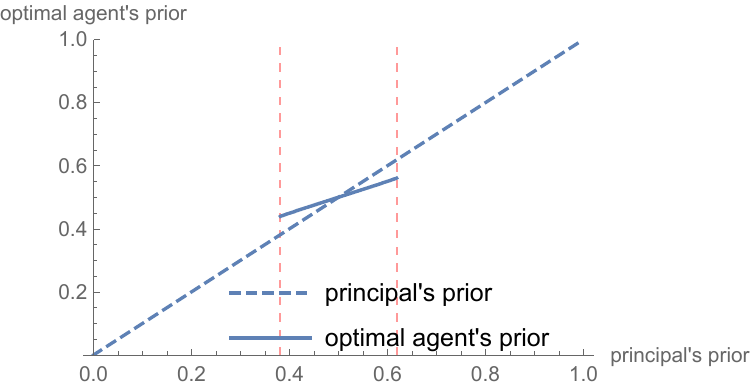}
	}
	\caption{The optimal delegation strategy $\mu^*(\mu_p)$ with LLR cost function.}
	\label{fig:opt_deleg_strat_llr} 
\end{figure}

PTS explore a binary problem in their Sections 6.1 and 6.6 but only demonstrate an analytical solution to the agent's problem for the case $\mu = 0.5$. We have found the agent's problem to be analytically intractable for $\mu\neq 0.5$, and therefore solve both the agent's and the principal's problems numerically. Figure \ref{fig:opt_deleg_strat_llr} demonstrates our findings.

\begin{observation}
	Figure \ref{fig:opt_deleg_strat_llr} suggests that the solution of the principal's problem with log-likelihood ratio costs $c_L(\phi)$ satisfies the second property in Proposition \ref{prop:binarydelegationri} and the property in Corollary \ref{corr:nonmonmisal}.
\end{observation}

One can see that the principal's optimal delegation strategy plotted therein looks qualitatively the same as for the uniformly posterior-separable and channel capacity costs (Figures \ref{fig:opt_deleg_strat} and \ref{fig:opt_deleg_strat_cc}, respectively): if the principal's belief $\mu_p$ is not too extreme, the principal chooses an agent with belief $\mu$ between $\mu_p$ and $0.5$. Further, similarly to the setting with channel capacity costs in Figure \ref{fig:opt_deleg_strat_cc}, the principal delegates less for higher values of the information cost parameter $\lambda$.

\subsubsection{Symmetric cost functions.} \label{sec:symm}
Finally, we explore a family of simple ``symmetric'' cost functions, which restrict the agent to symmetric signal structures. This analysis highlights the importance of flexibility of the agent's learning technology for the trade-off we identify. In particular, suppose that instead of being able to choose an arbitrary signal structure $\phi: \Omega \to \varDelta(\mathcal{S})$, the agent is restricted to a binary signal space $\mathcal{S} \equiv \{ l,r \}$ and can only choose signal precision that we denote, abusing notation, by $\phi \equiv \phi(r|r) = \phi(l|l) \in \left[\frac{1}{2}, 1\right]$. Suppose further that the cost of information is then given by some function $c_S (\phi)$ that is strictly increasing, convex, differentiable in $\phi \in \left[\frac{1}{2}, 1\right]$, and satisfies $c_S \left(\frac{1}{2} \right) = 0$ and the Inada conditions $c_S' \left(\frac{1}{2} \right)=0$, $\lim\limits_{\phi \to 1} c_S'(\phi)=+\infty$.

Let $\phi^*_\mu$ denote the precision optimally chosen by an agent with prior belief $\mu$.
The agent only acquires information ($\phi^*_\mu > \frac{1}{2}$) if he intends to follow the signal realization ($\sigma(r)=R$ and $\sigma(l)=L$), since this trivially dominates doing the converse, and conditional on ignoring the signal realization, acquiring an uninformative signal structure $\phi=\frac{1}{2}$ is strictly cheaper. Hence if $\phi^*_\mu > \frac{1}{2}$, then 
\begin{align} \label{eq:agentprb_symm}
	\phi^*_\mu = \arg \max_{\phi} \left\{ \mu\phi + (1-\mu)\phi - c_S(\phi) \right\}.
\end{align} 
Let $\phi^{**}$ denote the candidate solution given by the FOC of \eqref{eq:agentprb_symm}: $c_S'(\phi^{**}) = 1$. Note that $\phi^{**}$ does not depend on the agent's belief $\mu$. The agent's expected utility from acquiring no information ($\phi=\frac{1}{2}$) and taking the ex ante optimal action is given by $\max \{\mu, 1-\mu\}$. The expected utility from choosing $\phi = \phi^{**}$ is given by $\phi^{**} - c_S(\phi^{**}) \in \left[\frac{1}{2}, 1\right]$.\footnote{Function $\phi-c_S(\phi)$ is strictly concave in $\phi$ due to the assumptions made, evaluates to $\frac{1}{2}$ when $\phi=\frac{1}{2}$ (hence $\phi^{**} - c_S(\phi^{**}) \geq \frac{1}{2}$), and from $c_S(\phi) \geq 0$ we have that $\phi - c_S(\phi) \leq 1$ for all $\phi \leq 1$.} 
Consider then the agents who are indifferent between the two options and denote them by $\bar{\mu}_R \equiv \phi^{**} - c_S(\phi^{**})$ and $\bar{\mu}_L \equiv 1 - \left( \phi^{**} - c_S(\phi^{**}) \right)$, respectively. Then we can describe the agent's optimal choice of precision by
\begin{align*}
	\phi^*_\mu = \begin{cases}
		\phi^{**} & \text{ if } \mu \in \left[ \bar{\mu}_L, \bar{\mu}_R \right],
		\\
		\frac{1}{2} & \text{ otherwise}.
	\end{cases}
\end{align*}

Moving on to the principal's problem (and maintaining the assumption that $\mu_p > \frac{1}{2}$), the principal's payoff is given by $\phi^{**}$ if $\phi^*_\mu = \phi^{**}$, by $\mu_p$ if $\phi^*_\mu = \frac{1}{2}$ and $\mu > \frac{1}{2}$, and by $1-\mu_p$ if $\phi^*_\mu = \frac{1}{2}$ and $\mu < \frac{1}{2}$. Therefore, the principal's preferred agent is
\begin{align} \label{eq:binarydeleg_symm}
	\mu^*(\mu_p) = \begin{cases}
		\mu \in \left[ \bar{\mu}_L, \bar{\mu}_R \right] & \text{ if } \mu_p \leq \phi^{**},
		\\
		\mu \in \left( \bar{\mu}_R, 1 \right] & \text{ if } \mu_p > \phi^{**}.
	\end{cases}
\end{align}
Notably, if $\mu_p \in (\bar{\mu}_R, \phi^{**})$, then the principal strictly prefers a misaligned agent, whose prior belief is more uncertain than the principal's. Further, there exists a selection from \eqref{eq:binarydeleg_symm} that supports the following proposition (which is proved by the preceding argument):
\begin{proposition} \label{prop:binarydeleg_symm}
	Given a symmetric information cost function $c_S(\phi)$, there exists an equilibrium in which the principal always delegates to a more uncertain agent: for any $\mu_p > \frac{1}{2}$, $\mu^*(\mu_p) \in \left[ \frac{1}{2}, \mu_p \right)$.
\end{proposition}
It is evident, however, that the statement of Proposition \ref{prop:binarydeleg_symm} is quite weak. Symmetric information cost leads to the principal actually being indifferent between all agents $\mu \in \left[ \bar{\mu}_L, \bar{\mu}_R \right]$, as well as between all agents $\mu \in \left( \bar{\mu}_R, 1 \right]$. The reason is that without flexible learning technology, the agent is not engaging in confirmatory learning which the principal seeks to either amend or exploit. Consequently, if $\mu_p \notin (\bar{\mu}_R, \phi^{**})$, then hiring an aligned agent or, possibly, even a more certain agent, is just as optimal for the principal as hiring a more uncertain agent. Conversely, if the principal strictly prefers to hire a learning agent, then she might as well hire an agent with $\mu = \frac{1}{2}$, whose decisions would not be any more tilted than those of an agent with $\mu = \bar{\mu}_R$.

It is straightforward that the result above continues to hold for any weakly increasing $c_S(\phi)$, whereas all the other assumptions on $c_S(\phi)$ are not strictly necessary and were adopted to simplify the argument. For example, we could also consider a ``Pandora box'' cost function, under which the agent can either learn the state perfectly at a cost, or learn nothing. This can be captured as $c_S(\phi) = \lambda \cdot \mathbb{I}\{\phi > \frac{1}{2}\}$, where $\mathbb{I}\{\cdot\}$ is the indicator function. Under such a cost function, the agent would learn the state perfectly if he is sufficiently uncertain, and stick to his prior belief otherwise; and it is thus always weakly optimal for the principal to choose the most uncertain agent: $\mu^*(\mu_p) = \frac{1}{2}$.
Another learning technology that is also symmetric across states and signal realizations, but does not fall under the parametrization above, is the one used by \citet{szalay2005economics} and \citet{ball2021benefitting}. In their respective models, the agent selects an effort $e \in [0,1]$ subject to cost $c_F(e)$, which allows him to perfectly learn the state with probability $e$, and with the complementary probability $1-e$ the agent observes nothing. It is easy to see that under this technology, the learning effort $e^*_\mu$ is higher when $\mu$ is closer to $\frac{1}{2}$ (more uncertain agents learn more). However, same as with symmetric cost functions, there is no disadvantage to hiring a misaligned agent -- the principal would strictly prefer to hire the most uncertain agent, $\mu = \frac{1}{2}$.

\medskip 

The goal of this exercise is to demonstrate that to fully capture the trade-off that the principal faces -- that misalignment may improve the quality of decisions in one state, but comes at the expense of worse decisions in another state, -- a flexible learning technology is necessary. Flexibility enables the agent to design his learning in a way to exactly meet his required standards of proof. This leads to confirmatory learning: agent acquires more information about the state he considers ex ante more likely. While the principal shares this preference to an extent, she would like to tame the agent's confirmatory learning somewhat. Without flexible learning, none of these effects are present. At the same time, the examples with the channel capacity and the log-likelihood ratio cost functions demonstrate that our results are not specific to UPS cost functions -- that it is indeed the flexibility of the agent's learning technology that drives our results, and not any specific features of our chosen cost function.

\subsection{Proofs for the Online Appendix} 

\subsubsection{Proof of Theorem \ref{prop:misaligned_prefs_bride}}

Part 2 of the statement follows immediately from Proposition 3 of \cite{MATVEENKO2021105356}.

To show part 1, we invoke Theorem 1 from \cite{MM} stated in \eqref{eq:choice_prob_cond_N}, which claims that in the contracting problem, the $\pi^*_\mu: \Omega \to \varDelta(\mathcal{A})$ that solves the agent's problem \eqref{eq:agents_problem_biasprefs} is given by
\begin{align}
	\pi^*_\mu(a_i|\omega_j) &= \frac{ \beta(a_i) e^{\frac{u(a_i,\omega_j) + \tau(a_i)}{\lambda}} }{ \sum_{k=1}^N \beta(a_k) e^{\frac{u(a_k,\omega_j) + \tau(a_k)}{\lambda}} }
	\nonumber
	\\
	&= \frac{ \beta'(a_i) e^{\frac{u(a_i,\omega_j)}{\lambda}} }{ \sum_{k=1}^N \beta'(a_k) e^{\frac{u(a_k,\omega_j)}{\lambda}} },
	\label{eq:p5_1}
	\\
	\text{where } 
	\beta(a_i) &= \sum_{j=1}^N \mu(\omega_j) \pi^*_\mu(a_i|\omega_j).
	\nonumber
	\\
	\text{and }
	\beta'(a_i) &\equiv \frac{\beta(a_i) e^\frac{\tau(a_i)}{\lambda}}{\sum_{k=1}^N \beta(a_k) e^\frac{\tau(a_k)}{\lambda}}.
	\nonumber
\end{align}
Since $\beta'$ is a valid probability distribution on $\mathcal{A}$, representation \eqref{eq:p5_1} together with \eqref{eq:choice_prob_cond_N} imply that such a collection of conditional probabilities $\pi^*_\mu$ is a valid solution to the agent's problem \eqref{eq:ri_problem_full} when the agent's payoff function (not including the information costs) is given by $u(a_i,\omega_j)$. That is, the principal can implement the desired conditional choice probabilities $\pi^*_\mu$ by choosing an agent with unbiased preferences and some belief $\mu$, such that the unconditional choice probabilities selected by this agent are given by $\beta'$. Lemma \ref{lem:equivalence} implies that such a belief $\mu \in \varDelta(\Omega)$ does indeed exist.

\subsubsection{Proof of Proposition \ref{prop:transfers_n_beliefs}}

From \eqref{eq:choice_prob_cond_N} and \eqref{eq:p3_2}, the agent-optimal conditional choice probabilities in the binary states model from Section \ref{sec:binary} with Shannon entropy cost are given by
\begin{equation}
	\label{eq:precisions_binary}
	\begin{aligned}
		\pi(R|r) &= \frac{ \left( \mu e^\frac{1}{\lambda} - (1-\mu) \right) e^\frac{1}{\lambda} }{ \left( e^\frac{2}{\lambda} - 1 \right) \mu },
		\\
		\pi(L|l) &= \frac{ \left( (1-\mu) e^\frac{1}{\lambda} - \mu \right) e^\frac{1}{\lambda} }{ \left( e^\frac{2}{\lambda} - 1 \right) (1-\mu) },
	\end{aligned}
\end{equation}
cropped to $[0,1]$.
Plugging \eqref{eq:opt_deleg_binary} into \eqref{eq:precisions_binary} yields the principal-optimal conditional choice probabilities for the binary model, given by
\begin{equation} \label{eq:alf_mu_opt}
	\begin{aligned} 
		\pi^* (R|r) &= \left(e^{\frac{2}{\lambda}} - 1\right)^{-1} e^{\frac{1}{\lambda}} \left(e^{\frac{1}{\lambda}} - \sqrt{ \frac{1-\mu_{p}}{\mu_{p}} }\right),
		\\
		\pi^* (L|l) &= \left(e^{\frac{2}{\lambda}} - 1\right)^{-1} e^{\frac{1}{\lambda}} \left(e^{\frac{1}{\lambda}} - \sqrt{\frac{\mu_{p}}{1-\mu_{p}}}\right),
	\end{aligned}
\end{equation}
cropped to $[0,1]$.

The agent's preferences only depend on the difference $\tau(R)-\tau(L)$. Assuming all $\tau(R) \in \mathbb{R}$ are available to the principal (no limited liability), it is without loss to set $\tau(L) = 0$ in this proof. Both transfers can then be increased by a constant if needed to satisfy limited liability. 
The agent's problem is given by \eqref{eq:agents_problem_biasprefs}. Solving it given $\tau = (\tau(R), 0)$ yields
\begin{equation} \label{eq:cond_t_R}
	\begin{aligned}
		\pi (R|r) &= 1 - \frac{e^{\frac{2}{\lambda}} (1-\mu) - e^{\frac{1+\tau(R)}{\lambda}} + \mu }{ \left(e^{\frac{2}{\lambda}}-1\right) \left( e^{\frac{1+\tau(R)}{\lambda}} - 1 \right) \mu},
		\\
		\pi (L|l) &= \frac{e^{\frac{1}{\lambda}} \left( e^{\frac{2}{\lambda}}(1-\mu) - e^{\frac{1+\tau(R)}{\lambda}} + \mu \right)}{\left(e^{\frac{2}{\lambda}}-1\right)\left(e^{\frac{1}{\lambda}}-e^{\frac{\tau(R)}{\lambda}}\right) (1-\mu)},
	\end{aligned}
\end{equation}
cropped to $[0,1]$.

The principal's contracting problem \eqref{eq:principal_problem_biasprefs} in the binary setting with $\rho=0$ is similar to \eqref{eq:principal_problem_binary}:
\begin{equation} \label{eq:t_R_choice_OP}
	\begin{aligned}
		\max_{\tau(R)} &\left\{ \mu_{p} \pi(R|r) + \left(1-\mu_{p}\right)\pi(L|l) \right\}
		\\
		\text{s.t. } &\pi(R|r), \pi(L|l) \text{ are given by \eqref{eq:cond_t_R}}.
	\end{aligned}
\end{equation}
Assuming the probabilities in \eqref{eq:cond_t_R} are interior, the F.O.C. for \eqref{eq:t_R_choice_OP} yields the candidate solution $\tau(R)$ given by
\begin{align} \label{eq:t_R_opt}
	\tau^*(R) &= \lambda \ln \left[ \frac{
		\frac{1-\mu}{\mu} e^{\frac{1}{\lambda}} + \sqrt{\frac{1-\mu_p}{\mu_p}}
	}{
		\frac{1-\mu}{\mu} + e^{\frac{1}{\lambda}} \sqrt{\frac{1-\mu_p}{\mu_p}} 
	} \right],
\end{align}
where the expression inside the $\ln(\cdot)$ is non-negative for any $\lambda,\mu_{p},\mu$, and thus the candidate $\tau(R)$ exists for any $\mu$ that yields interior probabilities \eqref{eq:cond_t_R}.

Plugging \eqref{eq:t_R_opt} into \eqref{eq:cond_t_R} yields, after some routine manipulations, the conditional choice probabilities that coincide with \eqref{eq:alf_mu_opt} (hence, probabilities \eqref{eq:cond_t_R} are interior given $\mu$ and $\tau^*(R)$ if and only if probabilities \eqref{eq:alf_mu_opt} are interior). 
Thus, condition \eqref{eq:t_R_opt} is not only necessary, but also sufficient. Hence,
for any $\mu_p$ for which \eqref{eq:alf_mu_opt} are interior, $\tau^*(R)$ as given by \eqref{eq:t_R_opt} solves the principal's problem \eqref{eq:t_R_choice_OP},
and this solution exists for any $\mu$. 

If $\lambda$ and $\mu_p$ are such that probabilities \eqref{eq:alf_mu_opt} are not interior, then the principal would like the agent to take the ex ante (principal-)preferred action (it can be verified that the expressions in \eqref{eq:alf_mu_opt} are such that $\pi^*(R|r) \geq 1 \iff \pi^*(L|l) \leq 0$ and $\pi^*(R|r) \leq 0 \iff \pi^*(L|l) \geq 1$). The candidate transfers \eqref{eq:t_R_opt} yield exactly such non-interior probabilities (when plugged into \eqref{eq:cond_t_R}), and hence they still solve the principal's problem \eqref{eq:t_R_choice_OP} for any respective $\mu$.\footnote{
	Note that $\tau^*(R)$ is not the unique solution in this case. If $\pi^*(R|r)=1, \pi^*(L|l)=0$, then any $\tau(R) \geq \lambda \ln \left( \mu + (1-\mu) e^{\frac{2}{\lambda}} \right) - 1$ yields the optimal choice probabilities, and if $\pi^*(R|r)=0, \pi^*(L|l)=1$, then any $\tau(R) \leq 1 - \lambda \ln \left( \mu e^{\frac{2}{\lambda}} + (1-\mu) \right)$ solves the principal's problem.
}
This concludes the proof of part 1 of the proposition.

To show part 2, consider \eqref{eq:t_R_opt} as a function of $\mu$. It is strictly decreasing in $\mu$ on $[0,1]$, and the equation $\tau^*(R)(\mu) = 0$ has a unique root in $[0,1]$ equal to
\begin{align*}
	\mu^{*}=\frac{\sqrt{\mu_{p}}}{\sqrt{\mu_{p}}+\sqrt{1-\mu_{p}}},
\end{align*}
meaning that $\tau(R)\geq 0 = \tau(L) \iff \mu \leq \mu^{*}$.

\subsubsection{Proof of Proposition \ref{prop:contracts}}

As argued in the text, it is immediate that $\underline{\tau}=0$.
The agent's problem \eqref{eq:agents_problem_contracts} given the incentive payment $\bar{\tau} \geq 0$ is solved by
\begin{equation} \label{eq:cond_prob_tbar}
	\begin{aligned}
		&\pi(R|r) = \min \left\{ 1, \max \left\{ 0, \pi_u(R|r)\right\} \right\} \text{ and }
		\pi(L|l) = \min \left\{ 1, \max \left\{ 0, \pi_u(L|l)\right\} \right\}, 
		\\
		&\text{where }
		\pi_u \left(R|r\right) = \frac{ e^{\frac{1+\bar{\tau}}{\lambda}} \left( e^{\frac{1+\bar{\tau}}{\lambda}} \mu - (1-\mu) \right) }{ \left( e^{\frac{2\left(1+\bar{\tau}\right)}{\lambda}}-1 \right) \mu }
		= \frac{ e^{\frac{1+\bar{\tau}}{\lambda}} }{ e^{\frac{2(1+\bar{\tau})}{\lambda}} - 1 } \left( e^{\frac{1+\bar{\tau}}{\lambda}} - \frac{1-\mu}{\mu} \right),
		\\
		&\hphantom{wherei}
		\pi_u \left(L|l\right) = \frac{ e^{\frac{1+\bar{\tau}}{\lambda}} \left( e^{\frac{1+\bar{\tau}}{\lambda}} (1-\mu) - \mu \right) }{ \left( e^{\frac{2\left(1+\bar{\tau}\right)}{\lambda}}-1 \right) (1-\mu)}
		= \frac{ e^{\frac{1+\bar{\tau}}{\lambda}} }{ e^{\frac{2(1+\bar{\tau})}{\lambda}} - 1 } \left( e^{\frac{1+\bar{\tau}}{\lambda}} - \frac{\mu}{1-\mu} \right).
	\end{aligned}
\end{equation}

The principal's full contracting problem \eqref{eq:principal_problem_contracts} can be rewritten as
\begin{equation} \label{eq:princ_OP_tbar}
	\begin{aligned}
		\max_{\bar{\tau} \in \mathbb{R}_+} &\left\{ \left(1-\rho\bar{\tau}\right) \big( \mu_{p} \pi(R|r) + (1-\mu_{p}) \pi(L|l) \big) \right\} ,
		\\
		\text{s.t. } &\pi(R|r), \pi(L|l) \text{ are given by \eqref{eq:cond_prob_tbar}}.
	\end{aligned}
\end{equation}
We use $\tau^{*}$ to denote the solution to this problem.

To begin with, note that $\tau^{*} \geq 0$ (due to limited liability) and $\tau^{*} < 1/\rho$ (otherwise the principal's payoff is zero or negative, hence such $\tau^{*}$ are dominated by $\bar{\tau}=0$).
Further, if $\tau^{*}>0$, then $\pi = \pi_u$, since otherwise the principal could reduce $\bar{\tau}$ without affecting the agent's choice.

Let us define the principal's \textbf{relaxed contracting problem} as
\begin{equation} \label{eq:princ_OP_tbar2}
	\begin{aligned}
		\max_{\bar{\tau} \in \mathbb{R}} &\left\{ \left(1-\rho\bar{\tau}\right) \big( \mu_{p} \pi_u(R|r) + (1-\mu_{p}) \pi_u(L|l) \big) \right\} ,
		\\
		\text{s.t. } &\pi_u(R|r), \pi_u(L|l) \text{ are given by \eqref{eq:cond_prob_tbar}}.
	\end{aligned}
\end{equation}
It differs from the full problem \eqref{eq:princ_OP_tbar} in that it ignores the boundary constraints $\bar{\tau} \geq 0$ and $\pi(R|r), \pi(L|l) \in [0,1]$. We use $\tau^{**}$ to denote the interior solution of this relaxed problem, whenever it exists.
So far, we can conclude that the principal's problem \eqref{eq:princ_OP_tbar} is solved by $\tau^{*} \in \{ 0, \tau^{**} \}$. 
The local maximizer $\tau^{**}$ is optimal if it satisfies all of the following three properties (and $\tau^{*}=0$ otherwise):\footnote{Feasibility and preferability should be self-explanatory. Effectiveness means that the incentive payment is effective at inducing the agent to acquire a non-trivial amount of information. Note that $\bar{\tau}=0$ is effective when the agent acquires information in the absence of a transfer.}
\begin{description}
	\item[Feasibility:] $\tau^{**}$ exists and $\tau^{**} \in [0, 1/\rho]$.\footnote{Note that $\tau^{**} \leq 1/\rho$ is not an exogenous restriction, but is rather implied by preferability, as established previously. It is, however, convenient to include this an explicit restriction.}
	\item[Effectiveness:] $\tau^{**}$ generates $\pi = \pi_u$.
	\item[Preferability:] $\tau^{**}$ is preferred to $\bar{\tau} = 0$.
\end{description}

The FOC of problem \eqref{eq:princ_OP_tbar2} (that must be solved by $\tau^{**}$) is given by
\begin{equation} \label{eq:FOC_tbar}
	\mu_p \frac{1-\mu}{\mu} + (1-\mu_p) \frac{\mu}{1-\mu} = e^{\frac{1+\bar{\tau}}{\lambda}} \cdot \frac{ \lambda \rho \left( e^{2 \frac{1+\bar{\tau}}{\lambda}} - 1 \right) + 2(1-\rho\bar{\tau}) }{ \lambda \rho \left( e^{2 \frac{1+\bar{\tau}}{\lambda}} - 1 \right) + \left( e^{2 \frac{1+\bar{\tau}}{\lambda}} + 1 \right) (1-\rho\bar{\tau}) }.
\end{equation}
Let $\gamma(\mu, \mu_p)$ denote the LHS and $\chi(\bar{\tau})$ the RHS of \eqref{eq:FOC_tbar}, respectively. Note that $\chi(\bar{\tau})$ is continuous in $\bar{\tau}$ and only depends on $\bar{\tau}$, $\lambda$, and $\rho$, but not on $\mu$ or $\mu_p$. Further, Lemma \ref{lem:chi} below shows that if $\rho \geq \min \left\{1, \frac{1}{2\lambda}\right\}$ then for all $\lambda$, $\chi(\bar{\tau})$ is increasing in $\bar{\tau} \in [0, 1/\rho]$ (recall that $\tau^{**} > 1/\rho$ obviously violates preferability, hence we drop this case). We maintain this restriction on $\rho$ throughout the rest of the proof. Monotonicity implies that a feasible $\tau^{**}$ exists for given $\mu, \mu_p, \lambda, \rho$ if and only if $\chi(0) \leq \gamma(\mu,\mu_p) \leq \chi(1/\rho)$, where the ``if'' part follows from the intermediate value theorem, and the ``only if'' part follows from $\tau \geq 1/\rho$ never being optimal. The strict monotonicity of $\chi(\bar{\tau})$ also means that the objective function in \eqref{eq:princ_OP_tbar} is strictly concave in $\bar{\tau}$, so if $\tau^{**}$ exists, then it is unique and it is a local maximizer of \eqref{eq:princ_OP_tbar}. 

\begin{lemma} \label{lem:chi}
	Function $\chi(\bar{\tau})$ is continuous and increasing in $\bar{\tau} \in \left[0, 1/\rho \right)$ for all $\lambda$ and all $\rho \geq \min \left\{1, \frac{\lambda}{2} \right\}$.
\end{lemma}
\begin{proof}
	Denote $\xi = \xi(\tau,\lambda) \equiv e^{\frac{1+\tau}{\lambda}}$.
	For sake of brevity we drop the arguments of $\xi(\tau,\lambda)$ and the bar from $\bar{\tau}$ throughout the proof of this lemma. Then we can rewrite
	\begin{equation*}
		\chi(\tau) = \xi \frac{\lambda \rho (\xi^2-1) + 2(1- \rho\tau) }{ \lambda \rho(\xi^2-1) + (\xi^2+1)(1- \rho\tau) }.
	\end{equation*}
	This function is trivially continuous and differentiable in $\tau \in [0,1/\rho]$. Hence it suffices to show that $\frac{d \chi(\tau)}{d \tau} > 0$:
	\begin{align*}
		\frac{d \chi(\tau)}{d \tau}
		=& \frac{ \left( \lambda \rho (3\xi^2 - 1 ) + 2(1-\rho\tau) \right) \frac{\partial \xi}{\partial \tau} - 2\rho\xi }{ \lambda \rho(\xi^2-1) + (\xi^2+1)(1-\rho\tau) } 
		\\
		&- \left[ 2\xi (\lambda\rho + 1-\rho\tau) \frac{\partial \xi}{\partial \tau} - \rho(\xi^2+1) \right] \cdot \frac{ \xi \left[ \lambda\rho (\xi^2-1) + 2(1-\rho\tau) \right] }{ \left[ \lambda\rho(\xi^2-1) + (\xi^2+1)(1-\rho\tau) \right]^2 }
		\\ 
		=& \frac{\xi \left( \xi^2 -1 \right) }{\lambda} \cdot \frac{ 2 \left( \xi^2 - 1 + 2 \frac{1-\rho\tau}{\lambda\rho} \right) + \frac{1-\rho\tau}{\lambda\rho} \cdot \left( \xi^2 - 1 - 2 \frac{1-\rho\tau}{\lambda\rho} \right) }{ \left[ (\xi^2-1) + (\xi^2+1)\frac{1-\rho\tau}{\lambda\rho} \right]^2 }.
	\end{align*}
	The latter expression is strictly positive for $\tau \in \left[0, 1/\rho \right)$ if and only if
	\begin{equation} \label{eq:chi_mon1}
		2 \left( \xi^2 - 1 + 2 \frac{1-\rho\tau}{\lambda\rho} \right) + \frac{1-\rho\tau}{\lambda\rho} \cdot \left( \xi^2 - 1 - 2 \frac{1-\rho\tau}{\lambda\rho} \right) > 0.
	\end{equation}
	The first term is strictly positive (since $\xi > 1$ and $\tau \leq 1/\rho$). The second term is nonnegative for the given range of $\tau$ if $\xi^2 - 1 \geq 2\frac{1-\rho\tau}{\lambda\rho}$. Note that $\xi^2 - 1 \geq 2\frac{1+\tau}{\lambda}$, hence \eqref{eq:chi_mon1} holds if $\frac{1+\tau}{\lambda} \geq \frac{1-\rho\tau}{\lambda\rho}$ for all $\tau \in \left[0, 1/\rho \right)$, which holds if $\rho \geq 1$.
	
	Alternatively, we can rewrite \eqref{eq:chi_mon1} as
	\begin{equation*}
		\left(\xi^2 - 1\right) \left( 2 + \frac{1-\rho\tau}{\lambda\rho} \right) + 2\frac{1-\rho\tau}{\lambda\rho} \left( 2 - \frac{1-\rho\tau}{\lambda\rho} \right) \geq 0.
	\end{equation*}
	In the above expression, the first term is again always strictly positive; the second term is nonnegative if $\lambda \rho \geq \frac{1}{2}$ (since $\tau \leq 1/\rho$).
	
	We thus conclude that if either $\rho \geq 1$, or $\rho \geq \frac{\lambda}{2}$, then $\frac{d \chi(\tau)}{d \tau} > 0$ for $\tau \in \left[0, 1/\rho \right)$, so $\chi(\tau)$ is indeed increasing in $\tau$ on that interval.
\end{proof}

Let us define the following cutoffs on $\mu$ that will prove helpful in establishing the properties of interest of $\tau^{**}$ (feasibility, effectiveness, and preferability):
\begin{enumerate}
	\item Let $\mu_{L1} \in (0, \mu^*)$ and $\mu_{R1} \in (\mu_p, 1)$ be such that $\gamma(\mu_{L1},\mu_p) = \gamma(\mu_{R1},\mu_p) = \chi(0)$. Lemma \ref{lem:mu1} below establishes that these cutoffs exist.
	
	\item Let $\mu_{L2} \equiv \max \left\{ \mu: \pi_{\mu}^{0}(L|l)=1\right\}$, $\mu_{R2} \equiv \min \left\{ \mu : \pi_{\mu}^{0}(R|r)=1\right\}$, where $\pi_{\mu}^{0}(L|l)$ and $\pi_{\mu}^{0}(R|r)$ stand for the respective probabilities \eqref{eq:cond_prob_tbar} given $\mu$ and $\overline{\tau}=0$. In words, $\mu_{L2}$ and $\mu_{R2}$ are the most extreme beliefs $\mu$ for which the agent voluntarily acquires information in the absence of incentive payment. Closed-form expressions can be obtained from \eqref{eq:cond_prob_tbar}, with $\mu_{L2} = \frac{1}{1+e^{\frac{1}{\lambda}}}$ and $\mu_{R2} = \frac{e^{\frac{1}{\lambda}}}{1+e^{\frac{1}{\lambda}}}$.
	
	\item Let $\mu_{L3} \equiv \inf \left\{ \mu : \tau^{*} > 0 \right\}$, $\mu_{R3} \equiv \sup \left\{ \mu : \tau^{*} > 0 \right\}$. In words, these denote the most extreme beliefs $\mu$ up to which the principal is willing to offer incentive contracts. 
	Lemma \ref{lem:mu3} below shows that $\mu_{L3}$ and $\mu_{R3}$ are always well-defined (i.e., that the set $\left\{ \mu : \tau^{*} > 0 \right\}$ is nonempty).
	
	\item Let $\mu_{L4} \equiv \max \left\{ \mu : \pi_{\mu}^{*}(L|l)=1\right\}$, $\mu_{R4} \equiv \min \left\{ \mu : \pi_{\mu}^{*}(R|r)=1\right\}$, where $\pi_{\mu}^{*}(L|l)$ and $\pi_{\mu}^{*}(R|r)$ stand for the respective probabilities \eqref{eq:cond_prob_tbar} given $\mu$ and $\overline{\tau}=\tau^{**}$. In words, $\mu_{L4}$ and $\mu_{R4}$ are the most extreme beliefs $\mu$ for which the agent acquires information given the candidate-optimal incentive $\tau^{**}$. Closed-form expressions can be obtained, with $\mu_{L4} = \frac{1}{1+e^{\frac{1+\tau^{**}}{\lambda}}}$ and $\mu_{R4} = \frac{e^{\frac{1+\tau^{**}}{\lambda}}}{1+e^{\frac{1+\tau^{**}}{\lambda}}}$.
	
	\item Let $\mu_{L5} \in (0, \mu^*)$ and $\mu_{R5} \in (\mu^*, 1)$ be such that $\gamma(\mu_{L5},\mu_p) = \gamma(\mu_{R5},\mu_p) = \chi(1/\rho)$. These cutoffs exists due to the properties of $\gamma(\mu,\mu_p)$ established for $\mu_{L1}$, as well as the fact that $\chi(1/\rho) = e^{\frac{1+1/\rho}{\lambda}} > 1$. Closed-form expressions can be obtained, with
	\begin{equation*}
		\mu_{L5},\mu_{R5} = \frac{ e^{\frac{1+1/\rho}{\lambda}} + 2\mu_{p} \mp \sqrt{e^{2\frac{1+1/\rho}{\lambda}} +4\mu_{p}(\mu_{p}-1)}}{2 \left( 1+e^{\frac{1+1/\rho}{\lambda}} \right)}.
	\end{equation*}
	As $\chi(0) < \chi(1/\rho)$ for all $\lambda$ (see Lemma \ref{lem:chi}), it follows that $\mu_{L5} < \mu_{L1}$ and $\mu_{R5} > \mu_{R1}$.
\end{enumerate}

We now proceed to establishing the conditions on $\mu$ for which the three properties of $\tau^{**}$ (feasibility, effectiveness, preferability) do or do not hold.
To begin with, as was previously claimed, a \emph{feasible} $\tau^{**}$ exists if and only if $\chi(0) \leq \gamma(\mu,\mu_p) \leq \chi(1/\rho)$, which, due to the monotonicity of $\chi(\bar{\tau})$ in $\bar{\tau}$, is equivalent to
\begin{equation} \label{eq:cond_both}
	\mu \in [\mu_{L5}, \mu_{L1}] \cup [\mu_{R1}, \mu_{R5}].
\end{equation}
As shown by construction above, $\mu_{L5}, \mu_{R5}$ are always well-defined and are located to the outside of $\mu_{L1}$ and $\mu_{R1}$, respectively. It thus remains to verify that $\mu_{L1}$ and $\mu_{R1}$ are also well-defined, which is done by the following lemma.

\begin{lemma} \label{lem:mu1}
	For all $\lambda$, if $\rho \geq \min \left\{1, \frac{\lambda}{2} \right\}$, then $\mu_{L1}$ and $\mu_{R1}$ exist.
\end{lemma}
\begin{proof}
	Function $\chi(\bar{\tau})$ is independent of $\mu$. Function $\gamma(\mu,\mu_p)$ is single-dipped in $\mu$, with $\min_{\mu} \gamma(\mu, \mu_p) = 2\sqrt{\mu_{p}\left(1-\mu_{p}\right)} < 1$ achieved at $\mu = \mu^*(\mu_p)$ as given by \eqref{eq:opt_deleg_binary}, and $\sup_{\mu} \gamma(\mu, \mu_p) = +\infty$ achieved by $\mu \to \{0,1\}$.
	Hence a sufficient condition for the cutoffs of interest to exist is 
	\begin{equation}
		\chi(0) \geq \gamma(\mu^{*}(\mu_{p}),\mu_{p}).
		\label{eq:rho_tilde}
	\end{equation}
	In the inequality above, only $\chi(0)$ depends on $\rho$. Note further that $\frac{d\chi(0)}{d\rho} > 0$. Therefore, if \eqref{eq:rho_tilde} holds for some $\tilde{\rho}$, then it also holds -- and, consequently, $\mu_{L1}$ and $\mu_{R1}$ exist -- for all $\rho \geq \tilde{\rho}$.
	
	Observe that $\chi(0) \geq 1$ for $\rho = \frac{1}{2\lambda}$: denoting $\xi = \xi(\lambda) \equiv e^{\frac{1}{\lambda}}$, we have
	\begin{align}
		\chi(0)
		= \xi \frac{\lambda \rho (\xi^2-1) + 2 }{ \lambda \rho(\xi^2-1) + (\xi^2+1) } &\geq 1
		\nonumber
		\\ \iff 
		(\xi -1)^2 \left( \lambda \rho (\xi+1) - 1 \right) &\geq 0
		\label{eq:mu1_suffcond1} 
	\end{align} 
	Since $\xi = e^{\frac{1}{\lambda}} \geq 1$, a sufficient condition for \eqref{eq:mu1_suffcond1} is given by 
	\begin{align} \label{eq:mu1_suffcond2}
		\rho &\geq \frac{1}{\lambda(\xi+1)},
	\end{align}
	which obviously holds if $\rho \geq \frac{1}{2\lambda}$.
	Further, $e^{\frac{1}{\lambda}} \geq 1 + \frac{1}{\lambda} \iff \lambda (\xi + 1) \geq 1$, hence the RHS of \eqref{eq:mu1_suffcond2} is weakly smaller than $1$, so the inequality also holds for all $\rho \geq 1$.
	
	We conclude that if $\rho \geq \min \left\{1, \frac{1}{2\lambda}\right\}$, then $\chi(0) \geq 1$, and hence \eqref{eq:rho_tilde} is satisfied and the relevant cutoffs exist.
\end{proof}

Moving on to \emph{effectiveness}, from \eqref{eq:precisions_binary}, it is immediate  that for a feasible $\tau^{**}$ to yield interior choice probabilities \eqref{eq:cond_prob_tbar}, it must be that $\mu \in [\mu_{L4}, \mu_{R4}]$. The following lemma establishes the location of $\mu_{L4},\mu_{R4}$ relative to other cutoffs.
\begin{lemma} 
	Cutoffs $\mu_{L4}$ and $\mu_{R4}$ are such that $\mu_{L4} \in [\mu_{L5}, \mu_{L2}]$ and $\mu_{R4} \in [\mu_{R2}, \mu_{R5}]$.
\end{lemma}
\begin{proof}
	Denoting $\mu_L(\bar{\tau}) \equiv \max \left\{ \mu : \pi_{\mu}(L|l,\bar{\tau})=1\right\} = \frac{1}{1+e^{\frac{1+\overline{\tau}}{\lambda}}}$ and observing that it is strictly decreasing in $\overline{\tau}$, we get $\mu_L(0) \geq \mu_L(\tau^{**}) \geq \mu_L(1/\rho)$, which is equivalent to $\mu_{L2} \geq \mu_{L4} \geq \mu_L(1/\rho)$.
	Routine calculations using the closed-form expression for $\mu_{L5}$ then demonstrate that $\mu_L(1/\rho) \geq \mu_{L5}$, implying that in the end, $\mu_{L4} \in [\mu_{L5}, \mu_{L2}]$.
	The result for $\mu_{R4}$ is shown analogously.
\end{proof}

Finally, we need to establish when the principal \emph{prefers} a feasible $\bar{\tau}=\tau^{**}$ to $\bar{\tau}=0$, which is done by the following lemma. 

\begin{lemma} \label{lem:mu3}
	The principal weakly prefers a feasible $\bar{\tau} = \tau^{**}$ to $\bar{\tau}=0$ if and only if $\mu \in [\mu_{L3}, \mu_{R3}]$. Further, these cutoffs satisfy $\mu_{L3} \in [\mu_{L4},\mu_{L2}]$ and $\mu_{R3} \in [\mu_{R2},\mu_{R4}]$.
\end{lemma}
\begin{proof}
	For $\mu \in [\mu_{L4}, \mu_{L2}]$, the
	principal compares his payoff from choosing $\overline{\tau}=0$,
	given by $1-\mu_{p}$, and his payoff from choosing $\overline{\tau}=\tau^{**}$. Thus, $\mu_{L3}$ satisfies the following indifference condition
	\begin{equation}
		\mathbb{E} [u(a,\omega) \mid \mu_p, \tau^{**}] \equiv \big( 1-\rho \tau^{**}(\mu) \big) \big(\mu_{p}\pi_{\mu}^{*}(R|r)+(1-\mu_{p})\pi_{\mu}^{*}(L|l)\big)=1-\mu_{p}.\label{eq:mu_prime_eqn}
	\end{equation}
	
	The LHS of \eqref{eq:mu_prime_eqn} is single-peaked in $\mu$:
	\begin{align*}
		\frac{d\mathbb{E} [u(a,\omega) \mid \mu_p, \tau^{**}] }{d\mu}
		&= \frac{\partial \mathbb{E} [u(a,\omega) \mid \mu_p, \tau^{**}] }{\partial \mu}
		\\
		&=
		\left(1-\rho\tau^{**}(\mu)\right) \left(\mu_{p}\frac{\partial\pi_{\mu}^{*}(R|r)}{\partial\mu}+(1-\mu_{p})\frac{\partial\pi_{\mu}^{*}(L|l)}{\partial\mu}\right)
		\\
		&= \left(1-\rho\tau^{**}(\mu)\right) \frac{ e^{2\frac{1+\bar{\tau}}{\lambda}} }{ e^{\frac{2\left(1+\bar{\tau}\right)}{\lambda}}-1 } 
		\left( \frac{\mu_p}{\mu^2} - \frac{1-\mu_p}{(1-\mu)^2} \right)
	\end{align*}
	where the first equality follows from the envelope theorem. The final expression is strictly positive for $\mu < \mu^*(\mu_p)$ and strictly negative for $\mu > \mu^*(\mu_p)$, hence the single-peakedness follows.
	
	Thus, $\frac{d}{d\mu} \mathbb{E} [u(a,\omega) \mid \mu_p, \tau^{**}] > 0$ for $\mu \in [\mu_{L4}, \mu_{L2}]$ (since also $\mu_{L2} < \frac{1}{2} \leq \mu^*$).
	Hence we can show that $\mu_{L3}\in[\mu_{L4},\mu_{L2}]$ by establishing that
	$$\mathbb{E} [u(a,\omega) \mid \mu_p, \tau^{**}; \mu=\mu_{L4}] \leq 1-\mu_{p} \leq \mathbb{E} [u(a,\omega) \mid \mu_p, \tau^{**}; \mu=\mu_{L2}]$$ 
	and applying the intermediate value theorem.
	The first inequality follows from the fact that at $\mu_{L4}$, $\tau^{**}$ is such that the agent does not acquire information, yet the principal pays a positive transfer to him (which is trivially dominated by $\bar{\tau}=0$). The second inequality follows from the fact that given $\mu=\mu_{L2}$, \eqref{eq:FOC_tbar} holds for all $\bar{\tau} \in \left[ 0, 1/\rho \right]$, so if $\tau^{**}$ exists, it is preferred to $\bar{\tau}=0$.
	We conclude that $\mu_{L3}\in[\mu_{L4},\mu_{L2}]$, and the mirror argument can establish that $\mu_{R3}\in[\mu_{R2},\mu_{R4}]$.
	
	Finally, the single-peakedness of $\mathbb{E} [u(a,\omega) \mid \mu_p, \tau^{**}]$ in $\mu$ implies that $\bar{\tau} = \tau^{**}$ is preferred to $\bar{\tau}=0$ for all $\mu \in [\mu_{L3}, \mu_{R3}]$.
\end{proof} 

To summarize, the principal's problem \eqref{eq:principal_problem_contracts} is solved by $\tau^{*} \in \{ 0, \tau^{**}\}$, with $\tau^{**}$ being the solution if and only if it is feasible, effective, and preferable. It is feasible if and only if \eqref{eq:cond_both} holds; effective if and only if $\mu \in [\mu_{L4}, \mu_{R4}]$, and preferable if and only if $\mu \in [\mu_{L3}, \mu_{R3}]$. Further, we have established that $\mu_{L5} \leq \mu_{L4} \leq \mu_{L3} \leq \mu_{L2}$ (and the converse holds for the other set of cutoffs), as well as $\mu_{L1} < \mu^* \leq \mu_p < \mu_{R1}$. Therefore, $\tau^{*} = \tau^{**}$ if and only if $\mu \in [\mu_{L3}, \mu_{L1}] \cup [\mu_{R1}, \mu_{R3}]$, whenever these intervals are non-empty.
After denoting $\hat{\mu}_{L} \equiv \mu_{L3}, \hat{\mu}_{R}\equiv \mu_{R3}$, $\bar{\mu}_{L} \equiv \max \{\mu_{L3}, \mu_{L1}\}, \bar{\mu}_{R} \equiv \min \{\mu_{R1}, \mu_{R3}\}$ and excluding the endpoints, at which $\tau^{**}=0$, we obtain the statement of the Proposition.

\subsubsection{Proof of Proposition \ref{prop:bqp}}

The proof is largely analogous to that of Lemma \ref{lem:agents_soln_ups} and Proposition \ref{prop:PScost}, and proceeds in two main parts: we first solve the agent's problem in the binary-quadratic setting, and then we proceed to analyze the principal's problem.

\paragraph*{The agent's problem.}
We solve the agent's problem in the binary-quadratic problem using the posterior approach outlined in the proof of Lemma \ref{lem:agents_soln_ups}. In what follows, we suppress signal realizations $s$ and refer to them according to the posterior beliefs $\eta(s)$ that they induce.
It can be easily seen from \eqref{eq:c_entropy} that the cost of a direct signal structure $\phi$ can be expressed as
\begin{equation*}
	c(\phi,\mu) = \lambda\Big[ - \mathbb{E}[H(\eta)|\phi] + H(\mu) \Big],
\end{equation*}
where $H(\mu) \equiv \mu \ln \mu + (1-\mu) \ln (1-\mu)$ is the entropy of belief $\mu$.

Under the quadratic preferences $u(a,\omega) = -(a-\omega)^2$, given some posterior belief $\eta$, the agent's optimal choice rule is $\sigma^*(\eta)=\eta$. His interim expected payoff given posterior $\eta$, thus, equals $\mathbb{E}[u(\eta,\omega) | \eta] = -\eta(1-\eta)$. Therefore, the agent's problem is given by
\begin{equation}\label{eq:bq_agent_problem}
	\begin{gathered}
		\max_{\phi \in \Phi_{\mu}} \mathbb{E} \Big[-\eta(1-\eta)+\lambda H(\eta) - \lambda H(\mu) \mid \phi \Big].
	\end{gathered}
\end{equation}
Since the last term $-\lambda H(\mu)$ does not affect the maximization, it can be safely ignored.
Then we can define the agent's \emph{net utility} as $v(\eta) \equiv -\eta(1-\eta)+\lambda H(\eta)$. Problem \eqref{eq:bq_agent_problem} is then equivalent to $\max_{\phi \in \Phi_{\mu}} \mathbb{E} [v(\eta) | \phi]$, which can be solved using a concavification approach \citep{kamenica2011bayesian,caplin2022rationally}. 

\begin{figure}
	\centering 
	\subfloat[][$\lambda\geq\frac{1}{2}$]{
		\includegraphics[width=0.48\textwidth]{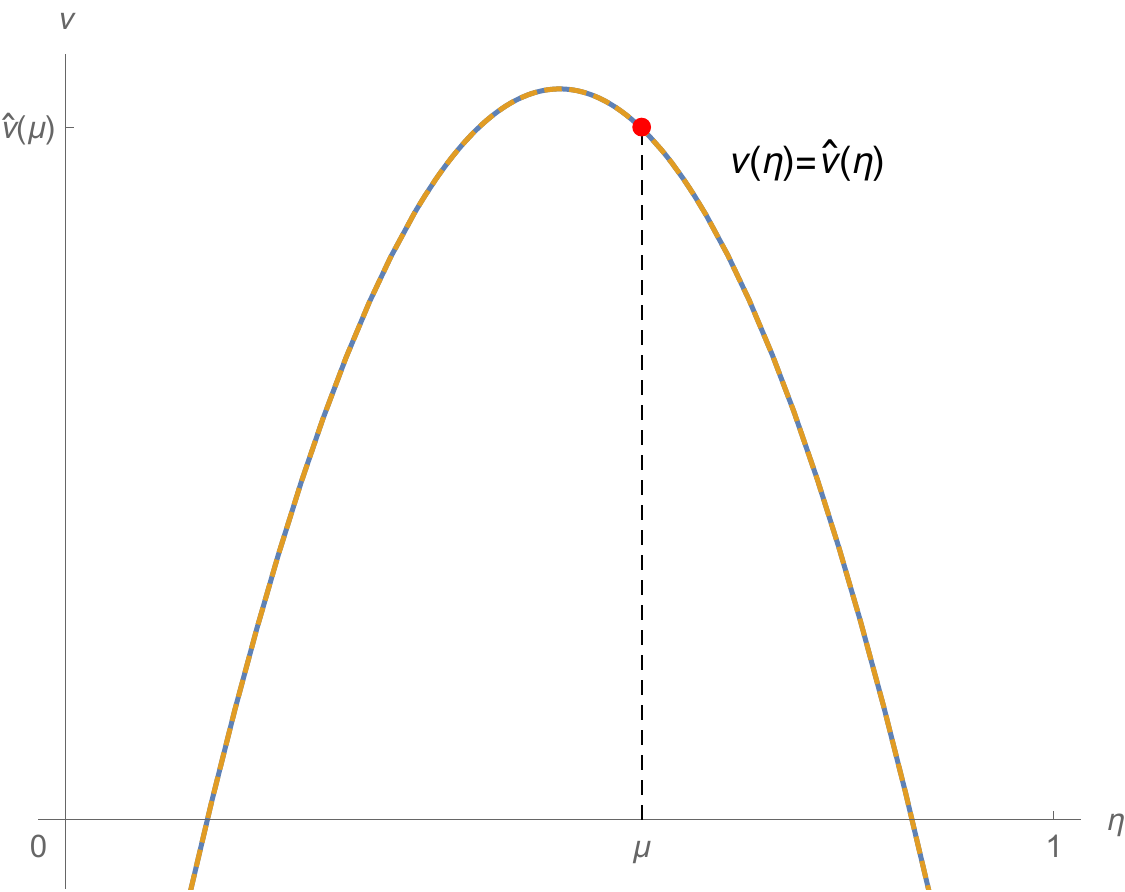}
	}
	\hfill 
	\subfloat[][$\lambda<\frac{1}{2}$]{
		\includegraphics[width=0.48\textwidth]{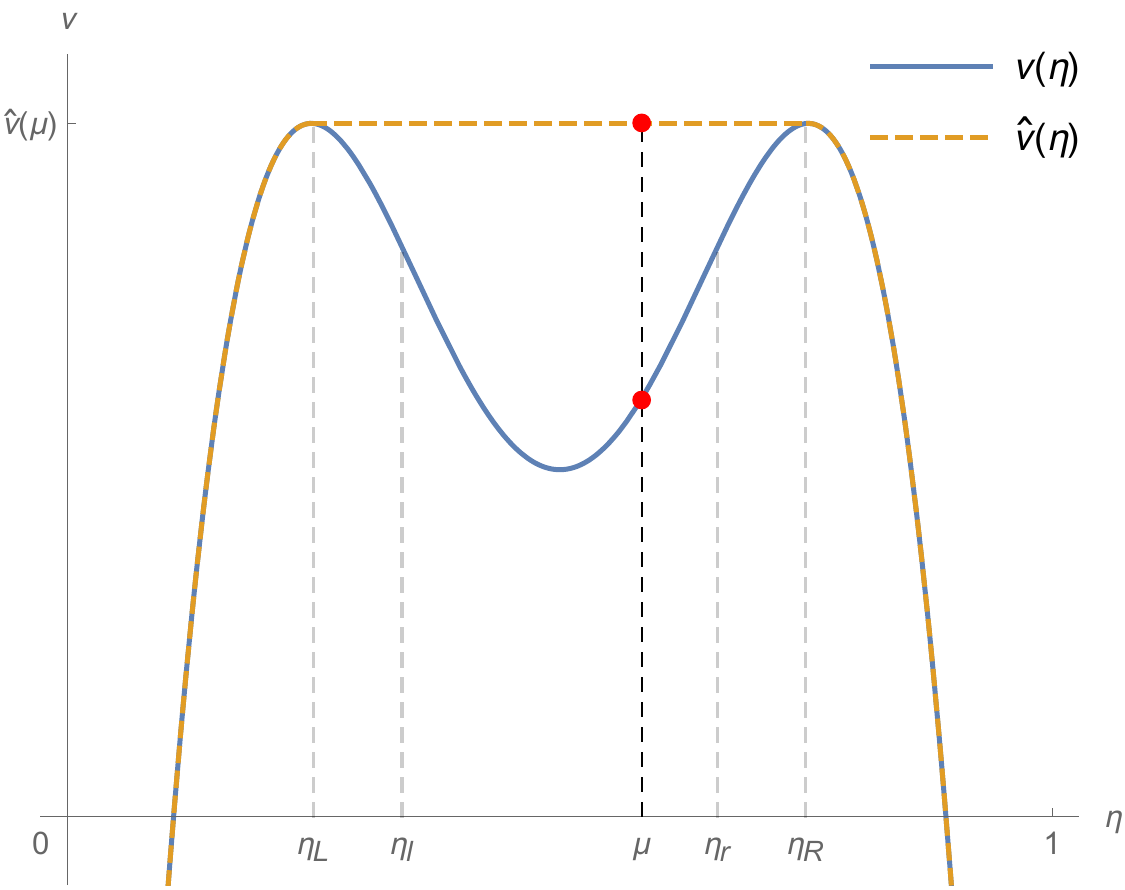}
	}
	\caption{The net utility function $v(\eta)$ and the concavified net utility function $\hat{v}(\eta)$.}
	\label{fig:concav} 
\end{figure}

In particular, let $\hat{v}(\eta)$ denote the \emph{concavified net utility function} (also known as the concave closure of $v(\eta)$; \citealp{rockafellar1970convex}), which is defined as the minimal concave function that majorizes all net utilities. We characterize $\hat{v}(\eta)$ by establishing some properties of $v(\eta)$ and its derivatives $v',v''$. Note first that $v(\eta)$ is symmetric around $0.5$: for any $\eta_1, \eta_2$ s.t. $\eta_1 + \eta_2 = 1$, it is true that $v(\eta_1) = v(\eta_2)$, $v'(\eta_1) = -v'(\eta_2)$, $v''(\eta_1) = v''(\eta_2)$. 
Further, simple algebra shows that $v''(\eta)=2-\lambda\frac{1}{\eta(1-\eta)}$. Two cases are then possible, as depicted in Figure \ref{fig:concav}.

Case 1: $\lambda \geq 0.5$.
Then $v''(\eta)\leq 0$ and function $v(\eta)$ is concave on $[0,1]$, meaning $\hat{v}(\mu)=v(\mu)$ for all $\mu$. The results of \cite{caplin2022rationally} then imply that the agent chooses degenerate distribution $\phi$ that puts all mass on $\eta=\mu$ and does not acquire any information, regardless of $\mu$. This concludes the analysis of the agent's problem in case 1, and the remainder of this section is devoted to the analysis of the case when $\lambda < 0.5$.

Case 2: $\lambda < 0.5$.
Then equation $v''(\eta)=0$ has two roots, denote them as $\eta_{l}, \eta_{r}$, with $\eta_{l}<\eta_{r}$. Function $v(\eta)$ is concave on $[0,\eta_{l}) \cup (\eta_{r},1]$ and convex on $(\eta_{l},\eta_{r})$, and so its derivative $v'(\eta)$ is decreasing on $[0,\eta_{l}) \cup (\eta_{r},1]$ and increasing on $(\eta_{l},\eta_{r})$. Simple algebra shows that $v'(0.5)=0$. Together with the fact that $\lim\limits_{\eta\to 0+} v(\eta) = -\infty = \lim\limits_{\eta\to 1-} v(\eta)$, it implies that there exist two other roots of $v'(\eta)=0$ and they belong to the intervals $[0,\eta_{l}]$ and $[\eta_{r},1]$. We denote these roots as $\eta_{L}, \eta_{R}$, respectively, and the symmetry of $v(\eta)$ implies that $\eta_{L} = 1 - \eta_{R}$. 

The behavior of $v'(\eta)$ suggests that $\eta_{L}, \eta_{R}$ are local and global maxima of $v(\eta)$. Therefore, $\hat{v}(\eta)=v(\eta)$ if $\eta \in [0,\eta_{L}] \cup [\eta_{R},1]$, and if $\eta \in (\eta_{L},\eta_{R})$ then $\hat{v}(\eta)$ lies on the straight line connecting points $(\eta_{L},v(\eta_{L}))$ and $(\eta_{R},v(\eta_{R}))$. Therefore, if $\mu\in [0,\eta_{L}] \cup [\eta_{R},1]$, tangent lines to $v(\eta)$ and $\hat{v}(\eta)$ coincide, and the agent does not acquire information; if $\mu\in (\eta_{L},\eta_{R})$, the agent chooses $\phi$ with two posterior beliefs in the support: $\eta_{L}$ and $\eta_{R}$.

In the latter case, we can derive the optimal signal structure $\phi^*$ in closed form. Conditional on $\eta_{L}$ and $\eta_{R}$, expressions \eqref{eq:bq_phistar_uncond}--\eqref{eq:bq_dphistar} apply in this context as well after replacing $\phi^*(\eta|1)$ and $\phi^*(\eta|0)$ with $\phi^*(\eta|r)$ and $\phi^*(\eta|l)$, respectively.

\paragraph*{The principal's problem.}
Begin by assuming $\lambda < 0.5$.
The principal's expected utility from hiring a learning agent with prior belief $\mu \in (\eta_L, \eta_R)$ is given by
\begin{align} \label{eq:bq_principal_obj}
	\mu_p \Big( \phi^*(\eta_R|1) \bar{u} + \phi^*(\eta_L|1) \underline{u} \Big) + (1-\mu_p) \Big( \phi^*(\eta_R|0) \underline{u} + \phi^*(\eta_L|0) \bar{u} \Big),
\end{align}
where $\bar{u} \equiv -\eta_L^2 = -(1-\eta_R)^2$ and $\underline{u} \equiv -\eta_R^2 = -(1-\eta_L)^2 < \bar{u}$.
If it is optimal to hire a learning agent, then the optimal agent's prior $\mu^*$ must maximize \eqref{eq:bq_principal_obj}, hence $\mu^*$ must solve the FOC given by
\begin{align*}
	\mu_p \frac{\eta_R \cdot \eta_L}{(\mu^*)^2 \cdot (\eta_R-\eta_L)} (\bar{u}-\underline{u}) &= (1-\mu_p) \frac{(1-\eta_R) \cdot (1-\eta_L)}{(1-\mu^*)^2 \cdot (\eta_R-\eta_L)} (\bar{u}-\underline{u})
	\\ \iff
	\frac{\mu^*}{1-\mu^*} &= \sqrt{ \frac{\mu_p}{1-\mu_p} },
\end{align*}
where the first line uses \eqref{eq:bq_dphistar}, and then the second line follows from $\eta_L = 1-\eta_R$. It is trivial to verify that the SOC also hold.
Representation \eqref{eq:opt_deleg_binary} therefore applies conditional on the principal hiring a learning agent.
To fully solve the principal's problem it is then left to characterize her choice between a learning and a non-learning agent.

Conditional on hiring a non-learning agent, it is trivially optimal for the principal to hire an aligned agent $\mu=\mu_p$ (since such an agent chooses $a \in \mathcal{A}$ to maximize $\mathbb{E}[u(a,\omega)|\mu_p]$, as opposed to some other function induced by an expectation w.r.t. another belief). 

Suppose w.l.o.g. $\mu_p \geq 0.5$ (the logic for $\mu_p < 0.5$ is analogous).
We first characterize the principal's optimal strategy for $\mu_p \in \left[ \eta_R, (\mu^*)^{-1}(\eta_R) \right)$ -- i.e., such $\mu_p$ that an agent with $\mu=\mu_p$ is not learning and an agent with $\mu^*(\mu_p)$ is learning.\footnote{Here $(\mu^*)^{-1}(\mu)$ is the inverse of function $\mu^*(\mu_p)$ given by \eqref{eq:opt_deleg_binary}.} 
We then show that when one of these is violated (either $\mu=\mu^*(\mu_p)$ does not learn, or $\mu=\mu_p$ does), an extrapolation of such a strategy is optimal.

If $\mu_p \in \left[ \eta_R, (\mu^*)^{-1}(\eta_R) \right)$, hiring an aligned agent $\mu=\mu_p$ yields the principal an expected payoff of $-\mu_p (1-\mu_p)$, while hiring agent $\mu^*(\mu_p)$ yields
\begin{multline*}
	-\mu_p \Big( \phi^*(\eta_R|1) (1-\eta_R)^2 + \phi^*(\eta_L|1) (1-\eta_L)^2 \Big) -(1-\mu_p) \Big( \phi^*(\eta_R|0) \eta_R^2 + \phi^*(\eta_L|0) \eta_L^2 \Big) =
	\\
	= -2 \eta_R(1-\eta_R) \sqrt{ \mu_p(1-\mu_p) },
\end{multline*}
where the equality uses \eqref{eq:bq_phistar_cond} and the fact that $\eta_R=1-\eta_L$.
The latter (hiring $\mu^*(\mu_p)$) is preferred if and only if
\begin{align}
	-2 \eta_R(1-\eta_R) \sqrt{ \mu_p(1-\mu_p) } & \geq -\mu_p(1-\mu_p)
	\nonumber
	\\ \iff \label{eq:quad_ic}
	2 \eta_R(1-\eta_R) & \leq \sqrt{ \mu_p(1-\mu_p) }.
\end{align}
Note that the RHS of the inequality above is strictly decreasing in $\mu_p \geq 0.5$, while the LHS does not depend on $\mu_p$. Furthermore,
\begin{itemize}
	\item if $\mu_p = \eta_R$, then \eqref{eq:quad_ic} holds: $2 \eta_R(1-\eta_R) \leq \sqrt{ \eta_R(1-\eta_R) } \iff \sqrt{ \eta_R(1-\eta_R) } \leq \frac{1}{2}$, which holds because $\eta_R(1-\eta_R) \leq \frac{1}{4}$ for any $\eta_R \in [0.5,1]$;
	
	\item if $\mu_p = (\mu^*)^{-1}(\eta_R) = \frac{\eta_R^2}{\eta_R^2 + (1-\eta_R)^2}$, then \eqref{eq:quad_ic} is equivalent to $\eta_R^2 + (1-\eta_R)^2 \leq \frac{1}{2} \iff 2 \left( \eta_R + \frac{1}{2} \right)^2 \leq 0,$ which fails to hold for any $\eta_R \in [0.5,1].$
\end{itemize}
By the intermediate value theorem, we conclude that there exists $\hat{\mu} \in \left[ \eta_R, (\mu^*)^{-1}(\eta_R) \right)$ such that \eqref{eq:quad_ic} is satisfied -- and delegation to a non-learning agent is preferred -- if and only if $\mu_p \geq \hat{\mu}$.

If $\mu_p \geq (\mu^*)^{-1}(\eta_R)$, then the optimal non-learning agent $\mu=\mu_p$ is still available, but the optimal learning agent $\mu=\mu^*(\mu_p)$ is not, hence (if the principal wants a learning agent, she has to deviate away from $\mu^*(\mu_p)$, and so) the principal has an even stronger preference towards hiring a non-learning agent than in the case above. The mirror logic also holds: if $\mu_p < \eta_R$, then the optimal learning agent is available, but the optimal non-learning agent is not, hence the principal has a stronger preference towards hiring a learning agent than in the case above.

We conclude that there exists $\hat{\mu} \in \left[ \eta_R, (\mu^*)^{-1}(\eta_R) \right)$ that solves \eqref{eq:quad_ic} w.r.t. $\mu_p$ such that the principal's optimal strategy is given by:
\begin{align*}
	\mu^{**}(\mu_p) = \begin{cases}
		\mu^*(\mu_p) & \text{ as given by \eqref{eq:opt_deleg_binary} if } \mu_p < \hat{\mu},
		\\
		\mu_p & \text{ if } \mu_p \geq \hat{\mu}.
	\end{cases}
\end{align*}
This concludes the proof for the case $\lambda < 0.5$. In case $\lambda \geq 0.5$, no agent acquires any information, hence by the logic above, hiring an agent $\mu = \mu_p$ is optimal. Setting $\hat{\mu} = 0.5$, we get the result in this case as well, which completes the proof.

\subsubsection{Proof of Proposition \ref{prop:impossibility}}

We provide an example for $N=3$. We use the same version of the Farkas' Lemma as in the proof of Lemma \ref{lem:equivalence}. To show that there is no prior belief that solves the system of the first-order conditions for the problem, it is sufficient to show that there is a solution to the following dual inequality system
\begin{equation} \label{eq:p8_1}
	\begin{cases}
		z_{1} e^\frac{u(a_{1},\omega_{1})}{\lambda} +z_{2} e^\frac{u(a_{1},\omega_{2})}{\lambda} + z_{3} e^\frac{u(a_{1},\omega_{3})}{\lambda} \geq 0, 
		\\
		z_{1} e^\frac{u(a_{2},\omega_{1})}{\lambda} +z_{2} e^\frac{u(a_{2},\omega_{2})}{\lambda} + z_{3} e^\frac{u(a_{2},\omega_{3})}{\lambda} \geq 0, 
		\\
		z_{1} e^\frac{u(a_{3},\omega_{1})}{\lambda} +z_{2} e^\frac{u(a_{3},\omega_{2})}{\lambda} + z_{3} e^\frac{u(a_{3},\omega_{3})}{\lambda} \geq 0, 
		\\
		z_{1}+z_{2}+z_{3} <0.
	\end{cases}
\end{equation}
Let us normalize $\lambda=1$ and consider payoffs given by the following matrix:
$$
\begin{pmatrix}
	u(a_{1},\omega_{1}) & u(a_{2},\omega_{1})  & u(a_{3},\omega_{1})\\
	u(a_{1},\omega_{2}) & u(a_{2},\omega_{2})  & u(a_{3},\omega_{2})\\
	u(a_{1},\omega_{3}) & u(a_{2},\omega_{3})  & u(a_{3},\omega_{3})
\end{pmatrix} = \begin{pmatrix}
	\ln 3 & 0 & \ln (2+\varepsilon)\\
	0 & \ln 3 & \ln (2+\varepsilon)\\
	0 & 0 & \ln (2+\varepsilon) 
\end{pmatrix}
$$
Notice that vector $(z_{1},z_{2},z_{3})=(-1-\delta,-1-\delta,2)$ for small enough $\delta,\varepsilon \geq 0$ solves system \eqref{eq:p8_1}: the two latter inequalities hold trivially for all such $z$, and the two former inequalities hold if $\varepsilon \geq 3^\frac{1+\delta}{2} - 2$. 
Therefore, there exists no $\mu$ that solves system \eqref{eq:t1_2} given $\beta \in \varDelta(\mathcal{A})$.

\end{document}